{}%disable hyper at arxiv
{}%use pdflatex at arxiv

\newif\ifpublic\publictrue
\newif\ifarxiv\arxivtrue

\documentclass[12pt,a4paper]{article}

\usepackage{amsmath,amssymb}
\ifpublic\else\usepackage{showkeys}\fi
\usepackage{graphicx}
\usepackage[bookmarks=true,hyperfigures=true]{hyperref}
\usepackage[usenames,dvipsnames]{color}
\usepackage[nosort]{cite}
\usepackage[bulletsep]{collref}
\usepackage[utf8]{inputenc}

\def\showkeysrefformat#1{{\normalfont\tiny\ttfamily#1}}
\makeatletter
\def\SK@@ref#1>#2\SK@{%
 {\@inlabelfalse\leavevmode\vbox to\z@{%
 \vss\SK@refcolor\rlap{\vrule\raise .75em%
  \hbox{\showkeysrefformat{#2}}}}}}
\makeatother

\usepackage[a4paper,text={160mm,247mm},centering]{geometry}
\RequirePackage{verbatim}

\makeatletter
\newwrite\bibinl@out
\newenvironment{bibtex}[1][\jobname]{%
  \immediate\openout\bibinl@out #1.bib
  \immediate\write\bibinl@out{\@percentchar generated from `\jobname' starting line \the\inputlineno^^J}%
  \def\verbatim@processline{\immediate\write\bibinl@out{\the\verbatim@line}}%
  \@bsphack\let\do\@makeother\dospecials\catcode`\^^M\active\verbatim@start
}%
{\immediate\closeout\bibinl@out\@esphack}
\makeatother

\allowdisplaybreaks[3]

\numberwithin{equation}{section}

\usepackage[font=small,labelfont=bf,width=0.85\textwidth]{caption}

\newcommand{\alignrel}[1][=]{\mathrel{\phantom{#1}}{}}
\makeatletter
\newcommand{\eqsep}{\@ifstar{\hspace{1em}}{\hspace{2em minus 1em}}}
\newcommand{\@eqjoin}[2]{\hspace{#1}#2\hspace{#1}}
\newcommand{\eqjoin}{\@ifstar{\@eqjoin{1em}}{\@eqjoin{2em minus 1em}}}
\makeatother

\renewcommand{\minalignsep}{2em}

\makeatletter
\expandafter\def\expandafter\calc@shift@align\expandafter
  {\expandafter\hook@align@sep\calc@shift@align}
\def\hook@align@sep{\ifnum\xatlevel@=\@ne
  \dimen@\displaywidth
  \advance\dimen@-\totwidth@
  \@tempcntb\maxfields@
  \divide\@tempcntb\tw@
  \@tempcnta\@tempcntb
  \advance\@tempcntb\m@ne
  \global\eqnshift@\dimen@
  \global\alignsep@\minalignsep\relax
  \global\advance\eqnshift@-\@tempcntb\alignsep@
  \global\divide\eqnshift@\tw@
  \ifdim\eqnshift@<\z@
   \global\eqnshift@\z@
  \fi
 \fi}
\makeatother

\makeatletter
\expandafter\def\expandafter\align\expandafter
  {\expandafter\hook@align@\align}
\expandafter\def\expandafter\math@cr@@@align\expandafter
  {\expandafter\hook@align@cr\math@cr@@@align}
\expandafter\def\expandafter\endalign\expandafter
  {\expandafter\hook@align@end\endalign}
\expandafter\def\expandafter\measure@\expandafter#\expandafter1\expandafter
  {\measure@{#1}\hook@align@}
\def\hook@align@{\gdef\hook@align@end{\donumber}%
  \gdef\hook@align@cr{\nonumber}}
\def\hook@align@cr{}
\def\hook@align@end{}

\newcommand{\donumber}{\gdef\hook@align@cr{\gdef\hook@align@cr{\nonumber}}}%
\newcommand{\numberhere}{\donumber\gdef\hook@align@end{}}%
\makeatother

\ifarxiv\PassOptionsToPackage{write=false,compile=false}{mpostinl}\fi
\ifpublic\else\PassOptionsToPackage{warnunused=true}{mpostinl}\fi
\ifpublic\PassOptionsToPackage{labelnames}{mpostinl}\fi
\PassOptionsToPackage{clean=false}{mpostinl}
\RequirePackage{graphbox}
\RequirePackage{mpostinl}

\newcommand{\mpostusemath}[2][]{\mathinner{\mpostuse[align,#1]{#2}}}

\begin{mposttex}
\usepackage{amsmath,amssymb}
\usepackage[utf8]{inputenc}
\usepackage[usenames, dvipsnames]{color}
\end{mposttex}

\begin{mpostdef}
def pensize(expr s)=withpen pencircle scaled s enddef;
def fillshape(expr p,c)=
  fill p withcolor c;
  draw p
enddef;
def drawdot(expr z,s)=
  fill fullcircle scaled s shifted z
enddef;
def filldot(expr z,s,ci)=
  fillshape(fullcircle scaled s shifted z, ci)
enddef;
def drawcross(expr z,s,r,t,c)=
  draw ((-0.5,-0.5)--(+0.5,+0.5)) scaled s rotated r shifted z pensize(t) withcolor c;
  draw ((+0.5,-0.5)--(-0.5,+0.5)) scaled s rotated r shifted z pensize(t) withcolor c;
enddef;
\end{mpostdef}

\begin{mpostdef}
pair vpos[];
numeric nums[];
path paths[];
newinternal numeric xu,yu;
xu:=1cm;
ahlength:=4pt;
\end{mpostdef}

\begin{mpostdef}
def midarrowperc (expr p, t) =
  fill arrowhead subpath(0,arctime(t*arclength(p)+0.5ahlength) of p) of p;
enddef;
def midarrow (expr p, t) =
  fill arrowhead subpath(0,arctime(arclength(subpath (0,t) of p)+0.5ahlength) of p) of p
enddef;
\end{mpostdef}

\begin{mpostdef}
def lambdapath(suffix i)(expr a,b,s)(text f)=
for i:=a step s until (b+0.5s): if i<>a: .. fi f endfor
enddef;
\end{mpostdef}

\begin{mposttex}[dual]
\newcommand{\abstraction}[1]{#1}
\end{mposttex}

\ifpublic\else
\begin{mposttex}[dual]
\definecolor{abstract}{rgb}{0,0.4,0}
\renewcommand{\abstraction}[1]{\textcolor{abstract}{#1}}
\end{mposttex}
\fi

\begin{mposttex}[dual]

\usepackage{mathfixs}
\ProvideMathFix{autobold,greekcaps,frac,root}
\ProvideMathFix{multskip}

\RequirePackage[extdef]{delimset}

\ProvideMathFix{vfrac,rfrac}
\newcommand{\iunit}{\abstraction{\mathring{\imath}}}
\newcommand{\eunit}{\mathrm{e}}
\newcommand{\half}{\rfrac{1}{2}}
\newcommand{\ihalf}{\rfrac{\iunit}{2}}
\newcommand{\quarter}{\rfrac{1}{4}}

\newcommand{\Real}{\mathbb{R}}
\newcommand{\Complex}{\mathbb{C}}
\newcommand{\ComplexComplete}{\bar{\mathbb{C}}}
\newcommand{\Sphere}{\mathrm{S}}
\newcommand{\Integer}{\mathbb{Z}}

\newcommand{\vdot}{\mathord{\cdot}}
\newcommand{\vtimes}{\mathord{\times}}
\newcommand{\vecsp}[1]{\vec{\mkern0mu#1\mkern1mu}}
\makeatletter
\def\vect{\@ifstar{\vecsp}{\vec}}
\makeatother
\newcommand{\trans}{{\scriptscriptstyle\mathsf{T}}}
\newcommand{\conj}{{*}}
\newcommand{\hconj}{{\dagger}}

\newcommand{\grp}[1]{\mathrm{#1}}

\newcommand{\pauli}{\abstraction{\sigma}}
\newcommand{\paulivec}{\vect{\pauli}}

\newcommand{\spinvar}{\abstraction{S}}
\newcommand{\spinvec}{\vect{\spinvar}}

\newcommand{\spincp}{\abstraction{\zeta}}

\newcommand{\kdvfld}{\abstraction{\phi}}

\newcommand{\laxval}{\abstraction{\tau}}
\newcommand{\laxvec}{\abstraction{\Psi}}
\newcommand{\laxfunc}{\abstraction{\psi}}
\newcommand{\laxcp}{\abstraction{\varpi}}
\newcommand{\laxconn}{\abstraction{A}}
\newcommand{\laxtransport}{\abstraction{W}}
\newcommand{\laxmono}{\abstraction{T}}
\newcommand{\laxscat}{\abstraction{S}}
\newcommand{\laxpar}{\abstraction{u}}
\newcommand{\laxzpar}{\abstraction{z}}
\newcommand{\laxqmom}{\abstraction{q}}
\newcommand{\ptsqrt}[1]{\abstraction{\hat{#1}}}
\newcommand{\ptcrit}[1]{\abstraction{#1_*}}
\newcommand{\ptsol}[1]{\abstraction{\breve{#1}}}
\newcommand{\ptdiv}[1]{\abstraction{\mathring{#1}}}
\DeclareMathOperator{\pathordleft}{\overset{\leftarrow}{P}}
\newcommand{\parfreq}{\abstraction{\omega}}
\newcommand{\parwnum}{\abstraction{\kappa}}
\newcommand{\parvel}{\abstraction{v}}
\newcommand{\parscl}{\abstraction{\alpha}}
\newcommand{\paroff}{\abstraction{\beta}}
\newcommand{\parsolb}{\abstraction{b}}
\newcommand{\parsolc}{\abstraction{c}}
\newcommand{\parsold}{\abstraction{d}}
\newcommand{\parsoleta}{\abstraction{\eta}}
\newcommand{\parpos}{\abstraction{y}}
\newcommand{\parang}{\abstraction{\phi}}
\newcommand{\parqper}{\abstraction{\lambda}}
\newcommand{\winbeg}{\abstraction{x_-}}
\newcommand{\winend}{\abstraction{x_+}}
\newcommand{\winbegend}{\abstraction{x_\mp}}
\newcommand{\winbegL}{\abstraction{x_{-,\fldlen}}}
\newcommand{\winendL}{\abstraction{x_{+,\fldlen}}}

\newcommand{\fldlen}{\abstraction{L}}
\newcommand{\potshift}{\abstraction{Q}}
\newcommand{\totmom}{\abstraction{P}}
\newcommand{\eng}{\abstraction{E}}
\newcommand{\actvar}{\abstraction{I}}
\newcommand{\modenum}{\abstraction{n}}
\newcommand{\angmom}{\abstraction{J}}

\newcommand{\contacyc}{\abstraction{\mathcal{A}}}
\newcommand{\contbcyc}{\abstraction{\mathcal{B}}}

\newcommand{\ismmom}{\abstraction{k}}
\newcommand{\ismscat}{\abstraction{S}}
\newcommand{\ismtrans}{\abstraction{\tau}}
\newcommand{\ismrefl}{\abstraction{\rho}}
\newcommand{\ismsing}{\abstraction{\ptsol\ismmom}}
\newcommand{\ismres}{\abstraction{\mu}}

\DeclareMathOperator{\sech}{sech}
\DeclareMathOperator{\arcosh}{arcosh}
\DeclareMathOperator{\arsinh}{arsinh}

\newcommand{\ellS}{\text{s}}

\newcommand{\ellN}{\text{n}}

\DeclareMathOperator{\ellSN}{sn}
\DeclareMathOperator{\ellCN}{cn}
\DeclareMathOperator{\ellDN}{dn}
\DeclareMathOperator{\ellNS}{ns}
\DeclareMathOperator{\ellCS}{cs}

\DeclareMathOperator{\ellDC}{dc}

\DeclareMathOperator{\ellCD}{cd}
\DeclareMathOperator{\ellZN}{zn}
\newcommand{\ellKval}{{\operatorname{K}}}
\newcommand{\ellKvalc}{{\operatorname{K}'}}
\newcommand{\ellEval}{{\operatorname{E}}}

\newcommand{\ellmods}{\abstraction{m}}

\DeclareMathOperator{\ellK}{K}
\DeclareMathOperator{\ellE}{E}

\DeclareMathOperator{\ellTH}{\vartheta}
\newcommand{\ellTS}{\ellTH_\ellS}

\newcommand{\ellTN}{\ellTH_\ellN}
\newcommand{\der}{\mathrm{d}}
\newcommand{\diff}[1]{\mathinner{\der#1}}

\newcommand{\Order}{\mathcal{O}}

\let\Re\undefined\DeclareMathOperator{\Re}{Re}
\let\Im\undefined\DeclareMathOperator{\Im}{Im}
\DeclareMathOperator{\tr}{tr}
\DeclareMathOperator{\diag}{diag}
\DeclareMathOperator{\res}{res}
\DeclareMathOperator{\sign}{sign}

\end{mposttex}

\def\[#1\]{\begin{equation}#1\end{equation}}

\providecommand{\href}[2]{#2}

\makeatletter
\def\mr@ignsp#1 {\ifx\:#1\@empty\else #1\expandafter\mr@ignsp\fi}%
\newcommand{\multiref}[1]{\begingroup%\let\protect\string%
\xdef\mr@no@sparg{\expandafter\mr@ignsp#1 \: }%
\def\mr@comma{}%
\@for\mr@refs:=\mr@no@sparg\do{\mr@comma\def\mr@comma{,}\ref{\mr@refs}}%
\endgroup}
\renewcommand{\eqref}[1]{(\multiref{#1})}
\makeatother

\makeatletter
\newcommand{\namedref}[2]{\hyperref[#2]{#1~\ref*{#2}}}
\newcommand{\secref}{\@ifstar{\namedref{Section}}{\namedref{Sec.}}}
\newcommand{\appref}{\@ifstar{\namedref{Appendix}}{\namedref{App.}}}
\newcommand{\tabref}{\@ifstar{\namedref{Table}}{\namedref{Tab.}}}
\newcommand{\figref}{\@ifstar{\namedref{Figure}}{\namedref{Fig.}}}
\makeatother

\let\oldbib=\thebibliography
\def\thebibliography{\phantomsection\addcontentsline{toc}{section}{\refname}\oldbib}

\let\oldtoc=\tableofcontents
\def\tableofcontents{\phantomsection\addcontentsline{toc}{section}{\contentsname}\oldtoc}

\providecommand{\hypersetup}[1]{}

\hypersetup{plainpages=false}
\hypersetup{pdfpagemode=UseNone}
\hypersetup{bookmarksnumbered=true}
\hypersetup{pdfstartview=FitH}
\hypersetup{colorlinks=false}
\hypersetup{citebordercolor={.5 1 .5}}
\hypersetup{urlbordercolor={.5 1 1}}
\hypersetup{linkbordercolor={1 .7 .7}}
\ifarxiv\else\PassOptionsToPackage{draft}{metastr}\fi
\RequirePackage{metastr}

\usepackage{ifpdf}
\ifpdf\else\RequirePackage[active]{srcltx}\fi
\RequirePackage{color}

\newcommand{\hlcolor}{\color}
\renewcommand{\hlcolor}[2][]{}
\ifpublic\renewcommand{\hlcolor}[2][]{}\fi
\begin{mposttex}
\newcommand{\hlcolor}[2][]{}
\end{mposttex}

\newcommand{\remark}[2][]{{\normalfont\sffamily\hspace{1ex}%
  \def\emph{\textsl}\def\textbullet{$\bullet$}
  \def\tmparga{#1}%
  \color[rgb]{0.5,0,0}%
  \def\tmpargb{NB}\ifx\tmparga\tmpargb\color[rgb]{0,0,0.8}\fi%
  \def\tmpargb{KG}\ifx\tmparga\tmpargb\color[rgb]{0,0.5,0}\fi%
  \def\tmpargb{}\ifx\tmparga\tmpargb\color{red}\fi%
  \def\tmpargb{}\ifx\tmparga\tmpargb\else \textbf{#1:} \fi%
  #2\hspace{1ex}}}

\ifpublic\renewcommand{\remark}[2][]{}\fi

\begin{mpostfig}[label={KdVEllipticCuts}]
paths[1]:=unitsquare shifted (-0.5,-0.5) xscaled 6xu yscaled 2.5xu;

vpos[1]:=(-3xu,0);
vpos[2]:=(-1.5xu,0);
vpos[3]:=(-0.5xu,0);
vpos[4]:=(+1xu,0);

paths[11]:=fullcircle yscaled 1.5xu xscaled (0.6xu+abs(vpos[3]-vpos[4])) shifted 0.5[vpos[3],vpos[4]];
paths[21]:=fullcircle yscaled 1.5xu xscaled (0.6xu+abs(vpos[2]-vpos[3])) shifted 0.5[vpos[2],vpos[3]];

fill paths[1] withgreyscale 0.9;

draw subpath (0,4) of paths[21];
draw subpath (4,8) of paths[21] dashed evenly;

draw paths[11];

midarrow(paths[11], 2);
label.top(btex $\contacyc$ etex, point 2 of paths[11]);
midarrow(paths[21], 2);
label.top(btex $\contbcyc$ etex, point 2 of paths[21]);

drawarrow (point -0.5 of paths[1])--(point 1.5 of paths[1]) pensize(0.5pt);
drawarrow (point 0.15 of paths[1])--(point 2.85 of paths[1]) pensize(0.5pt);

draw vpos[1]--vpos[2] pensize(2pt);
draw vpos[3]--vpos[4] pensize(2pt);

label.bot(btex $\ptsqrt{\laxpar}_1$ etex, vpos[2]);
label.bot(btex $\ptsqrt{\laxpar}_2$ etex, vpos[3]);
label.bot(btex $\ptsqrt{\laxpar}_3$ etex, vpos[4]);
label.llft(btex $\laxpar$ etex, point 2 of paths[1]);
label.llft(btex $\Re \laxpar$ etex, point 1.5 of paths[1]);
label.llft(btex $\Im \laxpar$ etex, point 2.85 of paths[1]);

drawdot(vpos[2], 5pt);
drawdot(vpos[3], 5pt);
drawdot(vpos[4], 5pt);
\end{mpostfig}

\begin{mpostfig}[label={KdVEllipticDivisor}]
paths[1]:=unitsquare shifted (-0.5,-0.5) xscaled 6xu yscaled 2.5xu;

vpos[1]:=(-3xu,0);
vpos[2]:=(-1.5xu,0);
vpos[3]:=(-0.5xu,0);
vpos[4]:=(+1xu,0);

fill paths[1] withgreyscale 0.9;

paths[2]:=vpos[3]--vpos[4];
paths[3]:=reverse(paths[2] shifted (0,0.15xu))..(paths[2] shifted (0,-0.15xu))..cycle;

drawarrow (point -0.5 of paths[1])--(point 1.5 of paths[1]) pensize(0.5pt);
drawarrow (point 0.15 of paths[1])--(point 2.85 of paths[1]) pensize(0.5pt);

draw vpos[1]--vpos[2] pensize(2pt);
draw paths[2] pensize(2pt);
draw paths[3] dashed evenly;
midarrow(reverse(paths[3]),1.5);
midarrow(reverse(paths[3]),3.5);
drawcross(point 0.75 of paths[3], 5pt, 0, 1pt, black);
label.top(btex $\ptdiv{\laxpar}(x)$ etex, point 0.75 of paths[3]);

label.bot(btex $\ptsqrt{\laxpar}_1$ etex, vpos[2]);
label.bot(btex $\ptsqrt{\laxpar}_2$ etex, vpos[3] shifted (0,-0.1xu));
label.bot(btex $\ptsqrt{\laxpar}_3$ etex, vpos[4] shifted (0,-0.1xu));
label.llft(btex $\laxpar$ etex, point 2 of paths[1]);
label.llft(btex $\Re \laxpar$ etex, point 1.5 of paths[1]);
label.llft(btex $\Im \laxpar$ etex, point 2.85 of paths[1]);

drawdot(vpos[2], 5pt);
drawdot(vpos[3], 5pt);
drawdot(vpos[4], 5pt);
\end{mpostfig}

\begin{mpostfig}[label={KdVSolitonDivisor}]
paths[1]:=unitsquare shifted (-0.5,-0.5) xscaled 6xu yscaled 2.5xu;

vpos[1]:=(-3xu,0);
vpos[2]:=(-0.5xu,0);
vpos[3]:=(-0.5xu,0);
vpos[4]:=(+1xu,0);

fill paths[1] withgreyscale 0.9;

paths[2]:=vpos[3]--vpos[4];
paths[3]:=(paths[2] shifted (0,-0.15xu))..reverse(paths[2] shifted (0,0.15xu));

drawarrow (point -0.5 of paths[1])--(point 1.5 of paths[1]) pensize(0.5pt);
drawarrow (point 0.15 of paths[1])--(point 2.85 of paths[1]) pensize(0.5pt);

draw vpos[1]--vpos[2] pensize(2pt);
draw paths[2] pensize(2pt);
draw paths[3] dashed evenly;
midarrow(reverse(paths[3]),0.5);
midarrow(reverse(paths[3]),2.5);
drawcross(point 2.75 of paths[3], 5pt, 0, 1pt, black);
label.top(btex $\ptdiv{\laxpar}(x)$ etex, point 2.75 of paths[3]);

label.bot(btex $\ptsqrt{\laxpar}_{1,2}$ etex, vpos[3] shifted (0,-0.1xu));
label.bot(btex $\ptsqrt{\laxpar}_3$ etex, vpos[4] shifted (0,-0.1xu));
label.llft(btex $\laxpar$ etex, point 2 of paths[1]);
label.llft(btex $\Re \laxpar$ etex, point 1.5 of paths[1]);
label.llft(btex $\Im \laxpar$ etex, point 2.85 of paths[1]);

drawdot(vpos[3], 5pt);
drawdot(vpos[4], 5pt);
\end{mpostfig}

\begin{mpostfig}[label={KdVEllipticTorus}]
interim xu:=1cm;
interim ahlength:=5pt;

nums[1]:=2.0xu;
nums[3]:=2.5xu;

nums[2]:=1.5xu;
nums[4]:=2.0xu;

paths[1]:=unitsquare shifted (-0.5,-0.5) xscaled 2nums[3] yscaled 2nums[4];
paths[2]:=unitsquare shifted (-0.5,-0.5) xscaled 2nums[1] yscaled 2nums[2];

vpos[1]:=(0,0);
vpos[2]:=(0,nums[2]);
vpos[12]:=(0,-nums[2]);
vpos[3]:=(nums[1],0);
vpos[13]:=(-nums[1],0);
vpos[4]:=(nums[1],nums[2]);
vpos[14]:=(-nums[1],nums[2]);
vpos[24]:=(-nums[1],-nums[2]);
vpos[34]:=(nums[1],-nums[2]);

fill paths[1] withcolor 0.7[0.8white,white];
fill (subpath(1.5,3.5) of paths[2])--cycle withcolor 0.05[0.8white,blue];
fill (subpath(3.5,5.5) of paths[2])--cycle withcolor 0.05[0.8white,green];

drawarrow (point -0.5 of paths[1])--(point 1.5 of paths[1]) pensize(0.5pt);
drawarrow (point 0.5 of paths[1])--(point 2.5 of paths[1]) pensize(0.5pt);

draw ((-nums[3],0)--(+nums[3],0)) shifted vpos[2] pensize(2pt);
draw ((-nums[3],0)--(+nums[3],0)) pensize(2pt);
draw ((-nums[3],0)--(+nums[3],0)) shifted vpos[12] pensize(2pt);
draw ((0,-nums[4])--(0,+nums[4])) shifted vpos[3] pensize(0.25pt) dashed evenly;
draw ((0,-nums[4])--(0,+nums[4])) shifted vpos[13] pensize(0.25pt) dashed evenly;

paths[3]:=(point 2.125 of paths[2])--(point 0.875 of paths[2]);
paths[4]:=(point 3.2 of paths[2])--(point 1.8 of paths[2]);
draw paths[3] pensize(0.25pt);
draw paths[4] pensize(0.25pt);
midarrow (paths[3],0.35);
midarrow (paths[4],0.65);
label.lft(btex $\contbcyc$ etex, point 0.35 of paths[3]);
label.bot(btex $\contacyc$ etex, point 0.65 of paths[4]);

label.llft(btex $\laxzpar$ etex, point 2 of paths[1]);
label.urt(btex $+\iunit\ellKvalc$ etex, vpos[2]);
label.lrt(btex $-\iunit\ellKvalc$ etex, vpos[12]);
label.urt(btex $+\ellKval$ etex rotated 90, vpos[3]);
label.ulft(btex $-\ellKval$ etex rotated 90, vpos[13]);
label.urt(btex $\ptsqrt{\laxzpar}_0$ etex, vpos[1]);
label.ulft(btex $\ptsqrt{\laxzpar}_1$ etex, vpos[3]);
label.llft(btex $\ptsqrt{\laxzpar}_2$ etex, vpos[4]);
label.lrt(btex $\ptsqrt{\laxzpar}_3$ etex, vpos[2]);

drawdot(vpos[1], 5pt);
drawdot(vpos[2], 5pt);
drawdot(vpos[3], 5pt);
drawdot(vpos[4], 5pt);
\end{mpostfig}

\begin{mpostfig}[label={KdVEllipticQ}]
vpos[1]:=(-4xu,0);
vpos[2]:=(-1.5xu,0);
vpos[3]:=(-0.5xu,0);
vpos[4]:=(+1xu,0);
vpos[5]:=(+4xu,0);
vpos[6]:=0.4[vpos[3],vpos[4]];
vpos[9]:=(-2.5xu,0);
vpos[10]:=(0,1.5xu);
vpos[11]:=(0,0.6xu);

paths[1]:=vpos[1]--vpos[5];
paths[2]:=((0,0)--(0,3xu));

paths[3]:=
  vpos[1]--vpos[2]{dir 90}..
  lambdapath(i,0.1,0.5,0.2,vpos[2]+(i*xu,1.15*sqrt(i)*xu))
  ..{dir 90}(vpos[3]+vpos[10])--(vpos[4]+vpos[10]){dir 90}..
  lambdapath(i,2.9,5.5,0.2,vpos[2]+(i*xu,1.15*sqrt(i)*xu));
paths[4]:=
  lambdapath(i,-2.5,-0.1,0.2,vpos[2]+(i*xu,1.15*sqrt(-i)*xu))
  ..{dir -90}vpos[2]--vpos[3]{dir 90}..{dir 0}(vpos[6]+vpos[11])..{dir -90}(vpos[4])--vpos[5];

fill (paths[2] shifted vpos[1])--reverse(paths[2] shifted vpos[2])--cycle withgreyscale 0.9;
fill (paths[2] shifted vpos[3])--reverse(paths[2] shifted vpos[4])--cycle withgreyscale 0.9;

drawarrow paths[1] pensize(0.5pt);
drawarrow paths[2] shifted vpos[9] pensize(0.5pt);

draw paths[3] pensize(1.0pt) withcolor 0.4blue;
draw paths[4] pensize(1.5pt) dashed evenly withcolor 0.3green;
draw paths[1] shifted vpos[10] pensize(0.5pt) dashed evenly;
draw paths[2] shifted vpos[2] pensize(0.5pt) dashed evenly;
draw paths[2] shifted vpos[3] pensize(0.5pt) dashed evenly;
draw paths[2] shifted vpos[4] pensize(0.5pt) dashed evenly;
draw paths[2] shifted vpos[6] pensize(0.5pt) dashed withdots scaled 0.5;

draw ((-0.1xu,0)--(-0.6xu,0)) shifted ((0,-0.2xu)+vpos[9]+point 1 of paths[2]) pensize(1.0pt) withcolor 0.4blue;
draw ((-0.1xu,0)--(-0.6xu,0)) shifted ((0,-0.6xu)+vpos[9]+point 1 of paths[2]) pensize(1.5pt) dashed evenly withcolor 0.3green;
label.rt(btex $\Re \laxqmom$ etex, (-0.0xu,-0.2xu)+vpos[9]+point 1 of paths[2]);
label.rt(btex $\Im \laxqmom$ etex, (-0.0xu,-0.6xu)+vpos[9]+point 1 of paths[2]);

label.bot(btex $\ptsqrt{\laxpar}_1$ etex, vpos[2]);
label.bot(btex $\ptsqrt{\laxpar}_2$ etex, vpos[3]);
label.bot(btex $\ptsqrt{\laxpar}_3$ etex, vpos[4]);
label.ulft(btex $0$ etex, vpos[9]);
label.ulft(btex $\pi$ etex, vpos[9]+vpos[10]);
label.urt(btex $\ptcrit{\laxpar}$ etex, vpos[6]+vpos[11]);
label.llft(btex $\laxpar\vphantom{I}$ etex, point 1 of paths[1]);

drawdot(vpos[2], 4pt);
drawdot(vpos[3], 4pt);
drawdot(vpos[4], 4pt);
drawdot(vpos[3]+vpos[10], 4pt);
drawdot(vpos[4]+vpos[10], 4pt);
drawcross(vpos[6]+vpos[11], 5pt, 0, 0.5pt, black);
\end{mpostfig}

\begin{mpostfig}[label={KdVEllipticF}]
vpos[1]:=(-4xu,0);
vpos[2]:=(-1.5xu,0);
vpos[3]:=(-0.5xu,0);
vpos[4]:=(+1xu,0);
vpos[5]:=(+4xu,0);
vpos[6]:=0.4[vpos[3],vpos[4]];
vpos[7]:=(+3.7xu,0);
vpos[8]:=(-1.7xu,0);
vpos[9]:=(-2.5xu,0);
vpos[10]:=(0,1xu);
vpos[11]:=(0,-1.3xu);

paths[1]:=vpos[1]--vpos[5];
paths[2]:=((0,-2xu)--(0,2xu));

paths[3]:=
(vpos[8]+2*vpos[10])..
(vpos[2]+vpos[10])..(vpos[3]-vpos[10])..(vpos[6]+vpos[11]){dir 0}..(vpos[4]-vpos[10])..{dir 0}(vpos[7]+vpos[10])..(vpos[5]+0.95vpos[10]);

fill ((vpos[10]--(point 1 of paths[2])) shifted vpos[1])--(((point 1 of paths[2])--vpos[10]) shifted vpos[2])--cycle withgreyscale 0.9;
fill (((-vpos[10])--(point 0 of paths[2])) shifted vpos[3])--(((point 0 of paths[2])--(-vpos[10])) shifted vpos[4])--cycle withgreyscale 0.9;

drawarrow paths[1] pensize(0.5pt);
drawarrow paths[2] shifted vpos[9] pensize(0.5pt);
draw paths[1] shifted vpos[10] pensize(0.5pt) dashed evenly;
draw paths[1] shifted -vpos[10] pensize(0.5pt) dashed evenly;
draw paths[2] shifted vpos[2] pensize(0.5pt) dashed evenly;
draw paths[2] shifted vpos[3] pensize(0.5pt) dashed evenly;
draw paths[2] shifted vpos[4] pensize(0.5pt) dashed evenly;
draw paths[2] shifted vpos[6] pensize(0.5pt) dashed withdots scaled 0.5;

draw paths[3] pensize(1.0pt);

label.llft(btex $\cos\laxqmom$ etex, vpos[9]+point 1 of paths[2]);

label.llft(btex $\ptsqrt{\laxpar}_1$ etex, vpos[2]+vpos[10]);
label.llft(btex $\ptsqrt{\laxpar}_2$ etex, vpos[3]-vpos[10]);
label.lrt(btex $\ptsqrt{\laxpar}_3$ etex, vpos[4]-vpos[10]);
label.ulft(btex $0$ etex, vpos[9]);
label.ulft(btex $+1$ etex, vpos[9]+vpos[10]);
label.ulft(btex $-1$ etex, vpos[9]-vpos[10]);
label.lrt(btex $\ptcrit{\laxpar}$ etex, vpos[6]+vpos[11]);
label.llft(btex $\laxpar\vphantom{I}$ etex, point 1 of paths[1]);

drawdot(vpos[2]+vpos[10], 4pt);
drawdot(vpos[3]-vpos[10], 4pt);
drawdot(vpos[4]-vpos[10], 4pt);
drawdot(vpos[7]+vpos[10], 4pt);
drawcross(vpos[6]+vpos[11], 5pt, 0, 0.5pt, black);
\end{mpostfig}

\begin{mpostfig}[label={KdVContinuumF}]
nums[1]:=1xu;
nums[2]:=1.3;
nums[3]:=120;
nums[4]:=80;
nums[5]:=-180/nums[3]*xu;
nums[6]:=-nums[4]/nums[3]*xu;
nums[7]:=angle(1,sqrt(nums[2]*nums[2]-1))/nums[3]*xu;
nums[10]:=-4.5xu;
nums[11]:=3.7xu;
nums[12]:=1.7xu;

paths[1]:=(nums[10],0)--(nums[11],0);
paths[2]:=((0,-nums[12])--(0,nums[12]));

paths[3]:=
  lambdapath(i,nums[10]/xu,nums[11]/xu,0.1,(i*xu,nums[2]*nums[1]*cosd(i*nums[3]+nums[4])));

drawarrow paths[1] pensize(0.5pt);
drawarrow paths[2] shifted (nums[10],0) pensize(0.5pt);
draw paths[1] shifted (0,nums[1]) pensize(0.5pt) dashed evenly;
draw paths[1] shifted (0,-nums[1]) pensize(0.5pt) dashed evenly;
for i:=-2 upto 2:
fill ((((0,nums[1])--(point 1 of paths[2])) shifted (-nums[7],0))
    --(((point 1 of paths[2])--(0,nums[1])) shifted (nums[7],0))--cycle)
        shifted (i*nums[5]+nums[6],0) yscaled cosd(180*i) withgreyscale 0.9;
draw paths[2] shifted (i*nums[5]+nums[6]-nums[7],0) pensize(0.5pt) dashed evenly;
draw paths[2] shifted (i*nums[5]+nums[6]+nums[7],0) pensize(0.5pt) dashed evenly;
draw ((-nums[7],0)--(+nums[7],0)) shifted (i*nums[5]+nums[6],0) pensize(1.5pt);
drawdot ((i*nums[5]+nums[6]-nums[7],0),4pt);
drawdot ((i*nums[5]+nums[6]+nums[7],0),4pt);
endfor
draw paths[2] shifted (nums[6],0) pensize(0.5pt) dashed withdots scaled 0.5;

draw paths[3] pensize(1.0pt);

label.llft(btex $\cos\laxqmom$ etex, (nums[10],0)+point 1 of paths[2]);

label.lft(btex $0$ etex, (nums[10],0));
label.lft(btex $+1$ etex, (nums[10],nums[1]));
label.lft(btex $-1$ etex, (nums[10],-nums[1]));
label.llft(btex $\ismmom$ etex, point 1 of paths[1]);

label.top(btex $\ptsqrt{\ismmom}_{n}^-$ etex, (nums[6]-nums[7],nums[12]));
label.top(btex $\ptsqrt{\ismmom}_{n}^+$ etex, (nums[6]+nums[7],nums[12]));
label.bot(btex $(\pi n-\arg\ismtrans_n)/\fldlen$ etex, (-0.5xu,-nums[12]));

paths[4]:=((-nums[7],0)--(nums[7],0)) shifted (nums[6],nums[12]-0.2xu);
drawdblarrow paths[4];

\end{mpostfig}

\begin{mpostfig}[label={KdVPinchedF}]
vpos[1]:=(-5xu,-7xu);
vpos[2]:=(-1.88xu,0);
vpos[3]:=(-1.85xu,0);
vpos[5]:=(+1.55xu,0);
vpos[6]:=(+1.64xu,0);
vpos[8]:=(+4xu,0);
vpos[4]:=0.4[vpos[3],vpos[5]]+(0,3.3xu);
vpos[7]:=0.3[vpos[6],vpos[8]]+(0,-1.3xu);
vpos[20]:=(0,0.25xu);

paths[1]:=vpos[8]--(-4xu,0);
paths[2]:=(0,-2xu)--(0,2xu);

paths[3]:=vpos[1]{dir 0}
  ..(vpos[2]-vpos[20])..(vpos[3]+vpos[20])..vpos[4]{dir 0}
  ..(vpos[5]+vpos[20])..(vpos[6]-vpos[20])..vpos[7]{dir 0}
  ..(vpos[8]-vpos[20]){dir 0};

fill (((point 1 of paths[1])--vpos[2]) shifted -vpos[20])--((vpos[2]--(point 1 of paths[1])) shifted (0,ypart(point 0 of paths[2])))--cycle withgreyscale 0.9;
fill ((vpos[3]--vpos[5]) shifted vpos[20])--((vpos[5]--vpos[3]) shifted (0,ypart(point 1 of paths[2])))--cycle withgreyscale 0.9;
fill ((vpos[6]--vpos[8]) shifted -vpos[20])--((vpos[8]--vpos[6]) shifted (0,ypart(point 0 of paths[2])))--cycle withgreyscale 0.9;

drawarrow paths[1] pensize(0.5pt);
drawarrow paths[2] shifted vpos[8] pensize(0.5pt);
draw paths[1] shifted vpos[20] pensize(0.5pt) dashed evenly;
draw paths[1] shifted -vpos[20] pensize(0.5pt) dashed evenly;
draw paths[2] shifted vpos[2] pensize(0.5pt) dashed evenly;
draw paths[2] shifted vpos[3] pensize(0.5pt) dashed evenly;
draw paths[2] shifted vpos[5] pensize(0.5pt) dashed evenly;
draw paths[2] shifted vpos[6] pensize(0.5pt) dashed evenly;
draw paths[2] shifted vpos[8] pensize(0.5pt) dashed evenly;

draw paths[3] pensize(2.0pt);

clip currentpicture to unitsquare shifted (-0.5,-0.5) xscaled 8.2xu yscaled 4xu;

label.llft(btex $\cos\laxqmom\ $ etex, vpos[8]+point 1 of paths[2]);

label.urt(btex $+1$ etex, vpos[8]+vpos[20]);
label.lrt(btex $-1$ etex, vpos[8]-vpos[20]);
label.lft(btex $\Im \ismmom$ etex, point 1 of paths[1]);

label.llft(btex $\ptsqrt{\ismmom}_{n-1}^+$ etex, vpos[2]-vpos[20]);
label.urt(btex $\ptsqrt{\ismmom}_{n}^-$ etex, vpos[3]+vpos[20]);
label.ulft(btex $\ptsqrt{\ismmom}_{n}^+$ etex, vpos[5]+vpos[20]);
label.lrt(btex $\ \ptsqrt{\ismmom}_{n+1}^-$ etex, vpos[6]-vpos[20]);

drawdot(vpos[2]-vpos[20], 4pt);
drawdot(vpos[3]+vpos[20], 4pt);
drawdot(vpos[5]+vpos[20], 4pt);
drawdot(vpos[6]-vpos[20], 4pt);
drawdot(vpos[8]-vpos[20], 4pt);
drawcross(vpos[7], 5pt, 0, 0.5pt, black);
\end{mpostfig}

\begin{mpostfig}[label={KdVContinuumDivisor}]
nums[1]:=1xu;
nums[2]:=1.3;
nums[3]:=120;
nums[4]:=80;
nums[5]:=-180/nums[3]*xu;
nums[6]:=-nums[4]/nums[3]*xu;
nums[7]:=angle(1,sqrt(nums[2]*nums[2]-1))/nums[3]*xu;

paths[1]:=unitsquare shifted (-0.5,-0.5) xscaled 9xu yscaled 1.0xu;
paths[2]:=(-nums[7],0)--(+nums[7],0);
paths[3]:=reverse(paths[2] shifted (0,0.15xu))..(paths[2] shifted (0,-0.15xu))..cycle;

fill paths[1] withgreyscale 0.9;

for i:=-3 upto 2:
draw paths[2] shifted (i*nums[5]+nums[6],0) pensize(1.5pt);
drawdot ((i*nums[5]+nums[6]-nums[7],0),4pt);
drawdot ((i*nums[5]+nums[6]+nums[7],0),4pt);
drawcross(point (i*0.4+1.9) of paths[3] shifted (i*nums[5]+nums[6],0), 5pt, 0, 1pt, black);
endfor

label.llft(btex $\ismmom$ etex, point 2 of paths[1]);

\end{mpostfig}

\begin{mpostfig}[label={KdVLimitCuts}]

paths[1]:=unitsquare shifted (-0.5,-0.5) xscaled 14xu yscaled 3.5xu;

fill paths[1] withgreyscale 0.9;

drawarrow (point 3.5 of paths[1])--(point 1.5 of paths[1]);

vpos[1]:=(-6xu,0);
vpos[11]:=(0.07xu,0);
vpos[2]:=(-4.5xu,0);
vpos[12]:=(0.09xu,0);
vpos[3]:=(0xu,0);

draw (point 3.5 of paths[1])--(vpos[1]-vpos[11]) pensize(2pt);
draw (vpos[1]+vpos[11])--(vpos[2]-vpos[12]) pensize(2pt);
draw (vpos[2]+vpos[12])--vpos[3] pensize(2pt);

fill (subpath (0,4) of fullcircle)--cycle xscaled 4.4xu yscaled 1xu shifted (4xu,0) withgreyscale 0.7;
label.top(btex $\abs{\ismrefl}$ etex, (4xu,0.5xu));
label.bot(btex $\Delta\sim1/\fldlen$ etex, (4xu,0xu));

for i:=-5 upto 5:
draw ((-1,0)--(+1,0)) scaled (sqrt(36-(i**2))*0.01xu) shifted (4xu+i*0.4xu,0) pensize(2pt);
endfor

drawdot (vpos[3],4pt);

label.top(btex $\kdvfld_0$ etex, vpos[3]);
label.top(btex $\kdvfld_0+\ismsing_1^2$ etex, vpos[1]);
label.top(btex $\kdvfld_0+\ismsing_2^2$ etex, vpos[2]);
label.bot(btex $\Delta\sim\eunit^{-\ast \fldlen}$ etex, vpos[2]);

label.llft(btex $\laxpar$ etex, point 2 of paths[1]);

\end{mpostfig}

\begin{mpostfig}[label={KdVEllipticState}]
interim xu:=1.5cm;
interim yu:=1.3cm;
paths[1]:=(-0.5xu,1.yu)..(-0.48xu,0.999639yu)..(-0.46xu,0.998532yu)..(-0.44xu,0.996601yu)
..(-0.42xu,0.993713yu)..(-0.4xu,0.98967yu)..(-0.38xu,0.984198yu)..(-0.36xu,0.976931yu)
..(-0.34xu,0.967396yu)..(-0.32xu,0.954986yu)..(-0.3xu,0.938942yu)..(-0.28xu,0.918333yu)
..(-0.26xu,0.892044yu)..(-0.24xu,0.858787yu)..(-0.22xu,0.817146yu)..(-0.2xu,0.765686yu)
..(-0.18xu,0.703144yu)..(-0.16xu,0.62875yu)..(-0.14xu,0.542668yu)..(-0.12xu,0.446547yu)
..(-0.1xu,0.344063yu)..(-0.08xu,0.241247yu)..(-0.06xu,0.146327yu)..(-0.04xu,0.0688093yu)
..(-0.02xu,0.0178145yu)..(0xu,0yu)..(0.02xu,0.0178145yu)..(0.04xu,0.0688093yu)
..(0.06xu,0.146327yu)..(0.08xu,0.241247yu)..(0.1xu,0.344063yu)..(0.12xu,0.446547yu)
..(0.14xu,0.542668yu)..(0.16xu,0.62875yu)..(0.18xu,0.703144yu)..(0.2xu,0.765686yu)
..(0.22xu,0.817146yu)..(0.24xu,0.858787yu)..(0.26xu,0.892044yu)..(0.28xu,0.918333yu)
..(0.3xu,0.938942yu)..(0.32xu,0.954986yu)..(0.34xu,0.967396yu)..(0.36xu,0.976931yu)
..(0.38xu,0.984198yu)..(0.4xu,0.98967yu)..(0.42xu,0.993713yu)..(0.44xu,0.996601yu)
..(0.46xu,0.998532yu)..(0.48xu,0.999639yu)..(0.5xu,1.yu);
fill unitsquare xscaled 10xu yscaled 1yu withcolor 0.9white;
draw (0,0)--(10xu,0) dashed withdots scaled 0.5;
draw (0,1yu)--(10xu,1yu) dashed withdots scaled 0.5;
paths[2]:=for i:=0 upto 9:  if i>0: .. fi (paths[1] shifted (i*xu+0.5xu,0)) endfor;
paths[3]:=((0,0)--(0.4xu,0)) shifted (1.5xu+0.2cm,-0.25cm);
drawarrow paths[3];
label.rt(btex $\parvel$ etex, point 1 of paths[3]);
paths[4]:=((0,0)--(1xu,0)) shifted (2.5xu,-0.1cm);
drawdblarrow paths[4];
label.bot(btex $\fldlen$ etex, point 0.5 of paths[4]);
draw paths[2] pensize(1.5pt) withcolor 0.6blue;
currentpicture:=currentpicture transformed (identity shifted (-1.1xu,0));
clip currentpicture to unitsquare shifted (0,-0.5) xscaled 10xu yscaled 10yu;
xu:=1cm;
clip currentpicture to unitsquare shifted (-0.5,-0.5) xscaled 8xu yscaled 10xu;
paths[3]:=(0,-0.75xu)--(1xu,-0.75xu);
drawarrow (-0.2xu,-0.5xu)--(4xu,-0.5xu);
label.top(btex $x$ etex, (4xu,-0.5xu));
drawarrow (0,-0.6xu)--(0,2xu);
label.lrt(btex $\kdvfld$ etex, (0,2xu));
\end{mpostfig}

\begin{mpostfig}[label={KdVSolitonState}]
interim xu:=1.0cm;
interim yu:=1.4cm;
paths[2]:=for i:=0 upto 60: if i>0:.. fi (i*0.1xu,(1-(2/(mexp(+30*(i-30))+mexp(-30*(i-30))))**2)*yu) endfor;
fill unitsquare xscaled 8xu yscaled 1yu withcolor 0.9white;
draw (0,0)--(6xu,0) dashed withdots scaled 0.5;
draw (0,1yu)--(6xu,1yu) dashed withdots scaled 0.5;
paths[3]:=((0,0)--(0.9xu,0)) shifted (3xu+0.2cm,-0.25cm);
drawarrow paths[3];
label.rt(btex $\parvel$ etex, point 1 of paths[3]);
draw paths[2] pensize(1.5pt) withcolor 0.6blue;
clip currentpicture to unitsquare shifted (0,-0.5) xscaled 6xu yscaled 10yu;
label.lft(btex $\kdvfld_0$ etex, (0,1yu));
paths[3]:=(0,-0.75xu)--(1xu,-0.75xu);
drawarrow (-0.2xu,-0.5xu)--(6xu,-0.5xu);
label.top(btex $x$ etex, (6xu,-0.5xu));
drawarrow (0,-0.6xu)--(0,2xu);
label.lrt(btex $\kdvfld$ etex, (0,2xu));
\end{mpostfig}

\begin{mpostfig}[label={KdVSolitonDivisorPlot}]
interim xu:=1.0cm;
interim yu:=1.4cm;
paths[1]:=for i:=0 upto 60: if i>0:.. fi (i*0.1xu,((2/(mexp(+30*(i-30))+mexp(-30*(i-30))))**2-1)*1.0*yu) endfor;
fill unitsquare shifted (0,-1) xscaled 8xu yscaled 1yu withcolor 0.9white;
draw (0,-1yu)--(6xu,-1yu) dashed withdots scaled 0.5;
draw (0,0)--(6xu,0) dashed withdots scaled 0.5;
draw subpath (0,30) of paths[1] pensize(1.5pt) withcolor 0.6blue;
draw subpath (30,60) of paths[1] pensize(1.5pt) withcolor 0.5green;
clip currentpicture to unitsquare shifted (0,-0.5) xscaled 6xu yscaled 10yu;
label.lft(btex $\kdvfld_0$ etex, (0,0yu));
label.lft(btex $\ptsol{\laxpar}$ etex, (0,-1yu));
paths[3]:=((-0.2xu,0)--(6xu,0)) shifted (0,-1.3yu);
paths[4]:=(0,-1.4yu)--(0,0.5yu);
drawarrow paths[3];
drawarrow paths[4];
label.top(btex $x$ etex, point 1 of paths[3]);
label.lrt(btex $\ptdiv{\laxpar}$ etex, point 1 of paths[4]);
\end{mpostfig}

\begin{mpostfig}[label={KdVTwoSolitonDivisor1}]
interim xu:=1.0cm;
interim yu:=1.4cm;
paths[1]:=(for i:=0 upto 60: if i>0:.. fi (i*0.1xu,((2/(mexp(+30*(i-15))+mexp(-30*(i-15))))**2-1)*1.0yu) endfor);
paths[2]:=(for i:=0 upto 60: if i>0:.. fi (i*0.1xu,((2/(mexp(+50*(i-45))+mexp(-50*(i-45))))**2-1)*1.6yu) endfor);
fill unitsquare shifted (0,-1) xscaled 8xu yscaled 1.6yu withcolor 0.8white;
fill unitsquare shifted (0,-1) xscaled 8xu yscaled 1.0yu withcolor 0.9white;
draw (0,-1.6yu)--(6xu,-1.6yu) dashed withdots scaled 0.5;
draw (0,-1yu)--(6xu,-1yu) dashed withdots scaled 0.5;
draw (0,0)--(6xu,0) dashed withdots scaled 0.5;
vpos[1]:=(xpart(paths[1] intersectionpoint paths[2]),-1yu);
draw (subpath (0,15) of paths[1]) pensize(1.5pt) withcolor 0.5[white,0.6blue];
draw (subpath (15,38) of paths[1])..vpos[1] pensize(1.5pt) withcolor 0.5[white,0.5green];
draw (subpath (0,38) of paths[2])..vpos[1] pensize(1.5pt) withcolor 0.7[white,0.6blue];
draw vpos[1]..(subpath (42,45) of paths[2]) pensize(1.5pt) withcolor 0.5[white,0.6blue];
draw vpos[1]..(subpath (42,48) of paths[1] shifted (0,-0.01yu))..(subpath (53,60) of paths[2]) pensize(1.5pt) withcolor 0.7[white,0.5green];
draw (subpath (45,48) of paths[2])..(subpath (53,60) of paths[1]) pensize(1.5pt) withcolor 0.5[white,0.5green];
draw paths[1] pensize(0.5pt) dashed evenly;
draw paths[2] pensize(0.5pt) dashed evenly;
clip currentpicture to unitsquare shifted (0,-0.5) xscaled 6xu yscaled 10yu;
label.lft(btex $\kdvfld_0$ etex, (0,0yu));
label.lft(btex $\ptsol{\laxpar}_2$ etex, (0,-1.0yu));
label.lft(btex $\ptsol{\laxpar}_1$ etex, (0,-1.6yu));
paths[5]:=((-0.2xu,0)--(6xu,0)) shifted (0,-1.9yu);
paths[6]:=(0,-2.1yu)--(0,0.5yu);
drawarrow paths[5];
drawarrow paths[6];
label.top(btex $x$ etex, point 1 of paths[5]);
label.lrt(btex $\ptdiv{\laxpar}$ etex, point 1 of paths[6]);
\end{mpostfig}

\begin{mpostfig}[label={KdVTwoSolitonDivisor2}]
interim xu:=1.0cm;
interim yu:=1.4cm;
paths[1]:=(for i:=0 upto 60: if i>0:.. fi (i*0.1xu,((2/(mexp(+30*(i-45))+mexp(-30*(i-45))))**2-1)*1.0yu) endfor);
paths[2]:=(for i:=0 upto 60: if i>0:.. fi (i*0.1xu,((2/(mexp(+50*(i-15))+mexp(-50*(i-15))))**2-1)*1.6yu) endfor);
fill unitsquare shifted (0,-1) xscaled 8xu yscaled 1.6yu withcolor 0.8white;
fill unitsquare shifted (0,-1) xscaled 8xu yscaled 1.0yu withcolor 0.9white;
draw (0,-1.6yu)--(6xu,-1.6yu) dashed withdots scaled 0.5;
draw (0,-1yu)--(6xu,-1yu) dashed withdots scaled 0.5;
draw (0,0)--(6xu,0) dashed withdots scaled 0.5;
vpos[1]:=(xpart(reverse(paths[1]) intersectionpoint reverse(paths[2])),-1yu);
draw (subpath (45,60) of paths[1]) pensize(1.5pt) withcolor 0.5[white,0.5green];
draw vpos[1]..(subpath (22,45) of paths[1]) pensize(1.5pt) withcolor 0.5[white,0.6blue];
draw vpos[1]..(subpath (22,60) of paths[2]) pensize(1.5pt) withcolor 0.7[white,0.5green];
draw (subpath (15,18) of paths[2])..vpos[1] pensize(1.5pt) withcolor 0.5[white,0.5green];
draw (subpath (0,7) of paths[2])..(subpath (12,18) of paths[1] shifted (0,-0.01yu))..vpos[1] pensize(1.5pt) withcolor 0.7[white,0.6blue];
draw (subpath (0,7) of paths[1])..(subpath (12,15) of paths[2]) pensize(1.5pt) withcolor 0.5[white,0.6blue];
draw paths[1] pensize(0.5pt) dashed evenly;
draw paths[2] pensize(0.5pt) dashed evenly;
clip currentpicture to unitsquare shifted (0,-0.5) xscaled 6xu yscaled 10yu;
label.lft(btex $\kdvfld_0$ etex, (0,0yu));
label.lft(btex $\ptsol{\laxpar}_2$ etex, (0,-1.0yu));
label.lft(btex $\ptsol{\laxpar}_1$ etex, (0,-1.6yu));
paths[5]:=((-0.2xu,0)--(6xu,0)) shifted (0,-1.9yu);
paths[6]:=(0,-2.1yu)--(0,0.5yu);
drawarrow paths[5];
drawarrow paths[6];
label.top(btex $x$ etex, point 1 of paths[5]);
label.lrt(btex $\ptdiv{\laxpar}$ etex, point 1 of paths[6]);
\end{mpostfig}

\begin{mpostfig}[label={KdVTwoSolitonDivisorC}]
interim xu:=1.0cm;
interim yu:=1.4cm;
paths[1]:=(for i:=0 upto 60: if i>0:.. fi (i*0.1xu,((2/(mexp(+30*(i-20))+mexp(-30*(i-20))))**2-1)*1.0yu) endfor);
paths[2]:=(for i:=0 upto 60: if i>0:.. fi (i*0.1xu,((2/(mexp(+30*(i-40))+mexp(-30*(i-40))))**2-1)*1.6yu) endfor);
paths[3]:=(for i:=0 upto 60: if i>0:.. fi (i*0.1xu,((2/(mexp(+30*(i-40))+mexp(-30*(i-40))))**2-1)*1.0yu) endfor);
paths[4]:=(for i:=0 upto 60: if i>0:.. fi (i*0.1xu,((2/(mexp(+30*(i-20))+mexp(-30*(i-20))))**2-1)*1.6yu) endfor);
fill unitsquare shifted (0,-1) xscaled 8xu yscaled 1.6yu withcolor 0.8white;
fill unitsquare shifted (0,-1) xscaled 8xu yscaled 1.0yu withcolor 0.9white;
draw (0,-1.6yu)--(6xu,-1.6yu) dashed withdots scaled 0.5;
draw (0,-1yu)--(6xu,-1yu) dashed withdots scaled 0.5;
draw (0,0)--(6xu,0) dashed withdots scaled 0.5;
vpos[1]:=(3xu,-1yu);
paths[5]:=(subpath (0,22) of paths[1])..vpos[1]..(subpath (40,60) of paths[4]);
paths[6]:=(subpath (0,20) of paths[2])..vpos[1]..(subpath (38,60) of paths[3]);
draw subpath(0,20) of paths[5] pensize(1.5pt) withcolor 0.5[white,0.6blue];
draw subpath(20,23) of paths[5] pensize(1.5pt) withcolor 0.5[white,0.5green];
draw subpath(23,100) of paths[5] pensize(1.5pt) withcolor 0.7[white,0.5green];
draw subpath(0,21) of paths[6] pensize(1.5pt) withcolor 0.7[white,0.6blue];
draw subpath(21,24) of paths[6] pensize(1.5pt) withcolor 0.5[white,0.6blue];
draw subpath(24,100) of paths[6] pensize(1.5pt) withcolor 0.5[white,0.5green];
clip currentpicture to unitsquare shifted (0,-0.5) xscaled 6xu yscaled 10yu;
label.lft(btex $\kdvfld_0$ etex, (0,0yu));
label.lft(btex $\ptsol{\laxpar}_2$ etex, (0,-1.0yu));
label.lft(btex $\ptsol{\laxpar}_1$ etex, (0,-1.6yu));
paths[7]:=((-0.2xu,0)--(6xu,0)) shifted (0,-1.9yu);
paths[8]:=(0,-2.1yu)--(0,0.5yu);
drawarrow paths[7];
drawarrow paths[8];
label.top(btex $x$ etex, point 1 of paths[7]);
label.lrt(btex $\ptdiv{\laxpar}$ etex, point 1 of paths[8]);
\end{mpostfig}

\begin{mpostfig}[label={PatchingOriginal}]
interim xu:=1.0cm;
interim yu:=0.7cm;
paths[1]:=unitsquare shifted (-0.5,-0.5) xscaled 3xu yscaled 2yu;
fill paths[1] withcolor 0.9white;
paths[2]:=(lambdapath(i,-45,45,1,(i*0.05xu,(-0.2+0.3*sind(10i-223)+0.2*sind(20i-313)+0.2*sind(30i-13))*yu*(1*mexp(-1.6*((i-10)**2))+0.7*mexp(-2*((i+10)**2)))))) xscaled 0.4;
draw (-1.5xu,0)..paths[2]..(1.5xu,0) pensize(1.5pt) withcolor 0.6blue;
drawarrow (-1.5xu,0xu)--(1.5xu,0xu);
label.ulft(btex $x\vphantom{_x}$ etex, (1.5xu,0xu));
drawarrow (0xu,-1yu)--(0xu,1yu);
label.lrt(btex $\kdvfld$ etex, (0xu,1yu));
\end{mpostfig}

\begin{mpostfig}[label={PatchingShort}]
interim xu:=1.0cm;
interim yu:=0.7cm;

paths[1]:=unitsquare shifted (-0.5,-0.5) xscaled 1.2xu yscaled 2yu;
fill paths[1] withcolor 0.9white;
draw subpath (1,2) of paths[1] dashed evenly;
draw subpath (3,4) of paths[1] dashed evenly;
draw subpath (3,4) of paths[1] shifted (-1.2xu,0) dashed evenly;
draw subpath (1,2) of paths[1] shifted (+1.2xu,0) dashed evenly;

drawdblarrow subpath (2,3) of paths[1] shifted (0,-0.1xu);
label.bot(btex $\fldlen$ etex, (point 2.5 of paths[1])+(0,-0.1xu));

paths[2]:=(lambdapath(i,-30,30,1,(i*0.05xu,
  (-0.2+0.3*sind(10i-223)+0.2*sind(20i-313)+0.2*sind(30i-13))*yu*(1*mexp(-1.6*((i-10)**2))+0.7*mexp(-2*((i+10)**2)))))) xscaled 0.4;
draw (paths[2] shifted (-1.2xu,0))..paths[2]..(paths[2] shifted (+1.2xu,0)) pensize(1.5pt) withcolor 0.6blue;
draw (-1.8xu,0xu)--(1.8xu,0xu);
\end{mpostfig}

\begin{mpostfig}[label={PatchingLong}]
interim xu:=1.0cm;
interim yu:=0.7cm;

paths[1]:=unitsquare shifted (-0.5,-0.5) xscaled 2xu yscaled 2yu;
fill paths[1] withcolor 0.9white;
draw subpath (1,2) of paths[1] dashed evenly;
draw subpath (3,4) of paths[1] dashed evenly;

drawdblarrow subpath (2,3) of paths[1] shifted (0,-0.1xu);
label.bot(btex $\fldlen$ etex, (point 2.5 of paths[1])+(0,-0.1xu));

paths[2]:=(lambdapath(i,-45,45,1,(i*0.05xu,
  (-0.2+0.3*sind(10i-223)+0.2*sind(20i-313)+0.2*sind(30i-13))*yu*(1*mexp(-1.6*((i-10)**2))+0.7*mexp(-2*((i+10)**2)))))) xscaled 0.4;
draw (paths[2] shifted (-2xu,0))..paths[2]..(paths[2] shifted (+2xu,0)) pensize(1.5pt) withcolor 0.6blue;
clip currentpicture to unitsquare shifted (-0.5,-0.5) xscaled 5xu yscaled 10yu;
draw (-2.5xu,0xu)--(2.5xu,0xu);
\end{mpostfig}

\begin{mpostfig}[label={KdVPinchedQ}]
vpos[1]:=(-4xu,0);
vpos[2]:=(-1.5xu,0);
vpos[3]:=(-1.3xu,0);
vpos[4]:=(+1xu,0);
vpos[5]:=(+4xu,0);
vpos[6]:=0.2[vpos[3],vpos[4]];
vpos[9]:=(-2.5xu,0);
vpos[10]:=(0,1.5xu);
vpos[11]:=(0,0.9xu);

paths[1]:=vpos[1]--vpos[5];
paths[2]:=((0,0)--(0,3xu));

paths[3]:=
  vpos[1]--vpos[2]{dir 90}
  ..{dir 90}(vpos[3]+vpos[10])--(vpos[4]+vpos[10]){dir 90}..
  lambdapath(i,0.1,3,0.2,vpos[4]+vpos[10]+(i*xu,0.7*sqrt(i)*xu));
paths[4]:=
  lambdapath(i,-5,-3.5,0.2,vpos[4]+(i*xu,0.7*sqrt(-i)*xu))
  ..{dir -90}vpos[2]--vpos[3]{dir 90}..{dir 0}(vpos[6]+vpos[11])
  ..lambdapath(i,-1.5,-0.1,0.2,vpos[4]+(i*xu,0.7*sqrt(-i)*xu))
  ..{dir -90}(vpos[4])--vpos[5];

fill (paths[2] shifted vpos[1])--reverse(paths[2] shifted vpos[2])--cycle withgreyscale 0.9;
fill (paths[2] shifted vpos[3])--reverse(paths[2] shifted vpos[4])--cycle withgreyscale 0.9;

drawarrow paths[1] pensize(0.5pt);
drawarrow paths[2] shifted vpos[9] pensize(0.5pt);

draw paths[3] pensize(1.0pt) withcolor 0.4blue;
draw paths[4] pensize(1.5pt) dashed evenly withcolor 0.3green;
draw paths[1] shifted vpos[10] pensize(0.5pt) dashed evenly;
draw paths[2] shifted vpos[2] pensize(0.5pt) dashed evenly;
draw paths[2] shifted vpos[3] pensize(0.5pt) dashed evenly;
draw paths[2] shifted vpos[4] pensize(0.5pt) dashed evenly;
draw paths[2] shifted vpos[6] pensize(0.5pt) dashed withdots scaled 0.5;

draw ((-0.1xu,0)--(-0.6xu,0)) shifted ((0,-0.2xu)+vpos[9]+point 1 of paths[2]) pensize(1.0pt) withcolor 0.4blue;
draw ((-0.1xu,0)--(-0.6xu,0)) shifted ((0,-0.6xu)+vpos[9]+point 1 of paths[2]) pensize(1.5pt) dashed evenly withcolor 0.3green;
label.rt(btex $\Re \laxqmom$ etex, (-0.0xu,-0.2xu)+vpos[9]+point 1 of paths[2]);
label.rt(btex $\Im \laxqmom$ etex, (-0.0xu,-0.6xu)+vpos[9]+point 1 of paths[2]);

label.llft(btex $\ptsqrt{\laxpar}_1$ etex, vpos[2]);
label.lrt(btex $\ptsqrt{\laxpar}_2$ etex, vpos[3]);
label.bot(btex $\ptsqrt{\laxpar}_3$ etex, vpos[4]);
label.ulft(btex $0$ etex, vpos[9]);
label.ulft(btex $\pi$ etex, vpos[9]+vpos[10]);
label.urt(btex $\ptcrit{\laxpar}$ etex, vpos[6]+vpos[11]);
label.llft(btex $\laxpar\vphantom{I}$ etex, point 1 of paths[1]);

drawdot(vpos[2], 4pt);
drawdot(vpos[3], 4pt);
drawdot(vpos[4], 4pt);
drawdot(vpos[3]+vpos[10], 4pt);
drawdot(vpos[4]+vpos[10], 4pt);
drawcross(vpos[6]+vpos[11], 5pt, 0, 0.5pt, black);
\end{mpostfig}

\begin{mpostfig}[label={CHMEllipticCuts}]
paths[1]:=unitsquare shifted (-0.5,-0.5) xscaled 6xu yscaled 4xu;

vpos[1]:=(0xu,1xu);
vpos[2]:=(0xu,-1xu);
vpos[3]:=(1.2xu,1.5xu);
vpos[4]:=(1.2xu,-1.5xu);

paths[11]:=fullcircle xscaled 0.5xu  yscaled (0.3xu+abs(vpos[1]-vpos[2])) shifted 0.5[vpos[1],vpos[2]];
paths[12]:=fullcircle xscaled 0.5xu  yscaled (0.3xu+abs(vpos[3]-vpos[4])) shifted 0.5[vpos[3],vpos[4]];
paths[21]:=fullcircle  xscaled (0.3xu+abs(vpos[3]-vpos[1])) yscaled 0.4xu rotated 25 shifted 0.5[vpos[1],vpos[3]];
paths[22]:=fullcircle  xscaled (0.3xu+abs(vpos[4]-vpos[2])) yscaled 0.4xu rotated -25 shifted 0.5[vpos[2],vpos[4]];

fill paths[1] withgreyscale 0.9;

drawarrow (point -0.5 of paths[1])--(point 1.5 of paths[1]) pensize(0.5pt);
drawarrow (point 0.15 of paths[1])--(point 2.85 of paths[1]) pensize(0.5pt);

draw vpos[1]{dir -75}..{dir -105}vpos[2] pensize(2pt);
draw vpos[3]{dir -80}..{dir -100}vpos[4] pensize(2pt);

label.llft(btex $\ptsqrt{\laxpar}_3$ etex, vpos[2]);
label.ulft(btex $\ptsqrt{\laxpar}_1$ etex, vpos[1]);
label.lrt(btex $\ptsqrt{\laxpar}_4$ etex, vpos[4]);
label.urt(btex $\ptsqrt{\laxpar}_2$ etex, vpos[3]);
label.llft(btex $\laxpar$ etex, point 2 of paths[1]);
label.llft(btex $\Re \laxpar$ etex, point 1.5 of paths[1]);
label.llft(btex $\Im \laxpar$ etex, point 2.85 of paths[1]);

drawdot(vpos[1], 4pt);
drawdot(vpos[2], 4pt);
drawdot(vpos[3], 4pt);
drawdot(vpos[4], 4pt);
\end{mpostfig}

\begin{mpostfig}[label={CHMEllipticCuts2}]
paths[1]:=unitsquare shifted (-0.5,-0.5) xscaled 6xu yscaled 4xu;

vpos[1]:=(0xu,1xu);
vpos[2]:=(0xu,-1xu);
vpos[3]:=(1.2xu,1.4xu);
vpos[4]:=(1.2xu,-1.4xu);

paths[11]:=fullcircle xscaled 0.5xu  yscaled (0.3xu+abs(vpos[1]-vpos[2])) shifted 0.5[vpos[1],vpos[2]];
paths[12]:=fullcircle xscaled 0.5xu  yscaled (0.3xu+abs(vpos[3]-vpos[4])) shifted 0.5[vpos[3],vpos[4]];
paths[21]:=fullcircle  xscaled (0.3xu+abs(vpos[3]-vpos[1])) yscaled 0.4xu rotated 25 shifted 0.5[vpos[1],vpos[3]];
paths[22]:=fullcircle  xscaled (0.3xu+abs(vpos[4]-vpos[2])) yscaled 0.4xu rotated -25 shifted 0.5[vpos[2],vpos[4]];

fill paths[1] withgreyscale 0.9;

drawarrow (point -0.5 of paths[1])--(point 1.5 of paths[1]) pensize(0.5pt);
drawarrow (point 0.15 of paths[1])--(point 2.85 of paths[1]) pensize(0.5pt);

paths[2]:=vpos[1]{dir -75}..{dir -105}vpos[2];
paths[3]:=vpos[3]{dir -80}..{dir -100}vpos[4];
paths[4]:=reverse(paths[2] shifted (-0.6xu,0))..(paths[2] shifted (0.6xu,0))..cycle;
paths[5]:=(point 0.85 of paths[2]){dir -10}..(point 0.78 of paths[3]){dir -10}
  ..((0.2xu,-0.5xu)+point 1 of paths[3]){dir -175}..{dir 160}((-0.5xu,-0.8xu)+point 1 of paths[2])..cycle;

draw paths[2] pensize(2pt);
draw paths[3] pensize(2pt);
draw paths[4];
draw subpath (1,4) of paths[5];
draw subpath (0,1) of paths[5] dashed withdots scaled 0.5;

label.llft(btex $\ptsqrt{\laxpar}_3$ etex, vpos[2]);
label.ulft(btex $\ptsqrt{\laxpar}_1$ etex, vpos[1]);
label.lrt(btex $\ptsqrt{\laxpar}_4$ etex, vpos[4]);
label.urt(btex $\ptsqrt{\laxpar}_2$ etex, vpos[3]);
label.llft(btex $\laxpar$ etex, point 2 of paths[1]);
label.llft(btex $\Re \laxpar$ etex, point 1.5 of paths[1]);
label.llft(btex $\Im \laxpar$ etex, point 2.85 of paths[1]);
midarrow(paths[4], 0.3);
label.lft(btex $\contacyc$ etex, point 0.3 of paths[4]);
midarrow(paths[5], 3.3);
label.lft(btex $\contbcyc$ etex, point 3.3 of paths[5]);

drawdot(vpos[1], 4pt);
drawdot(vpos[2], 4pt);
drawdot(vpos[3], 4pt);
drawdot(vpos[4], 4pt);
\end{mpostfig}

\begin{mpostfig}[label={CHMEllipticTorus}]
interim xu:=1cm;
interim ahlength:=5pt;

nums[1]:=2.0xu;
nums[3]:=2.5xu;

nums[2]:=1.5xu;
nums[4]:=2.0xu;

paths[1]:=unitsquare shifted (-0.5,-0.5) xscaled 2nums[3] yscaled 2nums[4];
paths[2]:=unitsquare shifted (-0.5,-0.5) xscaled 2nums[1] yscaled 2nums[2];

vpos[1]:=(0,0);
vpos[2]:=(0,nums[2]);
vpos[12]:=(0,-nums[2]);
vpos[3]:=(nums[1],0);
vpos[13]:=(-nums[1],0);
vpos[4]:=(nums[1],nums[2]);
vpos[14]:=(-nums[1],nums[2]);
vpos[24]:=(-nums[1],-nums[2]);
vpos[34]:=(nums[1],-nums[2]);

fill paths[1] withcolor 0.7[0.8white,white];
fill (subpath(1.5,3.5) of paths[2])--cycle withcolor 0.05[0.8white,blue];
fill (subpath(3.5,5.5) of paths[2])--cycle withcolor 0.05[0.8white,green];

drawarrow (point -0.5 of paths[1])--(point 1.5 of paths[1]) pensize(0.5pt);
drawarrow (point 0.5 of paths[1])--(point 2.5 of paths[1]) pensize(0.5pt);

draw ((-nums[3],0)--(+nums[3],0)) shifted vpos[2] pensize(2pt);
draw ((-nums[3],0)--(+nums[3],0)) pensize(2pt);
draw ((-nums[3],0)--(+nums[3],0)) shifted vpos[12] pensize(2pt);
draw ((0,-nums[4])--(0,+nums[4])) shifted vpos[3] pensize(0.25pt) dashed evenly;
draw ((0,-nums[4])--(0,+nums[4])) shifted vpos[13] pensize(0.25pt) dashed evenly;

paths[3]:=(point 2.125 of paths[2])--(point 0.875 of paths[2]);
paths[4]:=(point 3.2 of paths[2])--(point 1.8 of paths[2]);
draw paths[3] pensize(0.25pt);
draw paths[4] pensize(0.25pt);
midarrow (paths[3],0.35);
midarrow (paths[4],0.65);
label.lft(btex $\contbcyc$ etex, point 0.35 of paths[3]);
label.bot(btex $\contacyc$ etex, point 0.65 of paths[4]);

label.llft(btex $\laxzpar$ etex, point 2 of paths[1]);
label.urt(btex $+\iunit\ellKvalc$ etex, vpos[2]);
label.lrt(btex $-\iunit\ellKvalc$ etex, vpos[12]);
label.urt(btex $+\ellKval$ etex rotated 90, vpos[3]);
label.ulft(btex $-\ellKval$ etex rotated 90, vpos[13]);
label.urt(btex $\ptsqrt{\laxzpar}_3$ etex, vpos[1]);
label.ulft(btex $\ptsqrt{\laxzpar}_1$ etex, vpos[3]);
label.llft(btex $\ptsqrt{\laxzpar}_2$ etex, vpos[4]);
label.lrt(btex $\ptsqrt{\laxzpar}_4$ etex, vpos[2]);

drawdot(vpos[1], 5pt);
drawdot(vpos[2], 5pt);
drawdot(vpos[3], 5pt);
drawdot(vpos[4], 5pt);
\end{mpostfig}

\begin{mpostfig}[label={CHMTrigonometricDivisor}]
paths[1]:=unitsquare shifted (-0.5,-0.5) xscaled 5xu yscaled 3xu;

vpos[1]:=(0xu,0.8xu);
vpos[2]:=(0xu,-0.8xu);

paths[11]:=fullcircle xscaled 0.5xu  yscaled (0.3xu+abs(vpos[1]-vpos[2])) shifted 0.5[vpos[1],vpos[2]];

paths[2]:=vpos[1]{dir -80}..{dir -100}vpos[2];
paths[3]:=reverse(paths[2] shifted (0.5xu,0))..(paths[2] shifted (-0.5xu,0))..cycle;

fill paths[1] withgreyscale 0.9;

drawarrow (point -0.5 of paths[1])--(point 1.5 of paths[1]) pensize(0.5pt);
drawarrow (point 0.15 of paths[1])--(point 2.85 of paths[1]) pensize(0.5pt);

draw paths[2] pensize(2pt);
draw paths[3] dashed evenly;
midarrow(reverse(paths[3]),1.8);
midarrow(reverse(paths[3]),3.8);
drawcross(point 0.75 of paths[3], 4pt, 0, 1pt, black);
label.urt(btex $\ptdiv{\laxpar}(x)$ etex, point 0.75 of paths[3]);

label.bot(btex $\ptsqrt{\laxpar}^\conj$ etex, vpos[2]);
label.top(btex $\ptsqrt{\laxpar}$ etex, vpos[1]);
label.llft(btex $\laxpar$ etex, point 2 of paths[1]);
label.llft(btex $\Re \laxpar$ etex, point 1.5 of paths[1]);
label.lrt(btex $\Im \laxpar$ etex, point 2.85 of paths[1]);

drawdot(vpos[1], 4pt);
drawdot(vpos[2], 4pt);
\end{mpostfig}

\begin{mpostfig}[label={CHMLimitCuts}]

  paths[1]:=unitsquare shifted (-0.5,-0.5) xscaled 9xu yscaled 5xu;

  fill paths[1] withgreyscale 0.9;

  drawarrow (point 3.5 of paths[1])--(point 1.5 of paths[1]);

  vpos[10]:=(-1.85xu,-1.5xu);
  for i:=1 upto 2: vpos[10+i]:=vpos[10]+0.15xu*dir(50+180*i); endfor
  for i:=0 upto 2: vpos[20+i]:=vpos[10+i] yscaled -1; endfor

  vpos[30]:=(1.75xu,-2xu);
  for i:=1 upto 4: vpos[30+i]:=vpos[30]+0.25xu*dir(35+90*i); endfor
  for i:=0 upto 4: vpos[40+i]:=vpos[30+i] yscaled -1; endfor

  fill (subpath (0,4) of fullcircle)--cycle xscaled 8.5xu yscaled 1xu shifted (0,0) withgreyscale 0.7;
  label.top(btex $\abs{\ismrefl}$ etex, (0xu,0.5xu));
  label.urt(btex $\ptsol{\laxpar}$ etex, vpos[20]);
  label.urt(btex $\ptsol{\laxpar}$ etex, vpos[44]);
  label(btex $\Delta\sim1/\fldlen$ etex, (0xu,-0.5xu));
  label.rt(btex $\Delta\sim\eunit^{-\ast \fldlen}$ etex, vpos[34]);

  for i:=-10 upto 10:
  draw ((0,-1)--(0,1)) scaled (sqrt(100-(i**2))*0.02xu) shifted (0xu+i*0.4xu,0) pensize(2pt);
  endfor

  draw vpos[10]--vpos[20] pensize(2pt) withgreyscale 0.3;
  draw vpos[30]--vpos[40] pensize(2pt) withgreyscale 0.3;

  for i:=1 upto 2:
    drawdot(vpos[10+i], 4pt);
    drawdot(vpos[20+i], 4pt);
    draw vpos[10+i]--vpos[10] pensize(1.5pt);
    draw vpos[20+i]--vpos[20] pensize(1.5pt);
  endfor

  for i:=1 upto 4:
    drawdot(vpos[30+i], 4pt);
    drawdot(vpos[40+i], 4pt);
    draw vpos[30+i]--vpos[30] pensize(1.5pt);
    draw vpos[40+i]--vpos[40] pensize(1.5pt);
  endfor

  label.llft(btex $\laxpar$ etex, point 2 of paths[1]);

\end{mpostfig}

\begin{mpostfig}[label={CHMContinuumCuts}]
paths[1]:=unitsquare shifted (-0.5,-0.5) xscaled 9.25xu yscaled 2.5xu;
fill paths[1] withgreyscale 0.9;
label.llft(btex $\ismmom$ etex, point 2 of paths[1]);

drawarrow (point 3.5 of paths[1])--(point 1.5 of paths[1]);

for i:=-7 upto 7:
paths[i+10]:=((0,-1)--(0,1)) scaled ((1+0.2*sind(i*30))*sqrt(100-((i+3)**2))/100*3xu) shifted (0xu+i*0.6xu,0);
paths[i+30]:=fullcircle xscaled 0.4xu yscaled (0.4xu+2*ypart(point 1 of paths[i+10])) shifted (point 0.5 of paths[i+10]);
draw paths[i+10] pensize(2pt);
drawdot(point 0 of paths[i+10],4pt);
drawdot(point 1 of paths[i+10],4pt);
draw paths[i+30] pensize(0.25pt) withgreyscale 0.5;
drawcross(point (0.7i) of paths[i+30],5pt,0,0.5pt,black);
endfor
\end{mpostfig}

\begin{mpostfig}[label={CHMSolitonFlower}]
for i:=1 upto 10:
vpos[i]:=1xu*dir(i*36-96);
paths[i]:=(0.025xu*(13-i),-2xu)--(0.025xu*(13-i),0xu-((i-5)**2)*0.02xu){dir 90}..{vpos[i]}vpos[i];
drawdot(vpos[i],4pt);
draw paths[i];
endfor;
label.lrt(btex $n+9$ etex, vpos[1]);
label.rt(btex $n+8$ etex, vpos[2]);
label.rt(btex $n+7$ etex, vpos[3]);
label.urt(btex $n+6$ etex, vpos[4]);
label.top(btex $n+5$ etex, vpos[5]);
label.ulft(btex $n+4$ etex, vpos[6]);
label.lft(btex $n+3$ etex, vpos[7]);
label.lft(btex $n+2$ etex, vpos[8]);
label.lft(btex $n+1$ etex, vpos[9]);
label.llft(btex $n$ etex, vpos[10]);
\end{mpostfig}

\begin{mpostfig}[label={CHMSolitonTwirl}]
for i:=1 upto 10:
paths[i]:=lambdapath(j,0,1.75-i/10,0.1,0.4xu*mexp(j*256)*dir(i*36-96+j*360))
 ..{dir(-90)}(-2.2xu*mexp(-i/10*256),-1.5xu);
drawdot(point 0 of paths[i],4pt);
draw paths[i];
endfor;
\end{mpostfig}

\begin{mpostfig}[label={BranchCutTransition1}]
paths[1]:=unitsquare shifted (-0.5,-0.5) xscaled 4xu yscaled 3xu;
fill paths[1] withgreyscale 0.9;

paths[2]:=((0,-0.6xu)--(0,0.6xu)) shifted (-0.6xu,0);
paths[3]:=((0,-1xu)--(0,1xu)) shifted (+0.0xu,0);

draw paths[2] pensize(2pt) withcolor 0.8blue;
draw paths[3] pensize(2pt) withcolor 0.6green;
drawarrow subpath (0.2,0.8) of ((point 0.5 of paths[2])--(point 0.5 of paths[3])) withgreyscale 0.5;

label.ulft(btex $\modenum_j$ etex, point 0.5 of paths[2]);
label.llft(btex $\actvar_j$ etex, point 0.5 of paths[2]);
label.urt(btex $\modenum_{j+1}$ etex, point 0.5 of paths[3]);
label.lrt(btex $\actvar_{j+1}$ etex, point 0.5 of paths[3]);

drawdot(point 0 of paths[2], 4pt) withcolor 0.8blue;
drawdot(point 1 of paths[2], 4pt) withcolor 0.8blue;
drawdot(point 0 of paths[3], 4pt) withcolor 0.6green;
drawdot(point 1 of paths[3], 4pt) withcolor 0.6green;
\end{mpostfig}

\begin{mpostfig}[label={BranchCutTransition2}]
paths[1]:=unitsquare shifted (-0.5,-0.5) xscaled 4xu yscaled 3xu;
fill paths[1] withgreyscale 0.9;

paths[2]:=((0,-1xu)--(0,1xu)) shifted (-0.0xu,0);
paths[3]:=((0,-0.6xu)--(0,0.6xu)) shifted (+0.6xu,0);
drawarrow subpath (0.2,0.8) of ((point 0.5 of paths[2])--(point 0.5 of paths[3])) withgreyscale 0.5;

draw paths[2] pensize(2pt) withcolor 0.6green;
draw paths[3] pensize(2pt) withcolor 0.8blue;

label.ulft(btex $\modenum'_j$ etex, point 0.5 of paths[2]);
label.llft(btex $\actvar'_j$ etex, point 0.5 of paths[2]);
label.urt(btex $\modenum'_{j+1}$ etex, point 0.5 of paths[3]);
label.lrt(btex $\actvar'_{j+1}$ etex, point 0.5 of paths[3]);

drawdot(point 0 of paths[2], 4pt) withcolor 0.6green;
drawdot(point 1 of paths[2], 4pt) withcolor 0.6green;
drawdot(point 0 of paths[3], 4pt) withcolor 0.8blue;
drawdot(point 1 of paths[3], 4pt) withcolor 0.8blue;
\end{mpostfig}

\begin{mpostfig}[label={CHMCurve}]
paths[1]:=unitsquare shifted (-0.5,-0.5) xscaled 7xu yscaled 4xu;
fill paths[1] withgreyscale 0.9;
label.llft(btex $\laxpar$ etex, point 2 of paths[1]);

drawarrow (point 3.5 of paths[1])--(point 1.5 of paths[1]);
drawarrow (point 0.5 of paths[1])--(point 2.5 of paths[1]);

label.llft(btex $\Re \laxpar$ etex, point 1.5 of paths[1]);
label.llft(btex $\Im \laxpar$ etex, point 2.5 of paths[1]);

paths[11]:=(-2.5xu,0.8xu)..(-2.6xu,0xu)..(-2.5xu,-0.8xu);
paths[12]:=(-1.0xu,1.2xu)..(-1.2xu,0xu)..(-1.0xu,-1.2xu);
paths[13]:=(1.5xu,1.0xu)..(1.7xu,0xu)..(1.5xu,-1.0xu);

for i:=1 upto 3:
paths[20+i]:=fullcircle xscaled 1.0xu yscaled (0.6xu+2ypart(point 0 of paths[10+i]))
  shifted 0.5[point 0 of paths[10+i],point 2 of paths[10+i]];
paths[30+i]:=((point 0.43 of paths[1]){dir 90}..(0,0){dir 0}..{dir -90}(point 0.57 of paths[1])) shifted (point 1 of paths[10+i]);
endfor

vpos[11]:=point 4.41234 of paths[21];
vpos[12]:=point 3.11234 of paths[22];
vpos[13]:=point 0.31234 of paths[23];

for i:=1 upto 3:
draw paths[i+20];
draw subpath (0,1) of paths[i+30] dashed withdots scaled 0.4 pensize(0.25pt);
draw subpath (1,2) of paths[i+30] pensize(0.25pt);
draw paths[i+10] pensize(2pt);
drawdot(point 0 of paths[i+10],4pt);
drawdot(point 2 of paths[i+10],4pt);
drawcross(vpos[10+i],5pt,0,0.5pt,black);
midarrow(paths[20+i],6);
midarrow(reverse(paths[30+i]),0.2);
endfor

label.lrt(btex $\ptsqrt{\laxpar}_1$ etex, point 0 of paths[11]+(-0.06xu,0));
label.urt(btex $\ptsqrt{\laxpar}_1^\conj$ etex, point 2 of paths[11]+(-0.06xu,0));
label.lrt(btex $\ptsqrt{\laxpar}_2$ etex, point 0 of paths[12]+(-0.08xu,0));
label.urt(btex $\ptsqrt{\laxpar}_2^\conj$ etex, point 2 of paths[12]+(-0.08xu,0));
label.llft(btex $\ptsqrt{\laxpar}_3$ etex, point 0 of paths[13]+(0.1xu,0));
label.ulft(btex $\ptsqrt{\laxpar}_3^\conj$ etex, point 2 of paths[13]+(0.1xu,0.1xu));

label.bot(btex $\contacyc_1$ etex, point 6 of paths[21]);
label.bot(btex $\contacyc_2$ etex, point 6 of paths[22]);
label.bot(btex $\contacyc_3$ etex, point 6 of paths[23]);

label.rt(btex $\contbcyc_1$ etex, point 1.8 of paths[31]);
label.rt(btex $\contbcyc_2$ etex, point 1.8 of paths[32]);
label.rt(btex $\contbcyc_3$ etex, point 1.8 of paths[33]);

label.llft(btex $\ptdiv{\laxpar}_1$ etex, vpos[11]);
label.ulft(btex $\ptdiv{\laxpar}_2$ etex, vpos[12]);
label.urt(btex $\ptdiv{\laxpar}_3$ etex, vpos[13]);
\end{mpostfig}

\begin{mpostfig}[label={CHMSpinSoliton3D}]
paths[4]:=unitsquare shifted (-0.5,-0.5) rotated 90 slanted -0.3 rotated -90 xscaled 0.4 scaled 2.5xu shifted (-0.8xu,0);
drawarrow center(paths[4])--(point 2.5 of paths[4]);
drawarrow center(paths[4])--(point 1.5 of paths[4]);
label.ulft(btex $\Im\spincp$ etex, point 2.5 of paths[4]);
label.top(btex $\Re\spincp$ etex, point 1.5 of paths[4]);
draw paths[4] dashed evenly withgreyscale 0.5;
paths[5]:=(-5xu,0)--(+5xu,0);
draw paths[5] withgreyscale 0.5;
midarrow(paths[5],0.71);
label.bot(btex $x$ etex, point 0.71 of paths[5]);
paths[1]:=lambdapath(i,-50,50,1,((i*0.1xu,0)+((dir (i*20+30)) rotated 90 slanted -0.3 rotated -90 xscaled 0.4)*3xu*0.5* (2/(mexp(-25*i)+mexp(25*i)))));
draw paths[1] pensize(2pt);
midarrow(paths[1],45.5);
midarrow(paths[1],55.5);
vpos[1]:=point 50 of paths[1];
drawdot(vpos[1],4pt);
label.urt(btex $\parpos$ etex, vpos[1]);
paths[3]:=((0,0)--(1xu,0)) shifted point 50 of paths[1];
drawarrow paths[3];
label.top(btex $\parvel$ etex, point 0.6 of paths[3]);
paths[2]:=subpath(0,-2) of fullcircle scaled 2xu rotated 90 slanted -0.3 rotated -90 xscaled 0.4;
paths[2]:=paths[2] shifted (-point 0 of paths[2]) shifted (point 0.5 of paths[3]);
drawarrow paths[2];
label.lrt(btex $\parang$ etex, point 0 of paths[2]);
label.rt(btex $\parfreq$ etex, point 1.0 of paths[2]);
\end{mpostfig}

\begin{mpostfig}[label={CHMScatteringData}]
paths[1]:=unitsquare shifted (-0.5,0) xscaled 3.9xu yscaled 2.0xu;
paths[3]:=unitsquare shifted (-0.5,-1) xscaled 3.9xu yscaled 0.7xu;
vpos[1]:=(-0.7xu,0.7xu);
vpos[2]:=(0.8xu,0.3xu);
vpos[3]:=(-0.2xu,1.3xu);
vpos[4]:=(0,1.0xu);
vpos[5]:=(-1.8xu,0);
fill paths[1] withgreyscale 0.9;
fill paths[3] withgreyscale 0.8;
label.llft(btex $\ismmom$ etex, point 2 of paths[1]);
drawarrow subpath (3,2) of paths[3];
drawarrow (point 0.5 of paths[3])--(point 2.5 of paths[1]);
draw subpath (0,1) of paths[1] pensize(1.5pt) withcolor 0.6green;
drawdot (vpos[1],5pt) withcolor 0.6green;
drawdot (vpos[2],5pt) withcolor 0.6green;
drawdot (vpos[3],5pt) withcolor 0.6green;
label.llft(btex $\Im \ismmom$\,etex, point 2.5 of paths[1]);
label.lft(btex $\ptsol{\ismmom}_n,\ismres_n$ etex, vpos[1]);
label.ulft(btex $\Re \ismmom_{}$etex, point 1 of paths[1]);
label.bot(btex $\ismrefl(\ismmom)$ etex, point 0.25 of paths[1]);
\end{mpostfig}

\mpostdone

\metaset{title}{Infinite-Length Limit of Spectral Curves and Inverse Scattering}
\metaset[print]{title}{Infinite-Length Limit of\\Spectral Curves and Inverse Scattering}
\metaset{author}{Niklas Beisert, Kunal Gupta}
\metaset[print]{author}{Niklas Beisert\textsuperscript{1}, Kunal Gupta\textsuperscript{2}}

\begin{document}

\iftrue
\pdfbookmark[1]{Title Page}{title}
\thispagestyle{empty}

\begingroup\raggedleft\footnotesize\ttfamily
%\arxivlink{yymm.nnnnnn}
UUITP-26/24
\par\endgroup

\vspace*{2cm}
\begin{center}%
\begingroup\Large\bfseries\metapick[print]{title}\par\endgroup
\vspace{1cm}

\begingroup\scshape
\metapick[print]{author}
\endgroup
\vspace{5mm}

\textsuperscript{1} 
\textit{Institut für Theoretische Physik,\\
Eidgenössische Technische Hochschule Zürich,\\
Wolfgang-Pauli-Strasse 27, 8093 Zürich, Switzerland}
\par\vspace{1mm}
\texttt{nbeisert@itp.phys.ethz.ch}
\vspace{5mm}

\textsuperscript{2}
\textit{Department of Physics and Astronomy,\\
Uppsala University,\\
Box 516, 751 20 Uppsala, Sweden}
\par\vspace{1mm}
\texttt{kunal.gupta@physics.uu.se}
\vspace{8mm}

\vfill

\ifarxiv\else
\begingroup\bfseries\Large DRAFT\endgroup
\par
\vfill
\fi

\textbf{Abstract}\vspace{5mm}

\begin{minipage}{12.7cm}
Integrability equips models of theoretical physics
with efficient methods for the exact construction of useful states and their evolution.
Relevant tools for classical integrable field models in one spatial dimensional
are spectral curves in the case of periodic fields
and inverse scattering for asymptotic boundary conditions.
Even though the two methods are quite different in many ways,
they ought to be related by taking the periodicity length of closed boundary conditions to infinity.

Using the Korteweg-de Vries equation and the continuous Heisenberg magnet
as prototypical classical integrable field models,
we discuss and illustrate how data for spectral curves transforms into asymptotic scattering data.
In order to gain intuition and also for concreteness,
we review how the elliptic states of these models degenerate into solitons at infinite length.
\end{minipage}

\vspace*{4cm}

\end{center}

\ifpublic
\newpage
\tableofcontents
\fi

\newpage
\fi

%%%%%%%%%%%%%%%%%%%%%%%%%%%%%%%%%%%%%%%%%%%%%%%%%%%%%%%%%%%%%%%%%%%%%%%%%%%%%%%%
%%%%%%%%%%%%%%%%%%%%%%%%%%%%%%%%%%%%%%%%%%%%%%%%%%%%%%%%%%%%%%%%%%%%%%%%%%%%%%%%
\section{Introduction and Overview}
\label{sec:intro}

The classical inverse scattering method and
the method of classical spectral curves
(also known as finite-gap method or inverse spectral problem)
are two well-known and established techniques to analyse and construct general solutions
of the equations of motions for classical integrable chain and field models.

The inverse scattering method addresses classical integrable models
in one spatial dimension with asymptotic boundary conditions.
This method was devised by Gardner, Greene, Kruskal and Miura in \cite{Gardner:1967wc}
to exactly construct soliton solutions in the Korteweg de Vries (KdV) equation.
Subsequently, Lax developed the framework of Lax pairs in \cite{Lax:1968fm}
which established an algebraic foundation for multi-soliton states.
Integrability of the KdV model was established in
\cite{miura1968korteweg2,ZakharovFaddeev1971,gardner1971}.
Since then, the inverse scattering framework has been applied to many other systems
like the nonlinear Schrödinger equation (NLS) \cite{shabat1972exact}
and the continuous Heisenberg magnet (CHM) \cite{Takhtajan:1977rv,fogedby1980solitons}.
It is based on a scattering problem for an auxiliary linear problem
which transforms the physical phase space into
a set of abstract scattering data.
These scattering data consist of kinematical parameters
for discrete solitonic excitations as well as for a continuum of wave-like excitations.
Importantly, this map is bijective: it is inverted by solving the
Gel'fand--Levitan--Marchenko integral equation \cite{gelfand1951determination,marchenko1952}.
Consideration of the scattering data bears two advantages:
First, even though the scattering data is constructed by an abstract procedure,
it rather immediately represents physical parameters
such as amplitudes and wave numbers for a set of excitations.
Such a description of states is preferable in many situations (especially when it comes to quantisation).
Second, the time evolution of the scattering data is linear and thus readily solved.
In that sense, the map is analogous to a Fourier transformation in many regards
albeit it takes the non-linearity of the model properly and exactly into account.
The inverse scattering method for solving the equations of motion
thus consists in transforming a physical state into abstract scattering data,
performing a trivial time evolution in this space,
and then mapping back to the time-evolved physical state.
It is equally useful to skip the first step
by defining the initial state based on its scattering data.
The main practical difficulty of the inverse scattering method
is that the forward and inverse transforms require solving either differential or integral equations.

The method of spectral curves addresses classical integrable models
with a periodic spatial dimension and thus with closed boundary conditions.
It was devised for the KdV equation
by Novikov, Dubrovin, Its, Matveev \cite{novikov1974periodic,dubrovin1975InvPerShort,its1975Hills,dubrovin1975InvPerLong}
as well as Lax and Marchenko \cite{lax1975periodic,marchenko1974dokl}
based on techniques by Akhiezer \cite{Akhiezer1961},
see also \cite{dubrovin1976survey}.
The spectral curve method was extended to the integrable NLS model
by Kotlyarov and Its \cite{kotlyarov1976periodic,its1976explicit}
and the equivalence of the model and its integrable structure to CHM
was shown by Lakshmanan, Zakharov and Takhtajan \cite{LAKSHMANAN197753,zakharov1979equivalence}.
Here, the same auxiliary linear problem as mentioned above
gives rise to a characteristic matrix, the so-called monodromy,
which encodes all the relevant data contained in the state. %\cite{novikov1995periodic}.
In particular, the spectrum of eigenvalues is conserved
while the direction of eigenvectors evolves with time. %\cite{marchenko1974periodic,lax1975periodic,its1975schrodinger}.
Therefore the monodromy naturally provides a splitting of the degrees of freedom
into conserved and dynamical variables.
Moreover, these physical variables are naturally encoded by a Riemann surface,
the so-called spectral curve,
and certain meromorphic functions on it. %\cite{dubrovin1976schrodinger}.
In this form, the investigation of the physical system
is enhanced by the powerful tools of complex analysis.
In particular, the moduli of such spectral curves again qualitatively translate
to the amplitudes of excitations with well-defined wave numbers
which in this case are quantised due to the periodicity of space.
Time evolution is again simplified and stream-lined in this picture,
and a solution along the lines of the inverse scattering method
applies to the periodic case as well.
Analogously, the transformation between phase space and the abstract spectral curve data
remains difficult, making use of differential and integral equations
or special functions which encode these.

Without going into details, the above two constructive methods
are qualitatively similar, but they are formulated in terms of rather
different sets of abstract data as well as different flavours of mathematical concepts and equations.
Nonetheless, they ought to be related for the following reason:
Models on the infinite line can be understood as limiting cases of
models on a spatial domain of finite extent.
The matching between the closed boundary on the periodic interval
and asymptotic boundaries on the infinite line
may of course introduce some minor adjustments between the two models,
but the bulk degrees of freedom should directly transform into each other.
In that sense, the inverse scattering method
should arise as a limiting case of the method of spectral curves.
Consequently, the data and structures of the two methods should transform
into each other.

Aspects of this issue have been addressed in several works in the past.
The original work \cite{novikov1974periodic} on the spectral curve method for KdV
points out how a soliton is embedded into the genus-one spectral curve for the wave train.
In \cite{its1976problems} the spectral curve solution
is explicitly degenerated into multi-soliton solutions,
and similar results for the nonlinear Schrödinger equation
have been presented in \cite{its1983problems,bbeim1994}.
The works \cite{Osborne:1985,Osborne:1986ua}
address the relationship for small amplitudes
and analyse it for piecewise-constant approximations using numerical methods.
Features of the distribution of branch cuts for spectral curves
in the sine-Gordon model at large periodicity length
have been presented in \cite{ForestMcL1982}.

\medskip

The purpose of the present investigation is to work out
the relationship between the two integrability methods in detail
and how their corresponding sets of data transform into each other.
In particular, we observe how to meet the asymptotic boundary conditions in the infinite-length limit
(sometimes also called decompactification limit
as the compact closed circle turns into the non-compact infinite line)
and how this turns the states on the periodic domain into
a combination of discrete solitons and a continuum of wave-like solutions.

We shall perform the detailed analysis for the sample model
of the KdV equation.
In order to gather experience in taking the infinite-length limit,
we shall shall consider the spectral curve representation of the
elliptic wave train solution of the KdV model.
This class of states is particular suited for this purpose
because it can be written explicitly in terms of elliptic functions
and it has several parameters that are either used
to perform the infinite-length limit or that remain as tuneable parameters in the limit.

We shall then proceed to the general case.
Starting from a generic state on the infinite line
as described through its scattering data,
we introduce a one-parameter family of periodic states
with varying periodicity length.
We then approximate their spectral curve representations
using the scattering data.
This provides a clear picture of how to set up a sequence of spectral curves
such that their corresponding states have a well-defined infinite-length limit.

Finally, we shall redo the analysis for the CHM model
which is another prototypical integrable model of a one-dimensional field.
As such it shares many properties with the KdV model including
the applicability of an inverse scattering method and a spectral curve method,
and one may expect the infinite-length limit to work along the same lines.
While this will turn out to be true, some features of the abstract data in these two models are qualitatively different
because a different signature applies to the underlying structures of integrability.
This implies several adjustments to the precise formulation of the infinite-length limit
of the spectral curve method.

The present article is organised as follows: 
In \secref{sec:KdVElliptic} we use the example
of elliptic travelling wave solutions of the KdV model
to introduce and explore the infinite-length limit
for a well-tractable class of states.
In the subsequent \secref{sec:KdVFiniteLengthExtrapolation}
we discuss a scheme for extracting and relating data
in the infinite-length limit for generic states of the inverse scattering
and spectral curve methods for the KdV model.
We then apply the scheme to the CHM model in \secref{sec:CHM}
in order to understand its universality and the corresponding adjustments.
In \secref{sec:conclusions} we conclude and provide an outlook.
Finally, we point out that a qualitative summary and illustration of our results
is provided in \secref{KdVAsympLim} and \secref{CHMAsympLim}.

%%%%%%%%%%%%%%%%%%%%%%%%%%%%%%%%%%%%%%%%%%%%%%%%%%%%%%%%%%%%%%%%%%%%%%%%%%%%%%%%
%%%%%%%%%%%%%%%%%%%%%%%%%%%%%%%%%%%%%%%%%%%%%%%%%%%%%%%%%%%%%%%%%%%%%%%%%%%%%%%%
\section{KdV Elliptic States}
\label{sec:KdVElliptic}

We start by gaining some experience on the infinite-length limit for the KdV equation.
To that end, we consider the well-known class of fixed-profile states
which contains periodic as well as infinite-length solutions
and which is solved by elliptic functions.

%%%%%%%%%%%%%%%%%%%%%%%%%%%%%%%%%%%%%%%%%%%%%%%%%%%%%%%%%%%%%%%%%%%%%%%%%%%%%%%%
\subsection{Travelling Wave Solution}
\label{travellingkdv}

We will use the KdV equation for the field $\kdvfld(x,t)$
in the standard normalisation:
\[
\dot \kdvfld = 6\kdvfld \kdvfld' -\kdvfld'''.
\label{KdVeom}
\]
The class of elliptic solutions to this equation is well-known.
It can be obtained by an ansatz
describing a fixed profile $\kdvfld(x)$
moving at constant velocity $\parvel$:
\[
\kdvfld(x,t)=\kdvfld(x-\parvel t).
\label{KdVtravellingansatz}
\]
It reduces the KdV equation
to the ordinary differential equation
\[
6\kdvfld \kdvfld' -\kdvfld'''+\parvel  \kdvfld'=0.
\label{KdVdiffeqn1}
\]
Every solution to this equation
also solves the KdV equation at all $x$ and $t$.
This special feature is due to the system being homogeneous in both time and space.
\unskip\footnote{A combined spatial and temporal translation with speed $\parvel$
maps the coordinates $(x,t)$ to $(x-\parvel t,0)$,
and it therefore suffices that $\kdvfld(x-\parvel t)$ solves the equation at every $x-\parvel t$.}

The reduced KdV equation integrates straight-forwardly to
\[
3\kdvfld^2 -\kdvfld''+\parvel \kdvfld-\gamma_1=0.
\label{KdVdiffeqn2}
\]
Another exact integration is enabled upon multiplication by $\kdvfld'$:
\[
\kdvfld^3 -\half\kdvfld'^2+\half \parvel \kdvfld^2-\gamma_1 \kdvfld-\gamma_2=0.
\label{KdVdiffeqn3}
\]
The two coefficients $\gamma_1$ and $\gamma_2$ are integration constants.
Equation \eqref{KdVdiffeqn3} is further integrated by separation of variables,
and upon inversion of an elliptic integral one finds the full solution
\[
\kdvfld=-\rfrac{1}{6}\parvel-\rfrac{2}{3}\parscl^2(1-2\ellmods)-2\parscl^2\ellmods\ellCN(\parscl x+\paroff,\ellmods)^2.
\eqsep
\mpostusemath{KdVEllipticState}
\label{KdVcnoidalwave}
\]
%
%The parameters are related as
%%
%\begin{align}
%\gamma_1
%&=
%-\rfrac{1}{12}\parvel^2
%+\rfrac{4}{3}\parscl^4 (1-\ellmods+\ellmods^2)
%\\
%\gamma_2
%&=
%-\rfrac{1}{216}\parvel^3
%+\rfrac{2}{9}\parscl^4 \parvel (1-\ellmods+\ellmods^2)
%+\rfrac{8}{27}\parscl^6 (2-3\ellmods-3\ellmods^2+2\ellmods^3)
%\end{align}
%%
%
This is the well-known wave train or cnoidal wave solution
which was described in \cite{russell1845report}
and then analysed in \cite{boussinesq1877essai,kordeweg1895change}.
Here, $\ellCN(z,\ellmods)$ denotes the Jacobi elliptic cosine function.
\unskip\footnote{We shall omit the elliptic modulus $\ellmods$
in most occurrences of elliptic functions,
e.g.\ $\ellCN(z):=\ellCN(z,\ellmods)$.}
Next to the velocity $\parvel$, the solution has three degrees of freedom
as expected for a differential equation of third order:
These are the elliptic modulus $\ellmods$,
the inverse width parameter $\parscl$
as well as the offset $\paroff$.
All these parameters are real,
while the elliptic modulus belongs to the principal interval $0\leq \ellmods \leq 1$,
and we will assume $\parscl>0$.

%%%%%%%%%%%%%%%%%%%%%%%%%%%%%%%%%%%%%%%%%%%%%%%%%%%%%%%%%%%%%%%%%%%%%%%%%%%%%%%%
\subsection{Auxiliary Linear Problem}
\label{sec:KdVALP}

In order to formulate the above state in terms of spectral curves or inverse scattering data,
let us construct the corresponding solution of the auxiliary linear problem
\[
\laxvec' = \laxconn \laxvec
\eqjoin{\text{with}}
\laxconn = \begin{pmatrix} 0 & 1 \\ \kdvfld-\laxpar & 0 \end{pmatrix}
\eqjoin*{\text{and}}
\laxvec
= \begin{pmatrix} \laxfunc \\ \laxfunc' \end{pmatrix},
\label{KdVALP}
\]
where $\laxpar$ is called the spectral parameter.
Here, the first row of $\laxconn$ implies that the second component of the vector $\laxvec$
can be expressed universally as the spatial derivative of the first component
which we denote as $\laxfunc$.
By substituting this relationship in the auxiliary linear problem above,
we obtain a second-order linear differential equation for the function $\laxfunc$:
\[
\laxfunc''=(\kdvfld-\laxpar) \laxfunc.
\label{KdVAuxdiffeqn}
\]
Note that it takes the form of a Schrödinger equation
for a non-relativistic particle of mass $m=\hbar^2/2$,
potential $\kdvfld$ and energy $\laxpar$.

For the class of elliptic states $\kdvfld(x)$
introduced in \secref{travellingkdv},
the above differential equation takes the separated form
\[
\frac{\laxfunc''}{\laxfunc}
=
\kdvfld-\laxpar
=
-\rfrac{1}{6}\parvel
-\rfrac{2}{3}\parscl^2(1-2\ellmods)
-\laxpar
-2\parscl^2\ellmods\ellCN(\parscl x+\paroff)^2.
%=
%-\laxpar
%-\rfrac{1}{6}\parvel
%-\rfrac{2}{3}\parscl^2(\ellmods+1)
%+2\parscl^2\ellmods\ellSN(\parscl x+\paroff,\ellmods)^2.
%=
%-\rfrac{1}{6}\parvel
%+\rfrac{2}{3}\parscl^2(2-\ellmods)
%-\laxpar
%-2\parscl^2 \ellDN(\parscl x+\paroff,\ellmods)^2.
\label{KdVellipticauxdiffeqn}
\]
In order to solve it, we first collect some relevant features
of the function on the right-hand side
whose features must be matched by the left-hand side $\laxfunc''/\laxfunc$
with a proper $\laxfunc(x)$.

First, the function $\kdvfld(x)$ is elliptic of minimal degree two.
It is periodic with the length
\[
\fldlen=\frac{2\ellKval}{\parscl},
\eqsep
\ellKval:=\ellK(\ellmods).
\]
In addition, it has an imaginary periodicity in the complexified
spatial coordinate $x$ with length
\[
\fldlen'=\frac{2\iunit\ellKvalc}{\parscl},
\eqsep
\ellKvalc:=\ellK(1-\ellmods).
\]
Furthermore, there is a double pole at $\tilde x=\iunit\ellKvalc/\parscl$
with coefficient $2$, around which we expand \eqref{KdVellipticauxdiffeqn} to get
\[
\frac{\laxfunc''}{\laxfunc}
=
\frac{2}{(x-\tilde x)^2}
+
\frac{0}{x-\tilde x}
+
\ldots.
\]
Apart from this pole, the function has no further singularities.
The double pole on the right-hand side determines the critical behaviour of the function $\laxfunc$
according to
\[
\laxfunc\sim (x-\tilde x)^\kappa
\brk!{
1+0(x-\tilde x)+\ldots
}
\eqjoin*{\implies}
\frac{\laxfunc''}{\laxfunc}
=
\frac{\kappa(\kappa-1)}{(x-\tilde x)^2}
+\frac{0}{x-\tilde x}
+\ldots.
\]
Matching of coefficients implies that $\laxfunc$
must have a simple pole ($\kappa=-1$) or a double zero ($\kappa=2$) at $x=\tilde x$.
The former dominates over the latter,
and thus a pole is the rule whereas a double zero merely serves as an exception.
\unskip\footnote{In a linear combination of both solutions,
the residues of the pole may cancel leaving behind a double zero at this place.
In particular, periodic recurrences of a double zero will typically be poles
rather than further double zeros.}
The periodicities of the differential equation imply that
a shifted solution must again be a solution.
We point out that the solutions to the linear differential equation
span a two-dimensional space, and so the shifted solution can be
a different linear combination of the two basis solutions.
Periodicity is thus specified by $2\times 2$ matrices,
and we can choose solutions such that they are eigenfunctions of periodicity
\unskip\footnote{The genus-one surface has the fundamental group $\Integer\times\Integer$,
consequently the two periodicity matrices for shifts by $\fldlen$ and $\fldlen'$
commute and are diagonalised simultaneously.}
\begin{align}
\laxfunc(x+\fldlen)
&=
\exp(\iunit \laxqmom)\. \laxfunc(x),
\\
\laxfunc(x+\fldlen')
&=
\exp(\iunit \tilde\laxqmom)\. \laxfunc(x).
\end{align}
Altogether, the solutions $\laxfunc$ must be quasi-periodic functions in the sense of the above equation,
and they need to have double poles at $x=\tilde x$ with no further poles or multiple zeros.
Such functions are typically provided by balanced ratios of elliptic theta functions.
A suitable ansatz in terms of the Neville theta function $\ellTN(z,\ellmods)$
is as follows:
\[
\laxfunc
=
\exp \brk!{\iunit (\parscl x+\paroff)\laxqmom/2\ellKval}
\frac{\ellTN(\parscl x+\paroff+\laxzpar)}{\ellTN(\parscl x+\paroff)\ellTN(\laxzpar)}.
\label{KdVellipticpsiansatz}
\]
Here, $\laxqmom$ and $\laxzpar$ are parameters that need to be adjusted
to solve the above auxiliary linear differential equation.
Using the Jacobi zeta function $\ellZN(z,\ellmods)=(\partial/\partial z)\log\ellTN(z,\ellmods)$,
we find
\begin{align}
\laxpar
=
\laxpar(\laxzpar)
&:=
-\rfrac{1}{6}\parvel+\rfrac{1}{3}\parscl^2(\ellmods-2)-\parscl^2\ellCS(\laxzpar)^2,
\\
\laxqmom
=
\laxqmom(\laxzpar)
&:=
2\iunit \brk!{\ellZN(\laxzpar)+\ellCS(\laxzpar)\ellDN(\laxzpar)}\ellKval,
\\
\tilde\laxqmom
=
\tilde\laxqmom(\laxzpar)
&:=
-2\brk!{\ellZN(\laxzpar)+\ellCS(\laxzpar)\ellDN(\laxzpar)}\ellKvalc
-\frac{\pi}{\ellKval}\laxzpar.
\end{align}
Note that the first equation has two solutions $\pm\laxzpar$ for each value of the spectral parameter $\laxpar$
corresponding to the two distinct eigenfunctions.
It therefore makes sense to replace the spectral parameter $\laxpar=\laxpar(\laxzpar)$
by the uniformising parameter $\laxzpar$ on the universal cover of the elliptic surface
which is defined up to shifts by $2\ellKval$ and $2\iunit\ellKvalc$.
Such shifts leave the expressions invariant
up to some shift by integer multiples of $2\pi$:
\begin{align}
\laxfunc(\laxzpar+2\ellKval n+2\iunit\ellKvalc n')
&=
\laxfunc(\laxzpar),
\\
\laxqmom(\laxzpar+2\ellKval n+2\iunit\ellKvalc n')
&=
\laxqmom(\laxzpar)+2\pi n',
\\
\tilde\laxqmom(\laxzpar+2\ellKval n+2\iunit\ellKvalc n')
&=
\tilde\laxqmom(\laxzpar)-2\pi n.
\end{align}
Note that these relations render $\laxfunc$ as well as
the eigenvalues $\exp(\iunit\laxqmom)$ and $\exp(\iunit\tilde\laxqmom)$
properly periodic.

%%%%%%%%%%%%%%%%%%%%%%%%%%%%%%%%%%%%%%%%%%%%%%%%%%%%%%%%%%%%%%%%%%%%%%%%%%%%%%%%
\subsection{Spectral Curve}

Let us discuss the function $\der\laxqmom/\der\laxpar$
describing the spectral curve corresponding to the state $\kdvfld$.
Its form is in line with the expectation for a genus-one spectral curve of the KdV model:
\[
\frac{\der \laxqmom}{\der \laxpar}
=
\frac{\ellKval}{\parscl}
\frac{\laxpar-\ptcrit{\laxpar}}{\sqrt{\laxpar-\ptsqrt{\laxpar}_1}\sqrt{\laxpar-\ptsqrt{\laxpar}_2}\sqrt{\laxpar-\ptsqrt{\laxpar}_3}}.
\label{KdVgenusonespectralcurve}
\]
The square-root branch points $\ptsqrt{\laxpar}_k$
are the critical points of the map between $\laxpar$ and $\laxzpar$:
\begin{align}
\ptsqrt{\laxzpar}_0 &= 0,
&
\ptsqrt{\laxpar}_0 &= \infty,
\\
\ptsqrt{\laxzpar}_1 &= \ellKval,
&
\ptsqrt{\laxpar}_1 &= -\rfrac{1}{6}\parvel+\rfrac{1}{3}\parscl^2(\ellmods-2),
\\
\ptsqrt{\laxzpar}_2 &= \ellKval+\iunit\ellKvalc,
&
\ptsqrt{\laxpar}_2 &= -\rfrac{1}{6}\parvel+\rfrac{1}{3}\parscl^2(1-2\ellmods),
\\
\ptsqrt{\laxzpar}_3 &= \iunit\ellKvalc,
&
\ptsqrt{\laxpar}_3 &= -\rfrac{1}{6}\parvel+\rfrac{1}{3}\parscl^2(\ellmods+1).
\end{align}
We have ordered them such that $\ptsqrt{\laxpar}_1<\ptsqrt{\laxpar}_2<\ptsqrt{\laxpar}_3$.
Furthermore, the zero $\ptcrit{\laxpar}$ of $\der\laxqmom/\der\laxpar$
resides in the interval $\ptsqrt{\laxpar}_2<\ptcrit{\laxpar}<\ptsqrt{\laxpar}_3$,
and it is given by
\[
\ptcrit{\laxpar}=-\rfrac{1}{6}\parvel+\rfrac{1}{3}\parscl^2(\ellmods-2) + \frac{\ellEval}{\ellKval}\parscl^2,
\eqsep
\ellEval:=\ellE(\ellmods).
\]
We arrange the branch cuts to connect pairs of consecutive branch points along the real axis
starting with $\ptsqrt{\laxpar}_0=-\infty$.
Namely there is a semi-infinite interval from
$\ptsqrt{\laxpar}_0=-\infty$ to $\ptsqrt{\laxpar}_1$
and a finite interval extending from $\ptsqrt{\laxpar}_2$ to $\ptsqrt{\laxpar}_3$:
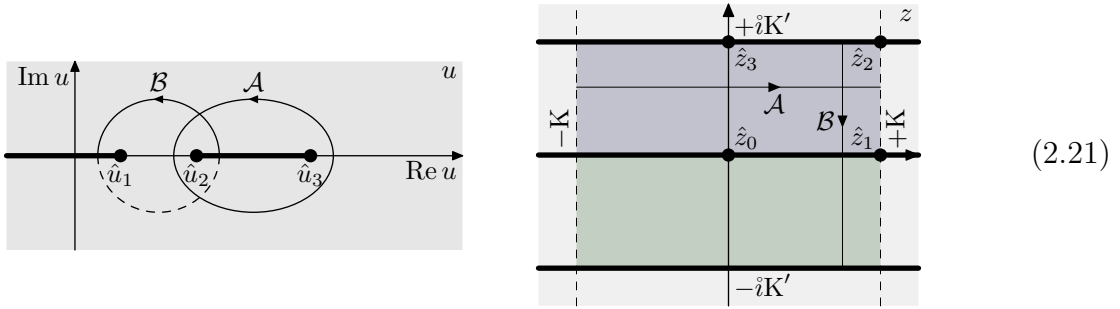
\[
\mpostusemath{KdVEllipticCuts}
\eqsep
\mpostusemath{KdVEllipticTorus}
\]
The shape of the complex function $\laxqmom(\laxpar)$ just above the real axis
is sketched as follows:
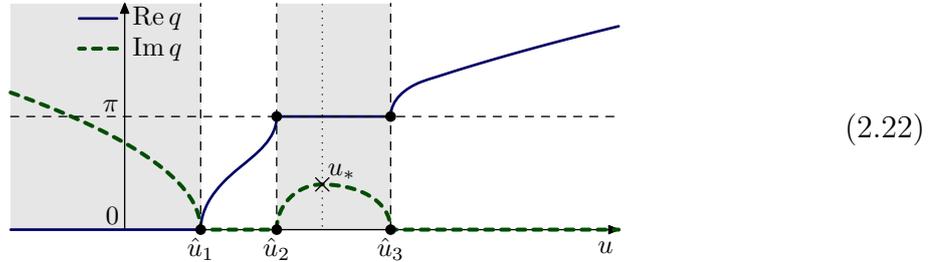
\[
\mpostusemath{KdVEllipticQ}
\]
Just below the real axis, the function takes the complex conjugate value.
Note that the real part resides at $0$ and $\pi$ on the two branch cuts
and monotonically increases otherwise.
The imaginary part vanishes away from the branch cuts
and takes a deformed semi-circle form on them.
The positions of the branch cuts are most clearly identified in the continuous function $\cos\laxqmom(\laxpar)$:
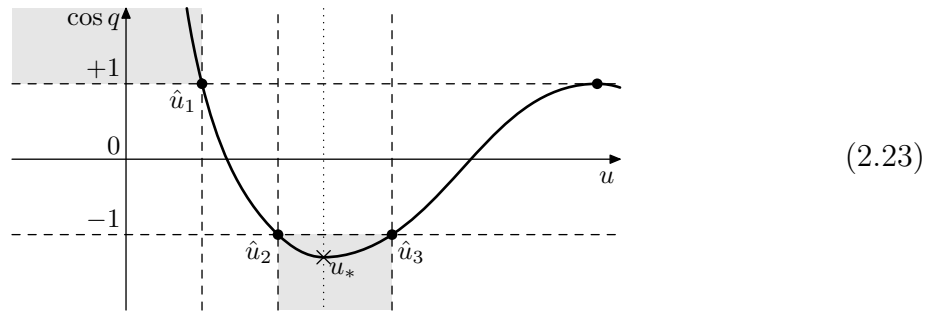
\[
\mpostusemath{KdVEllipticF}
\]
This function is real for real $\laxpar$, but it is not restricted to values between $-1$ and $+1$:
absolute values bounded by $1$ imply a real argument $\laxqmom$ of the cosine function
while absolute values greater than $1$ require the argument $\laxqmom$ to be complex and thus indicate a branch cut.
As $\laxpar$ increases starting from $-\infty$, the function $\cos\laxqmom(\laxpar)$ decreases exponentially from $+\infty$.
At some point $\ptsqrt{\laxpar}_1$ the function passes through the critical value $+1$,
later at $\ptsqrt{\laxpar}_2$ it undershoots the critical value $-1$ until the point $\ptsqrt{\laxpar}_3$.
Eventually, the function oscillates between the limiting values $+1$ and $-1$.
The branch cuts from $-\infty$ to $\ptsqrt{\laxpar}_1$
as well as from $\ptsqrt{\laxpar}_2$ to $\ptsqrt{\laxpar}_3$
coincide with the two forbidden zones where $\abs{\cos\laxqmom(\laxpar)}>1$.

%%%%%%%%%%%%%%%%%%%%%%%%%%%%%%%%%%%%%%%%%%%%%%%%%%%%%%%%%%%%%%%%%%%%%%%%%%%%%%%%
\subsection{Local Charges}

The KdV equation has an infinite tower of conserved local charges.
These represent useful physical information and they provide means for cross checks.
Let us therefore introduce and discuss them briefly.

The first few conserved local charges are
the potential shift $\potshift$, the momentum $\totmom$ and the energy $\eng$
defined by
\[
\potshift=\int \diff{x} \half \kdvfld,
\eqsep
\totmom=\int \diff{x} \half \kdvfld^2,
\eqsep
\eng=\int \diff{x} \brk[s]!{\half \kdvfld'^2+\kdvfld^3}.
\]
We can conveniently read off the values of these charges for the elliptic state
from the expansion of $\laxqmom(\laxpar)$ for large $\laxpar$:
\[
\laxqmom(\laxpar)=\fldlen\sqrt{\laxpar}
- \frac{\potshift}{u^{1/2}}
- \frac{\totmom}{4u^{3/2}}
- \frac{\eng}{16u^{5/2}}
+\ldots.
\]
One finds their explicit expressions in terms of complete elliptic integrals $\ellKval$ and $\ellEval$:
\begin{align}
\potshift&=
\brk[s]!{-\rfrac{1}{6}\parvel+\rfrac{2}{3}(2-\ellmods)\parscl^2}\frac{\ellKval}{\parscl}
-2\parscl\ellEval,
\\
\totmom&=
\brk[s]!{\rfrac{1}{36}\parvel^2-\rfrac{2}{9}(2-\ellmods)\parvel\parscl^2
+\rfrac{4}{9}(1-\ellmods+\ellmods^2)\parscl^4}\frac{\ellKval}{\parscl}
+\rfrac{2}{3}v\parscl\ellEval,
\\
\eng&=
\brk[s]!{\rfrac{1}{108}\parvel^3-\rfrac{1}{9}(2-\ellmods)\parvel^2\parscl^2+\rfrac{4}{9}(1-\ellmods+\ellmods^2)\parvel\parscl^4
-\rfrac{16}{135}(4-6\ellmods+12\ellmods^2-5\ellmods^3)\parscl^6}
\frac{\ellKval}{\parscl}
\\
&\alignrel{}
+\brk[s]!{-\rfrac{1}{3}\parvel^2-\rfrac{16}{15}(1-\ellmods+\ellmods^2)\parscl^4}\parscl\ellEval.
\end{align}

Furthermore, the states can be characterised in terms of action variables.
The main action variable is given by a period integral on the curve:
\[
\actvar=\frac{1}{\iunit\pi}\oint_{\contacyc}\laxpar\diff\laxqmom
=
\frac{4\parscl^2}{\pi}
\brk[s]!{
 -\ellEval^2
+\rfrac{2}{3} (2-\ellmods)\ellKval\ellEval
-\rfrac{1}{3} (1-\ellmods)\ellKval^2
}.
\]
An additional action variable corresponding to the average field value
is given by the potential shift $\potshift$.
The momentum and energy variables
expressed in terms of the action variables (assuming a constant length $\fldlen$)
satisfy two differential equations:
\[
\label{KdVEllChargeRel}
\der\totmom=\frac{4\potshift}{\fldlen}\diff{\potshift}+\frac{2\pi}{\fldlen}\diff{\actvar},
\eqsep
\der\eng=\frac{12\totmom}{\fldlen}\diff{\potshift}-\frac{2\pi \parvel}{\fldlen}\diff{\actvar}.
\]
Note that the coefficients of $\diff{\actvar}$
express the wave number and angular velocity
for the periodicities in space and time, respectively.

Unfortunately, the additional degree of freedom corresponding to the average field value
complicates our further treatment:
In principle, we should fix it to some specific value,
but there is no universally suitable way to do so.
Useful choices might be fixing the potential shift $\potshift=0$,
fixing the asymptotic field value $\kdvfld_0$ for infinite-length states,
or fixing one of the branch points $\ptsqrt{\laxpar}=0$.
These conditions are all different,
and they all have an impact on the dependencies of the charges on the action variables.
To resolve the issue, we can use a Galilean boost transformation
\[
\tilde\kdvfld(x,t)=\kdvfld(x+\tilde \parvel t,t)+\rfrac{1}{6}\tilde \parvel.
\]
This transforms the charges infinitesimally according to
$\delta \potshift=\mathinner{\delta \tilde \parvel} \fldlen/12$,
$\delta \totmom=\mathinner{\delta \tilde \parvel} \potshift/3$,
$\delta \eng=\mathinner{\delta \tilde \parvel} \totmom$,
and it allows us to construct Galilei-invariant combinations of the charges as
\[
\tilde\totmom:=\totmom-\frac{2\potshift^2}{\fldlen},
\eqsep
\tilde\eng:=\eng-\frac{12\totmom \potshift}{\fldlen}+\frac{16\potshift^3}{\fldlen^2}.
\]
Essentially, these yield the values of the charges in the inertial frame
defined by $\tilde\potshift=0$ with the relative velocity $\tilde \parvel=-12\potshift/\fldlen$.
The above relations then simplify to
\[
\der \tilde\totmom=\frac{2\pi}{\fldlen}\diff{\actvar},
\eqsep
\der \tilde\eng=
-\frac{2\pi}{\fldlen}\brk*{\parvel+\frac{12\potshift}{\fldlen}}\diff{\actvar}.
\]

%%%%%%%%%%%%%%%%%%%%%%%%%%%%%%%%%%%%%%%%%%%%%%%%%%%%%%%%%%%%%%%%%%%%%%%%%%%%%%%%
\subsection{Dynamical Divisor}

The above spectral curve encodes the time-independent data of the state,
in particular the momentum $\totmom$ and the energy $\eng$,
or equivalently the potential shift $\potshift$ and the action variable $\actvar$.
However, the state is further specified by one time-dependent variable $\paroff(t)$.
The time-dependent data is typically encoded into the dynamical divisor
related to the solution $\laxfunc$.
For the KdV model, the divisor consists of the points $\ptdiv{\laxzpar}_j$
where the logarithmic derivative $\laxcp:=\laxfunc'/\laxfunc$ equals a reference value.
\unskip\footnote{Note that the logarithmic derivative $\laxcp=\laxfunc'/\laxfunc$
describes the direction of the vector $\laxvec$
in the vector-valued auxiliary linear problem \eqref{KdVALP}.}
For the curve at hand,
the function $\laxcp$ is an elliptic function of minimal degree in $\laxzpar$,
\[
\laxcp:=\frac{\laxfunc'}{\laxfunc}
=-\parscl\brk[s]!{\ellCS(\laxzpar)\ellDN(\laxzpar)
+\ellmods\ellSN(\parscl x+\paroff+\laxzpar)\ellSN(\parscl x+\paroff)\ellSN(\laxzpar)},
\]
%\[
%\laxcp
%=
%-\parscl\frac{\ellCS(\laxzpar)\ellDN(\laxzpar)
%+\ellmods\ellSN(\laxzpar)^2\ellSN(\parscl x+\paroff)\ellCN(\parscl x+\paroff)\ellDN(\parscl x+\paroff)}
%{1-\ellmods \ellSN(\parscl x+\paroff)^2\ellSN(\laxzpar)^2}
%,
%\]
%
and therefore every value is attained precisely twice.
It makes sense to choose the reference value for $\laxcp$ to be $\infty$
in which case the divisor consists of the poles of $\laxcp$,
and one of the two poles is fixed at $\laxzpar=0$ corresponding to $\laxpar=\infty$.
The other pole resides at
\unskip\footnote{The dependence of the pole $\ptdiv{\laxzpar}$ on $x$ and $t$
is linear because $\laxzpar$ is a uniformising variable.
For more general variables,
the evolution equations in space and time
are first-order differential equations.
Therefore, the value of the pole at some point in space and time
is sufficient to specify the complete state.
In the following we shall disregard the dependence on $t$
as we are only interested in states on a given time slice.
However, we will keep the full dependence on $x$
because it allows us investigate and reconstruct the states without further ado.}
\[
\ptdiv{\laxzpar}(x,t)=\iunit\ellKvalc-\parscl x-\paroff(t),
\]
which clearly contains the desired information on the value of $\paroff(t)$.
In terms of the spectral parameter $\laxpar$, the pole is given by
\[
\ptdiv{\laxpar}(x)=-\rfrac{1}{6}\parvel
+\rfrac{1}{3}\parscl^2(1-2\ellmods)+\parscl^2\ellmods\ellCN(\parscl x+\paroff)^2.
%=-\rfrac{1}{4}v-\half \kdvfld.
\]
As $x$ or $\paroff$ changes, the pole oscillates between
the real values $\ptsqrt{\laxpar}_2$ and $\ptsqrt{\laxpar}_3$
which are the endpoints of one of the two branch cuts.
Formally, the pole is placed either just above or just below the finite branch cut
depending on its direction of motion.
In other words, the divisor pole moves around the branch cut along the real axis
precisely at its perimeter:
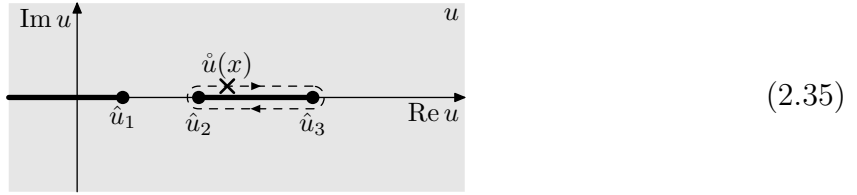
\[
\mpostusemath{KdVEllipticDivisor}
\]

Let us finally comment on the construction of states in terms of
the data of the spectral curve and divisor.
The properties of a suitable solution $\laxfunc$
with a given pole at $\ptdiv{\laxpar}$
determine it uniquely as a function of $x$ and $\laxpar$.
Given such a function, the original state $\kdvfld$ is then conveniently obtained from
the expansion of the logarithmic derivative $\laxcp$ at $\laxpar=\infty$:
\[
\laxcp=
\iunit\sqrt{\laxpar}-\frac{\iunit\kdvfld}{2\sqrt{\laxpar}}+\Order(\laxpar^{-1}).
\]
This follows from analysis of the auxiliary linear problem
$\laxcp'+\laxcp^2=\kdvfld-\laxpar$ at the singular value $\laxpar=\infty$.

\iffalse
\paragraph{monodromy}
\remark{might construct monodromy (if needed, but probably not)}
%
\[
\laxtransport(\winend,\winbeg)
=
\frac{1}{C}
\begin{pmatrix}
-\laxfunc_+(\winend)\laxfunc'_-(\winbeg)+\laxfunc_-(\winend)\laxfunc'_+(\winbeg)
& \laxfunc_+(\winend)\laxfunc_-(\winbeg)-\laxfunc_-(\winend)\laxfunc_+(\winbeg)
\\
-\laxfunc'_+(\winend)\laxfunc'_-(\winbeg)+\laxfunc'_-(\winend)\laxfunc'_+(\winbeg)
& \laxfunc'_+(\winend)\laxfunc_-(\winbeg)-\laxfunc'_-(\winend)\laxfunc_+(\winbeg)
\end{pmatrix}
\]
%
here $\laxfunc_\pm$ are the solutions with parameter $\pm z$ and $C$ is the constant
%
\[
C=\laxfunc_-\laxfunc'_+ - \laxfunc_+\laxfunc'_-
=-2\parscl\ellTN(\laxzpar)^2\ellCS(\laxzpar)\ellDN(\laxzpar).
\]
find
\[
\laxtransport(x+\fldlen,x)
=
\cos(\laxqmom)1+
\frac{\iunit \sin(\laxqmom)}{C}\laxfunc_+\laxfunc_-
\begin{pmatrix}
\parscl\ellmods\ellSN(\parscl x+\paroff)\ellSN(\laxzpar)
\brk[s]!{+\ellSN(\parscl x+\paroff+\laxzpar)-\ellSN(\parscl x+\paroff-\laxzpar)}
&
2
\\
-2\laxfunc'_+\laxfunc'_-/\laxfunc_+\laxfunc_-
&
\parscl\ellmods\ellSN(\parscl x+\paroff)\ellSN(\laxzpar)
\brk[s]!{-\ellSN(\parscl x+\paroff+\laxzpar)+\ellSN(\parscl x+\paroff-\laxzpar)}
\end{pmatrix}
\]
special case
\[
\laxtransport(-\parscl^{-1}\paroff+\fldlen,-\parscl^{-1}\paroff)
=
\cos(\laxqmom)1-
\frac{\iunit \sin(\laxqmom)}{\parscl\ellCS(\laxzpar)\ellDN(\laxzpar)}
\begin{pmatrix}
0
&
1
\\
\parscl^2 \ellCS(\laxzpar)^2\ellDN(\laxzpar)^2
&
0
\end{pmatrix}
\]
\fi

%%%%%%%%%%%%%%%%%%%%%%%%%%%%%%%%%%%%%%%%%%%%%%%%%%%%%%%%%%%%%%%%%%%%%%%%%%%%%%%%
\subsection{Infinite-Length Limit}
\label{KdVInfLen}

The class of states described above has an adjustable spatial periodicity of length $\fldlen=2\ellKval/\parscl$.
There are two ways to tune the parameters such that $\fldlen$ becomes infinite.
One is to set $\parscl=0$, but this trivialises the state to a constant $\kdvfld=-\parvel/6$.
The other is to let the modulus assume the upper critical value $\ellmods=1$
which leads to a non-trivial state.

For $\ellmods=1$ the state turns into a single exponentially localised excitation:
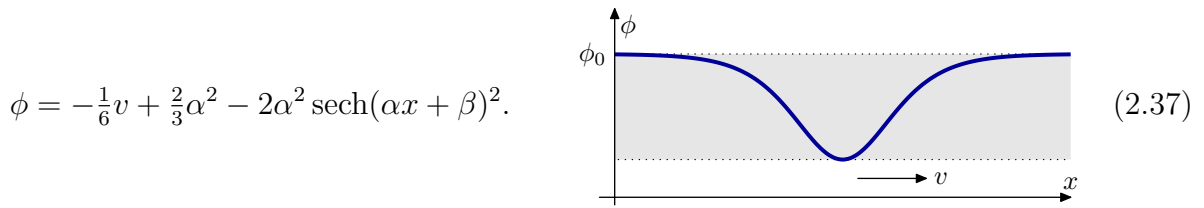
\[
\kdvfld=-\rfrac{1}{6}\parvel+\rfrac{2}{3}\parscl^2-2\parscl^2 \sech(\parscl x+\paroff)^2.
\eqsep
\mpostusemath{KdVSolitonState}
\]
This is the well-known soliton state but with a non-zero constant background
of height
\[
\kdvfld_0=-\rfrac{1}{6}\parvel+\rfrac{2}{3}\parscl^2.
\]
On the infinite line one typically demands the field to have a zero asymptotic value
in order to achieve a vanishing asymptotic energy density and moreover a finite overall energy.
For this, the width parameter $\parscl$ should be linked to the velocity $\parvel$ as $\parscl=\sqrt{\parvel}/2$,
but here we shall keep it arbitrary.

In the infinite-length limit, the solution of the auxiliary linear problem
reduces directly to
\[
\laxfunc
=
%\exp (-\iunit \ismmom x)
%\brk[s]!{\tanh(\parscl x+\paroff)+\iunit\parscl^{-1}\ismmom},
\exp \brk!{-(\parscl x+\paroff) \coth(\laxzpar)}
\brk[s]*{
1+\frac{\tanh(\parscl x+\paroff)}{\coth(\laxzpar)}
},
\eqsep
\laxpar
=-\rfrac{1}{6}\parvel+\rfrac{2}{3}\parscl^2-\parscl^2\coth(\laxzpar)^2
.
\]
The special points of the elliptic spectral curve reduce to
\[
\ptsqrt{\laxpar}_0 = -\infty,
\eqsep
\ptsqrt{\laxpar}_1 = \ptsqrt{\laxpar}_2 = \ptcrit{\laxpar} = -\rfrac{1}{6}\parvel-\rfrac{1}{3}\parscl^2,
\eqsep
\ptsqrt{\laxpar}_3 = -\rfrac{1}{6}\parvel+\rfrac{2}{3}\parscl^2=\kdvfld_0.
\]
Here, one observes that the branch points $\ptsqrt{\laxpar}_1$ and $\ptsqrt{\laxpar}_2$
as well as the critical point $\ptcrit{\laxpar}$ coincide \cite{novikov1974periodic}
and the genus of the spectral curve is reduced by one unit.
The points $\ptsqrt{\laxpar}_0=-\infty$ and $\ptsqrt{\laxpar}_3=\kdvfld_0$ serve as branch points
of the map between $\laxpar$ and $\coth(\laxzpar)$.
The divisor pole reduces to
\[
\ptdiv{\laxpar}(x)=-\rfrac{1}{6}\parvel-\rfrac{1}{3}\parscl^2+\parscl^2\sech(\parscl x+\paroff)^2.
\]
For almost all values of $x$ including the asymptotic regions at $x\to\mp\infty$
the divisor pole $\ptdiv{\laxpar}$ remains near the singular point $\ptsqrt{\laxpar}_1 = \ptsqrt{\laxpar}_2 = \ptcrit{\laxpar}$.
Conversely, in the region of the soliton where $\parscl x+\paroff$ is small,
$\ptdiv{\laxpar}$ moves towards $\ptsqrt{\laxpar}_3$
which is reached at the peak of the soliton function:
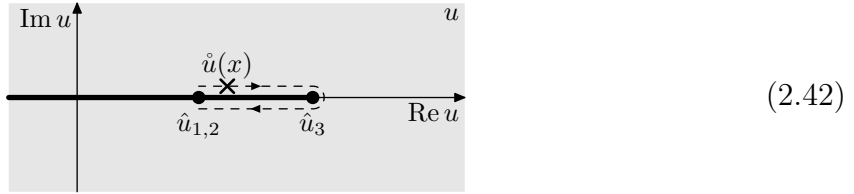
\[
\mpostusemath{KdVSolitonDivisor}
\]
The above solution to the auxiliary linear problem
evidently takes the form of Jost solutions of a scattering problem:
\[
\laxfunc_{\text{L/R}}
=
\exp (-\iunit \ismmom x)
\frac{\ismmom-\iunit\parscl\tanh(\parscl x+\paroff)}
{\ismmom\pm \iunit\parscl},
\]
where we have adjusted the normalisation such that the
solutions approach $\exp(-\iunit\ismmom x)$ in their respective asymptotic regions.
To this end, we have introduced the momentum parameter $\ismmom$ as
\[
\ismmom:=
-\iunit\parscl\coth(\laxzpar)
=
\sqrt{\laxpar+\rfrac{1}{6}\parvel-\rfrac{2}{3}\parscl^2}=\sqrt{\laxpar-\kdvfld_0}
.
\label{ismmomFromlaxpar}
\]
The two Jost solutions are related by the scattering relation
\[
\laxfunc_{\text{L}}(-\ismmom)=\ismtrans(\ismmom)\laxfunc_{\text{R}}(-\ismmom)+\ismrefl(\ismmom)\laxfunc_{\text{L}}(\ismmom)
\]
with the transmission and reflection coefficients
\[
\ismtrans(\ismmom)
=
\frac{\ismmom+\ismsing}{\ismmom-\ismsing},
\eqsep
\ismrefl(\ismmom)=0
\eqjoin{\text{where}}
\ismsing:=\iunit \parscl.
\]
Note that the zero and pole $\ismmom=\mp \ismsing$ of the transmission coefficient
map precisely to the singular point $\laxpar=\ptsqrt{\laxpar}_1=\ptsqrt{\laxpar}_2=\ptcrit{\laxpar}$ in the $\laxpar$-plane.
The values of the conserved charges take on particularly simple expressions:
\[
\potshift=\half \fldlen \kdvfld_0-2\parscl,
\eqsep
\tilde\totmom=\rfrac{8}{3} \parscl^3,
\eqsep
\tilde\eng=-\rfrac{32}{5} \parscl^5.
\]

Later on, we will be interested in a generalisation of this limit to other periodic solutions.
We observe that the present limit is well-defined and non-singular,
so it bears no trace of the length parameter $\fldlen$.
Let us therefore discuss in more detail, how the relevant quantities attain the limit $\fldlen\to\infty$.
We start by presenting a sketch of the quasi-momentum function $\laxqmom(\laxpar)$:
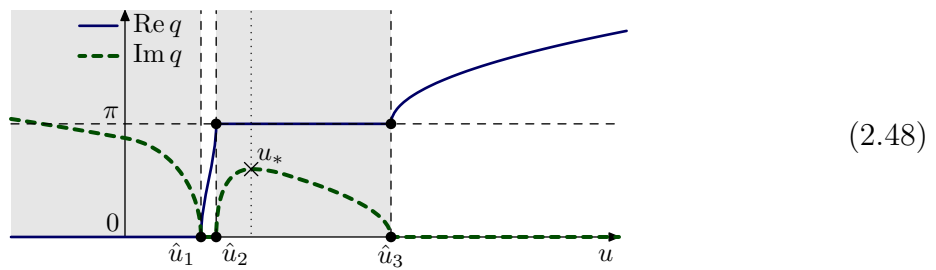
\[
\mpostusemath{KdVPinchedQ}
\]
In the limit, a relevant auxiliary variable is the momentum parameter $\ismmom$.
This variable expresses the spectral parameter $\laxpar$, but their functional relationship may depend on $\fldlen$,
and this dependency may introduce some arbitrariness in taking the limit.
It therefore makes sense to introduce this momentum parameter $\ismmom$ even at finite $\fldlen$
as a replacement for the parameter $\laxpar$ or $\laxzpar$.
A plausible choice is
\[
\ismmom:=\frac{\iunit\parscl}{\ellSN(\laxzpar)}
= \sqrt{\laxpar + \rfrac{1}{6}\parvel - \rfrac{1}{3}\parscl^2 (\ellmods+1)},
\]
which fixes the isolated branch point $\ptsqrt{\ismmom}_3=0$ to precisely zero where it should be.
The other two branch points are mapped to
$\ptsqrt{\ismmom}_1=\iunit\parscl$ and $\ptsqrt{\ismmom}_2=\iunit\parscl\sqrt{\ellmods}$.
Now the relationship $\fldlen=2\ellKval/\parscl$ and the logarithmic divergence
of the elliptic integral $\ellKval$ at $\ellmods=1$ leads to
$\ellmods$ being exponentially close to $1$:
\[
\ellmods = 1-16\eunit^{-\parscl \fldlen} +\ldots.
\]
Consequently, the two branch points are separated by exponentially small amounts:
\begin{align}
\ptsqrt{\ismmom}_2-\ptsqrt{\ismmom}_1
&=-\iunit\parscl\brk!{1-\sqrt{\ellmods}}
=-8\iunit\parscl \eunit^{-\parscl \fldlen}+\ldots,
\\
\ptsqrt{\laxpar}_2-\ptsqrt{\laxpar}_1
&=\parscl^2 (1-\ellmods)
=16\parscl^2 \eunit^{-\parscl \fldlen}+\ldots.
\end{align}
Two further quantities of interest for a spectral curve and its associated divisor
are the quasi-momentum $\laxqmom$ and the logarithmic derivative $\laxcp$.
By careful expansion we find
\begin{align}
\laxqmom
&=\ismmom \fldlen
+\iunit \log\frac{\ismmom-\iunit\parscl}{\ismmom+\iunit\parscl}
+\ldots,
\\
\laxcp
&=
\frac{\iunit \ismmom (\ismmom^2+\parscl^2)+\parscl^3\tanh(\parscl x+\paroff)\sech(\parscl x+\paroff)^2 }
{\ismmom^2+\parscl^2 \tanh(\parscl x+\paroff)^2}
+\ldots,
\end{align}
where further terms are exponentially suppressed.
It also makes sense to state the spectral curve equation \eqref{KdVgenusonespectralcurve}
for large $\fldlen$:
\[
\frac{\der\laxqmom}{\der\laxpar}
=
\frac{\fldlen}{2\sqrt{\laxpar-\kdvfld_0}}
-
\frac{\parscl}{\sqrt{\laxpar-\kdvfld_0}(\laxpar-\kdvfld_0+\parscl^2)}
+\ldots.
\]
Note that the leading term describes the asymptotic background
while the next-to-leading term combines two square roots into one pole describing the soliton shape.
Finally, it will be useful to extract the action variable
because it serves as a modulus of the spectral curve.
It turns out to scale with the length
\[
\actvar=
\frac{4\parscl^3 \fldlen}{3\pi}+\ldots,
\]
and it does satisfy the wave number and angular velocity relations
$\der\tilde\totmom=(2\pi/\fldlen)\der \actvar$ and $\der\tilde\eng=(-8\pi\parscl^2/\fldlen)\der \actvar$.

%%%%%%%%%%%%%%%%%%%%%%%%%%%%%%%%%%%%%%%%%%%%%%%%%%%%%%%%%%%%%%%%%%%%%%%%%%%%%%%%
\subsection{Small-Amplitude Limit}
\label{smallamplitudelimit}

It also makes sense to discuss the solution
when the modulus $\ellmods$ approaches the lower critical value $0$.
Here, the solution approaches a constant at $\ellmods=0$,
and we should discuss the leading perturbation around this constant solution.
As we move $\ellmods$ slightly away from $0$,
also the other degrees of freedom may receive explicit or implicit contributions from this shift
making the resulting perturbative expressions somewhat arbitrary.
In order to avoid such arbitrariness, we will fix the action variables $\potshift$ and $\actvar$
as well as the length $\fldlen$:
\[
v=-\frac{6\kappa \potshift}{\pi}-\kappa^2+\frac{3\actvar}{\pi}+\Order(\actvar^{3/2}),
\eqsep
\parscl=\frac{\kappa}{2}+\sqrt{\frac{\actvar}{2\pi}}+\frac{9\actvar}{4\pi\kappa}+\Order(\actvar^{3/2}).
\]
The action variable $\actvar$ is small for $\ellmods\to 0$ and it will be used as the perturbative parameter.
Furthermore, we have introduced the wave number $\kappa$ to match the periodicity:
\[
\ellmods=\frac{8}{\kappa}\sqrt{\frac{\actvar}{2\pi}}+\Order(\actvar),
\eqsep
\kappa:=\frac{2\pi}{\fldlen}.
\]
The state then reads
\[
\kdvfld=\frac{\kappa \potshift}{\pi}-2\kappa\sqrt{\frac{\actvar}{2\pi}}\cos(\kappa x+2\paroff)
-\frac{\actvar}{\pi}\cos(2\kappa x+4\paroff)+\ldots
.
\]
This state represents a plane wave with wave number $\kappa$
and a small amplitude.

Let us discuss some key parameters of the curve as well as conserved charges.
The distinguished points of the curve are given by
\[
\ptsqrt{\laxpar}_1 = \frac{\kappa \potshift}{\pi}+\Order(\actvar),
\eqsep
\ptcrit{\laxpar} = \frac{\kappa \potshift}{\pi}+\frac{\kappa^2}{4}+\Order(\actvar),
\eqsep
\ptsqrt{\laxpar}_{2,3} = \ptcrit{\laxpar}\mp \kappa\sqrt{\frac{\actvar}{2\pi}}+\Order(\actvar),
\]
We see that the infinite branch cut extends up to $\ptsqrt{\laxpar}_1$
which equals the mean value of the field $\kdvfld$.
The other branch cut is separated by $\kappa^2/4$
and has a small width proportional to $\sqrt{\actvar}$.
In fact, the function $\laxqmom$ is approximated by a semi-circle on the cut:
\[
\laxqmom=\pi+\frac{2\pi\iunit}{\kappa^2}\sqrt{\frac{\actvar}{2\pi}\kappa^2-(\laxpar-\ptcrit{\laxpar})^2}+\ldots.
\]
As such the resulting action variable equals
half the width times the height which amounts to $\actvar$ as it should.
Finally, one finds for the Galilei-invariant conserved charges
\[
\tilde\totmom=\kappa \actvar,
\eqsep
\tilde\eng=
\kappa^3 \actvar
+\Order(\actvar^2).
\]

%%%%%%%%%%%%%%%%%%%%%%%%%%%%%%%%%%%%%%%%%%%%%%%%%%%%%%%%%%%%%%%%%%%%%%%%%%%%%%%%
%%%%%%%%%%%%%%%%%%%%%%%%%%%%%%%%%%%%%%%%%%%%%%%%%%%%%%%%%%%%%%%%%%%%%%%%%%%%%%%%
\section{KdV Finite-Length Extrapolation}
\label{sec:KdVFiniteLengthExtrapolation}

In the following we shall approximate
a generic state on the infinite line $\kdvfld(x)$
with asymptotic boundary conditions
by a family $\kdvfld_{\fldlen}(x)$ of periodic states on an interval of length $\fldlen$:
\[
\mpostusemath{PatchingShort}
\eqsep*
\mpostusemath{PatchingLong}
\eqjoin*{\longrightarrow}
\mpostusemath{PatchingOriginal}
\label{window}
\]
We do so by constructing the family such that the function $\kdvfld_{\fldlen}$
periodically repeats a window of interest of $\kdvfld$ with given length $\fldlen$.
The periodic states are thus defined by
\[
\kdvfld_{\fldlen}(x+\fldlen\Integer)
:=
\kdvfld(x)
\eqjoin{\text{with}}
\winbegL<x<\winendL=\winbegL+\fldlen,
\]
where the window of interest in $\kdvfld(x)$ increases to the whole real line:
\[
\winbegL,\winendL\to\mp\infty
\eqjoin{\text{as}}
\fldlen\to\infty.
\]
By considering the spectral curves corresponding to $\kdvfld_{\fldlen}$ as $\fldlen\to\infty$,
we will see how the scattering data of $\kdvfld$ are approximated by spectral curves.

The procedure described above is reminiscent of
the truncation of a periodic function
onto its fundamental domain on an infinite line
which was used in \cite{Osborne:1985,Osborne:1986ua}
to relate monodromy and scattering matrix.
However, in order to properly define the infinite-length limit,
we take a fixed infinite-line profile as the starting point
and derive a sequence of periodic functions from it.

%%%%%%%%%%%%%%%%%%%%%%%%%%%%%%%%%%%%%%%%%%%%%%%%%%%%%%%%%%%%%%%%%%%%%%%%%%%%%%%%
\subsection{Extrapolation}
\label{KdVExtrapolation}

We start by assuming a state $\kdvfld$ on the infinite line
with the asymptotic boundary condition
\[
\lim_{x\to\pm\infty} \kdvfld(x) = \kdvfld_0.
\label{KdVasymptotics}
\]
For the inverse scattering method, it is convenient to set up
a Lax connection which becomes diagonal in the asymptotic regions.
Based on the Lax connection $\laxconn$ introduced in \eqref{KdVALP},
this is achieved by the transformation
\[
\label{KdVLaxTrans}
\laxconn\to M\laxconn M^{-1}
\eqjoin{\text{with}}
%\laxconn=
%\begin{pmatrix}
%0 & 1
%\\
%\kdvfld-\laxpar & 0
%\end{pmatrix}
%\eqsep*
M=
\begin{pmatrix}
-\iunit\ismmom & 1
\\
+\iunit\ismmom & 1
\end{pmatrix},
\eqsep*
\ismmom:=\sqrt{\laxpar-\kdvfld_0}.
\]
The transformed Lax connection clearly becomes diagonal
in the asymptotic regions where $\kdvfld\to\kdvfld_0$:
\[
\label{KdVLaxAlt}
\laxconn =
\begin{pmatrix}
-\iunit \ismmom& 0
\\
0 & \iunit \ismmom
\end{pmatrix}
+
\frac{\iunit(\kdvfld-\kdvfld_0)}{2\ismmom}\begin{pmatrix}
1 & -1
\\
1 & -1
\end{pmatrix}.
\]
The scattering matrix for the inverse scattering method is given by
\[
\ismscat=
\lim_{\winbeg,\winend\to\mp\infty}
\diag(\eunit^{\iunit \ismmom \winend},\eunit^{-\iunit \ismmom \winend})
\.\laxtransport(\winend,\winbeg)\.
\diag(\eunit^{-\iunit \ismmom \winbeg},\eunit^{\iunit \ismmom \winbeg}),
\label{ismscat}
\]
where the parallel transport $\laxtransport(\winend,\winbeg)$ of the Lax connection
is defined via the path-ordered exponential integral
\[
\laxtransport(\winend,\winbeg):=\pathordleft \brk[s]*{\exp \int_{\winbeg}^{\winend} \diff{x}\laxconn(x)}.
\label{LaxParallelTransport}
\]
The exponential factors in the above definition of the scattering matrix are precisely what it takes
to remove the expected oscillatory behaviour in the asymptotic regions.
We can thus invert their effect to obtain an approximation for the parallel transport:
\[
\laxtransport(\winend,\winbeg)
\sim
\diag(\eunit^{-\iunit \ismmom \winend},\eunit^{\iunit \ismmom \winend})\.
\ismscat\.
\diag(\eunit^{\iunit \ismmom \winbeg},\eunit^{-\iunit \ismmom \winbeg}).
\label{LaxParallelTransportApprox}
\]
It is common to express the auxiliary scattering matrix $\ismscat(\ismmom)$ in terms of transmission and reflection
coefficients $\ismtrans(\ismmom)$ and $\ismrefl(\ismmom)$ for the complementary scattering problem as
\[
\ismscat(\ismmom)
=
\begin{pmatrix}
1/\ismtrans(\ismmom) & \ismrefl(\ismmom)/\ismtrans(\ismmom)
\\
\ismrefl(-\ismmom)/\ismtrans(-\ismmom) & 1/\ismtrans(-\ismmom)
\end{pmatrix}.
\label{ismscatmatrix}
\]
We thus find for the asymptotic behaviour of the parallel transport
\[
\laxtransport(\winend,\winbeg)
\sim
\begin{pmatrix}
\eunit^{-\iunit \ismmom (\winend-\winbeg)}/\ismtrans(\ismmom)
& \eunit^{-\iunit \ismmom (\winend+\winbeg)}\ismrefl(\ismmom)/\ismtrans(\ismmom)
\\
\eunit^{\iunit \ismmom (\winend+\winbeg)}\ismrefl(-\ismmom)/\ismtrans(-\ismmom)
& \eunit^{\iunit \ismmom (\winend-\winbeg)}/\ismtrans(-\ismmom)
\end{pmatrix}.
\label{LaxParallelTransportApproxMatrix}
\]
This matrix can be used as an approximation to the monodromy
of a periodically identified state $\kdvfld_{\fldlen}$ with the following two approximations:
First, the above parallel transport is valid only asymptotically
and as far as all the relevant dynamics takes place on the interval $\winbeg<x<\winend$.
Second, the construction of the monodromy assumes a periodic state $\kdvfld_{\fldlen}$
whereas $\laxtransport(\winend,\winbeg)$ is based on a non-periodic interval
from the state $\kdvfld$ on the infinite line. Closing the period on the interval
generically introduces some discontinuity or disruption of smoothness
which would be represented by distributional terms in the parallel transport for $\kdvfld_{\fldlen}$.
Therefore, there will be a mismatch between $\laxtransport(\winend,\winbeg)$
and the corresponding monodromy for the periodic identification of the interval.
We assume both approximations to be asymptotically small,
and thus we will use $\laxtransport(\winend,\winbeg)$ as an approximation for the monodromy.

We can now extract the quasi-momentum function $\laxqmom$
via the monodromy trace:
\[
\cos\laxqmom = \half\tr \laxtransport(x+\fldlen,x) \approx
\frac{\eunit^{-\iunit \ismmom \fldlen}}{2\ismtrans(\ismmom)}
+ \frac{\eunit^{\iunit \ismmom \fldlen}}{2\ismtrans(-\ismmom)}.
\label{quasimomfun}
\]
We observe that $\laxqmom$ is given in terms of the transmission coefficient $\ismtrans$.
However, the scattering data for the inverse scattering method
is typically specified in terms of reflection coefficient $\ismrefl$
together with $N$ soliton momenta $\ismsing_n\in \iunit\Real^+$
and corresponding dynamical coefficients $\ismres_n\in\Real^+$.
Now the function $\ismtrans(\ismmom)$ can be reconstructed from these data
on the upper half complex plane including the real axis, $\Im \ismmom\geq 0$,
by the dispersion relation \cite{faddeev1964properties}
\[
\ismtrans(\ismmom)=\brk[s]*{\prod_{n=1}^N\frac{\ismmom+\ismsing_n}{\ismmom-\ismsing_n}}
\exp\brk[s]*
{
\frac{1}{2\pi\iunit}\int \frac{\diff{\ismmom'}}{\ismmom'-\ismmom-\iunit 0}\log\brk!{1-\abs{\ismrefl(\ismmom')}^2}
}.
\label{faddeevformula}
\]
Let us note a few remarks on the resulting transmission function:
First of all, $\ismtrans(\ismmom)$ is manifestly time-independent as it should
because it does not depend on the dynamical variables $\arg\ismrefl(\ismmom)$ and $\ismres_n$.
Second, for real $\ismmom$, the distributional identity
\[
\frac{1}{\ismmom'-\ismmom-\iunit 0}=\frac{1}{\ismmom'-\ismmom}+\iunit\pi\delta(\ismmom'-\ismmom)
\label{distribution}
\]
determines the magnitude of $\ismtrans(\ismmom)$ as
\[
\abs{\ismtrans(\ismmom)}=\sqrt{1-\abs{\ismrefl(\ismmom)}^2}.
\label{unitaryidentity}
\]
The resulting relation $\abs{\ismtrans}{}^2+\abs{\ismrefl}{}^2=1$ together
with the reality relations $\ismtrans(\ismmom)^\conj=\ismtrans(-\ismmom)$ and $\ismrefl(\ismmom)^\conj=\ismrefl(-\ismmom)$
is in agreement with unitarity of $\ismscat$ in \eqref{ismscatmatrix}
with regard to the indefinite hermitian form $\diag(+1,-1)$.
\unskip\footnote{Note that there exist two different notions of
scattering matrices in two dimensions with two different notions of unitarity.
See \appref{leftrightinout} for a (slightly off-topic) discussion.}
Third, the argument of $\ismtrans(\ismmom)$ interpolates
between $\ismtrans(0)=\pi N$ at $\ismmom=0$ and $\ismtrans(+\infty)=0$ at $\ismmom=+\infty$.

%%%%%%%%%%%%%%%%%%%%%%%%%%%%%%%%%%%%%%%%%%%%%%%%%%%%%%%%%%%%%%%%%%%%%%%%%%%%%%%%
\subsection{Continuum Cuts}

First, we would like to reconstruct the cuts of the spectral curve
for real $\ismmom$ (without loss of generality, we restrict to positive $\ismmom$).
For real momenta $\ismmom$, the scattering coefficients obey:
\[
\ismtrans(-\ismmom)=\ismtrans(\ismmom)^\conj,
\eqsep
\ismrefl(-\ismmom)=\ismrefl(\ismmom)^\conj,
\eqsep
\abs{\ismtrans(\ismmom)}^2+\abs{\ismrefl(\ismmom)}^2=1.
\]
Choosing a radial form for the transmission coefficient $\ismtrans=\abs{\ismtrans}\exp(\iunit\arg\ismtrans)$,
we obtain the following approximation for the quasi-momentum
\[
\label{KdVContF}
\cos \laxqmom(\ismmom)\approx
\frac{1}{\abs{\ismtrans(\ismmom)}}\cos\brk!{\ismmom \fldlen+\arg\ismtrans(\ismmom)}.
\]
As noted above, the branch cuts of the spectral curve
are the forbidden zones where $\abs{\cos \laxqmom}$ exceeds $1$:
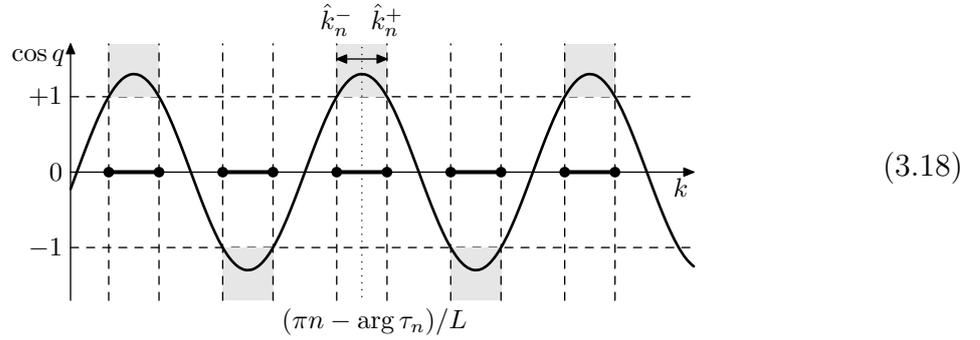
\[
\mpostusemath{KdVContinuumF}
\]
In order to find the solutions to $\cos \laxqmom(\ismmom)=\pm 1$,
we note that for large length $\fldlen$,
the term on the right-hand side of the equation represents a fast uniform oscillation in $\ismmom$
which is modulated in amplitude and phase by the relatively slow function $\ismtrans(\ismmom)$.
For our purposes we can assume $\ismtrans(\ismmom)\approx\ismtrans(\ismmom_n)=:\ismtrans_n$ to be approximately constant
in a neighbourhood of size $\Order(1/\fldlen)$ around the point $\ismmom_n:=\pi n/\fldlen$.
It is then straight-forward to approximate the branch points of the $n$-th cut where $\Re\laxqmom=\pi n$
\footnote{For these purposes, the phase $\arg\ismtrans(\ismmom)$ of the transmission coefficient
should be understood as a continuous function interpolating between $\arg\ismtrans(0)=\pi N$
and $\arg\ismtrans(+\infty)=0$ rather than as an angle bounded between $-\pi$ and $+\pi$.
We will see later that the first $N$ cuts will be used to represent the $N$ solitons,
and $n$ will consistently start at $N+1$ for branch cuts at real $\ismmom$.}
as
\[
\label{KdVContCutBranch}
\ptsqrt{\ismmom}_n^\pm
\approx
\frac{\pi n-\arg\ismtrans_n\pm\arccos\abs{\ismtrans_n}}{\fldlen}.
\]
We observe that the widths of the branch cuts are determined
by the magnitude of the reflection coefficient:
\[
\label{KdVContCutLen}
\ptsqrt{\ismmom}_n^+-\ptsqrt{\ismmom}_n^-
\approx
\frac{2}{\fldlen}\arccos\abs{\ismtrans_n}
=
\frac{2}{\fldlen}\arcsin\abs{\ismrefl_n}\leq \frac{\pi}{\fldlen}.
\]
As $\ismrefl_n$ increases from $0$ to $1$, the length of a cut increases from $0$ to $\pi/\fldlen$.
Note that the linear relationship between the length of the branch cuts
and the Fourier coefficients at small amplitudes
was already observed in \cite{Osborne:1985}.
Furthermore, the branch cuts are arranged on a lattice $(\pi/\fldlen)\Integer$
from which their centres are shifted by $-\arg\ismtrans_n/\fldlen$:
\unskip\footnote{It deserves mention that the notion of leading and sub-leading terms
in the following expression depends on the situation:
Superficially, the term $\ismmom_n$ is of order $1$
while the contribution from $\arg\ismtrans$ is of order $1/\fldlen$.
However, when the overall expression is multiplied by $\fldlen$,
the contribution $\ismmom_n \fldlen=\pi n$ is an integer multiple of $\pi$,
and as an angle it merely specifies a discrete parity rather than a continuous quantity.
In that case, the second term becomes the leading continuous contribution,
and this feature is what allowed us to determine it from the leading scattering data.
Nevertheless, we will have to be careful because in other places
there may be additional contributions whose order competes with the contribution from $\arg\ismtrans$.}
\[
\label{KdVContCutCentr}
\half(\ismmom_n^++\ismmom_n^-)
\approx
\frac{\pi n-\arg\ismtrans_n}{\fldlen}
=
\ismmom_n-
\frac{\arg\ismtrans_n}{\fldlen}.
\]

%%%%%%%%%%%%%%%%%%%%%%%%%%%%%%%%%%%%%%%%%%%%%%%%%%%%%%%%%%%%%%%%%%%%%%%%%%%%%%%%
\subsection{Continuum Divisor}
\label{KdVContDiv}

Next, we would like to reconstruct the dynamical divisor
which should correspond to the phase of the reflection coefficient $\ismrefl(\ismmom)$.
As discussed above, the divisor describes the poles $\ptdiv{\laxpar}=\laxpar(\ptdiv{\ismmom})$
of the logarithmic derivative $\laxcp$ of the function $\laxfunc(x)$.
In the original basis for the Lax connection in \eqref{KdVALP}, these are represented by the
points where the eigenvector $(\laxfunc,\laxfunc')$ aligns with the direction $(0,1)$.
For consideration of the auxiliary scattering problem we have transformed the basis for the Lax connection
using the matrix $M$ in \eqref{KdVLaxTrans}
such that the transformed reference direction for the divisor is $M(0,1)=(1,1)$.
The divisor therefore consists of the points $\ptdiv{\ismmom}$
where the eigenvector of the monodromy $\laxtransport(x+\fldlen,x)$ points in the direction $(1,1)$.
\unskip\footnote{Here, we assume the periodicity interval for the approximation to the monodromy
to range from $-\fldlen<x<0$ through $0<x+\fldlen<\fldlen$ so as to capture the range of interest.
Unfortunately, this turns out to be a slight contradiction in itself:
We are mainly interested in the region $\abs{x}\ll \fldlen$ for the monodromy eigenvector.
Modulo $\fldlen$ this region corresponds to either $x\approx -\fldlen$ for negative $x$
or $x\approx 0$ for positive $x$.
For both of these regions, $\laxtransport(x+\fldlen,x)$ misses an essential part of the region of interest,
and cannot be considered a good approximation for the monodromy.
This seemingly leads to self-contradictory approximations
which we may not be able to fully resolve.
Nevertheless, we shall arrive at conclusions which turn out self-consistent and physically reasonable.}

Using our approximation for the monodromy, we find the relation
\[
\frac{\eunit^{-\iunit \ptdiv{\ismmom} \fldlen}}{\ismtrans(\ptdiv{\ismmom})}
+\frac{\eunit^{-\iunit \ptdiv{\ismmom} (\fldlen+2x)}\ismrefl(\ptdiv{\ismmom})}{\ismtrans(\ptdiv{\ismmom})}
-\frac{\eunit^{\iunit \ptdiv{\ismmom} (\fldlen+2x)}\ismrefl(-\ptdiv{\ismmom})}{\ismtrans(-\ptdiv{\ismmom})}
-\frac{\eunit^{\iunit \ptdiv{\ismmom} \fldlen}}{\ismtrans(-\ptdiv{\ismmom})}
\approx 0.
\label{KdVdivisoreqn}
\]
Recalling that the contribution $\ptdiv{\ismmom} \fldlen$ to the exponents
has a significantly faster dependence on $\ptdiv{\ismmom}$ than all others,
we can treat the latter as constants.
\unskip\footnote{The contributions $2x\ptdiv{\ismmom}$ deserve special consideration
since next to the region $x\approx 0$ we are also interested in the region $x\approx -\fldlen$.
In the latter case, $2x\ptdiv{\ismmom}$ is approximated by $-2\pi n$
which can be discarded from the exponent.}
Furthermore, using the radial representation of the scattering data for real $\ismmom$,
we can express this equation near $\ismmom_n=\pi n/\fldlen$ as
\[
\sin\brk!{\ptdiv{\ismmom} \fldlen+\arg\ismtrans_n}
\approx
-\abs{\ismrefl_n}\sin\brk!{\ptdiv{\ismmom} \fldlen+\arg\ismtrans_n+2\pi n x/\fldlen-\arg\ismrefl_n}.
\label{KdVContDivEq}
\]
Clearly, there is no solution away from the branch cuts because the magnitude of the left-hand side is larger than $\abs{\ismrefl_n}$.
However, on a branch cut $\ptsqrt{\ismmom}_n^-<\ptdiv{\ismmom}<\ptsqrt{\ismmom}_n^+$,
the left-hand side sweeps the entire range between $-\abs{\ismrefl_n}$ and $+\abs{\ismrefl_n}$
while the right-hand side only covers a part of the interval.
Due to continuity, there is a solution on each branch cut,
and one can convince oneself that it is unique, $\ptdiv{\ismmom}=:\ptdiv{\ismmom}_n$.

Let us briefly discuss the dependence of the point $\ptdiv{\ismmom}_n$ on the variables.
Here it makes sense to introduce an effective phase for the $n$-th cut as
\[
\sigma_n:=
\frac{2\pi n x}{\fldlen}-\arg\ismrefl_n.
\label{effectivephase}
\]
Note that this phase is in agreement with results of \cite{Osborne:1986ua}.
As $\sigma_n$ shifts by $2\pi$,
the point $\ptdiv{\ismmom}_n$ moves back and forth along the branch cut once.
\unskip\footnote{The non-linearity of the equation results in
different slopes for the forward and backward motion especially for large amplitudes $\abs{\ismrefl}\to 1$
and depending on the position $x$.}
Consequently, the same holds for the phase $\arg \ismrefl_n$ of the reflection coefficient.
Furthermore, as the position $x$ shifts by one entire period $\fldlen$,
the point $\ptdiv{\ismmom}_n$ of the $n$-th cut oscillates precisely $n$ times.
\unskip\footnote{This property must hold by construction of the spectral curve
because the corresponding state is exactly periodic.}
This feature associates the $n$-th cut with the $n$-th periodic mode of the system.
Finally, as $n$ increases by one unit,
the effective phase shifts by the leading amount $2\pi x/\fldlen$.
Altogether, the distribution of the divisor poles on the array of cuts
can be sketched as follows:
\[
\mpostusemath{KdVContinuumDivisor}
\]

%%%%%%%%%%%%%%%%%%%%%%%%%%%%%%%%%%%%%%%%%%%%%%%%%%%%%%%%%%%%%%%%%%%%%%%%%%%%%%%%
\subsection{Soliton Intervals}
\label{KdVSolInt}

We now proceed to the case of cuts along the negative $\laxpar$-axis
which will encode the solitons.
Here, the momentum variable is positive imaginary $\ismmom\in\iunit\Real^+$,
and the dispersion relation fixes the upper left component $1/\ismtrans(\ismmom)$ of the scattering matrix.
The latter is a holomorphic function and its zeros or likewise the poles
of $\ismtrans(\ismmom)$ reside at the soliton positions $\ismsing_n\in\iunit\Real^+$.
We assume that there are $N$ solitons,
and we will order their positions such that their magnitude $\Im\ismsing_n$ decreases with $n$.
Unfortunately, the other three components of the scattering matrix are not universally determined.
There are several related indications and reasons why this is so,
let us briefly discuss them before we continue.

First and foremost, the auxiliary linear problem fixes the scattering matrix
based on asymptotic plane wave-like behaviour.
This works well for real momenta,
but introduces exponential amplification or attenuation for complex momenta.
In this case, only the solutions with the most dominant asymptotic attenuation
can be universally defined.
The other solution has lower attenuation,
and it can be supplemented with the solution of dominant attenuation
violating its prescribed asymptotic behaviour.
Such a change of basis can be introduced with arbitrary momentum-dependent coefficient
in order to adjust the reflection coefficients $\ismrefl(\ismmom)$ and $\ismrefl(-\ismmom)$
to any desired function of $\ismmom$ including even non-analytic ones.
In this sense, the reflection coefficient $\ismrefl(\ismmom)$
away from the real $\ismmom$-axis bears no significant information.
Eventually, the opposite transmission coefficient $\ismtrans(-\ismmom)$
is determined by the unimodularity relation
$\ismtrans(\ismmom)\ismtrans(-\ismmom)+\ismrefl(\ismmom)\ismrefl(-\ismmom)=1$,
which shows clearly that it is also ambiguously defined.
Fortunately, the upper left component $1/\ismtrans(\ismmom)$
of the scattering matrix is well-defined and even holomorphic on the upper half plane.
Therefore, the latter is the only component of the scattering matrix
that is at our disposal for the reconstruction of a suitable Lax monodromy.

A different approach to obtain the scattering matrix
is to solve the parallel transport operator of the auxiliary linear problem.
The result is defined uniquely for all values of $\ismmom$,
but only when the supporting interval is finite.
Here, the resulting matrix needs to be regularised properly
according to the expected asymptotic behaviour of its components.
In the infinite-length limit, the upper left component of the scattering matrix converges uniformly
ensuring the benign properties of $\ismtrans(\ismmom)$.
The other components may or may not even converge depending on the value of $\ismmom$,
e.g.\ for $\Im\ismmom$ exceeding the lowest soliton momentum.
To ensure convergence for all $\ismmom$ requires a non-trivial regularisation,
but this also introduces dependency of the resulting components
on higher-order corrections in the regularisation scheme.
Effectively, these considerations imply that all but the upper left component of the scattering matrix
are not universally defined.

Nevertheless, we need to find the approximate monodromy for a periodic system
in order to extract its branch points at finite $\fldlen$.
We make the following proposal for the vicinity of a soliton position $\ismsing_n$,
\[
\laxtransport(x+\fldlen,x)
\approx
\begin{pmatrix}
\eunit^{-\iunit \ismmom \fldlen}/\ismtrans(\ismmom) & B(\ismmom)
\\
C(\ismmom) & \eunit^{\iunit \ismmom \fldlen} D(\ismmom)
\end{pmatrix}
\label{solitonlaxtransportansatz}
\]
with some analytical functions $B,C,D$ of $\ismmom$ which may additionally depend on $x$.
The dependency of the diagonal components on the length $\fldlen$
follows from the regularisation of asymptotic states.
We can thus take the trace and obtain
\[
\cos\laxqmom = \half\tr \laxtransport(x+\fldlen,x) \approx
\frac{\eunit^{-\iunit \ismmom \fldlen}}{2\ismtrans(\ismmom)}.
\label{solitoncosq}
\]
The branch points of the quasi-momentum are where $\cos\laxqmom=\pm 1$.
As the function $\eunit^{-\iunit \ismmom \fldlen}/\ismtrans(\ismmom)$ has an exponentially steep slope at its zero,
there must be two branch points nearby:
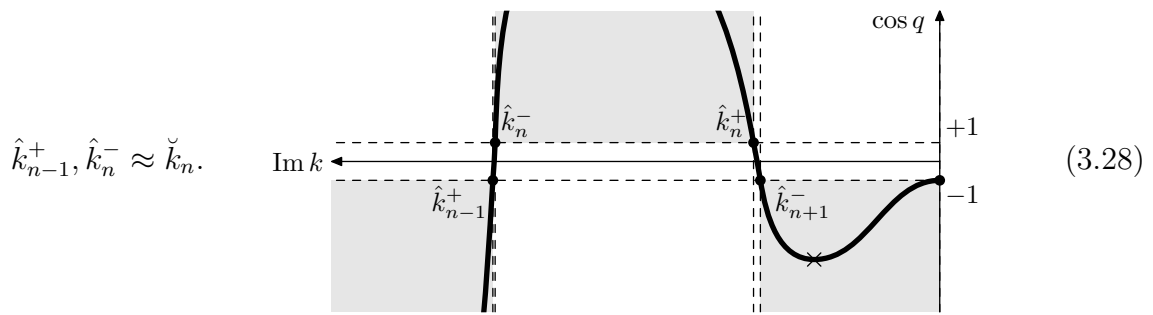
\[
\ptsqrt{\ismmom}_{n-1}^+,\ptsqrt{\ismmom}_n^-
\approx \ismsing_n.
\eqsep
\mpostusemath{KdVPinchedF}
\]
Their separation is determined by the slope and it is thus exponentially small:
\[\label{KdVSolSep}
\Im \ptsqrt{\ismmom}_{n-1}^+ - \Im\ptsqrt{\ismmom}_n^-
\approx 4 \Im \res \ismtrans(\ismsing_n) \exp(\iunit \ismsing_n \fldlen).
\]
This separation agrees with the above analysis of the elliptic solution for $\ellmods\to 1$
where $\res\ismtrans(\ismsing)=2\iunit\parscl=2\ismsing$.

%%%%%%%%%%%%%%%%%%%%%%%%%%%%%%%%%%%%%%%%%%%%%%%%%%%%%%%%%%%%%%%%%%%%%%%%%%%%%%%%
\subsection{Soliton Divisor}
\label{KdVSolDiv}

The dynamical divisor corresponding to the solitons is the most difficult element of the family of spectral curves
to recover from our method.
Effectively, the difficulty is due to the soliton positions dropping out completely
from the regularised monodromy in the infinite-length limit.
The corresponding variables $\ismres_n$ of the inverse scattering method at $\fldlen=\infty$
are thus not actually encoded into the scattering matrix,
but they can only be determined directly from correlators of the wave functions.
Nonetheless, the information on the soliton positions is of course
encoded into the dynamical divisor at finite length $\fldlen$.
We would thus like to understand at least their qualitative representation in the monodromy eigenvectors.
In particular, it is not self-evident whether and how the overall requirements
for the divisor of spectral curves
(on each cut there is one pole
and the one on the $n$-th cut encircles the latter $n$ times)
are compatible with the shape of a multi-soliton solution
($N$ bumps of different widths which almost do not interfere).

We assume that the limiting infinite-length state has $N$ solitons.
Each branch cut carries precisely one associated pole of the eigenvector function $\laxcp_{\fldlen}$
with one overall pole permanently fixed to the point $\ismmom=\infty$.
The above discussion showed that there are $N+1$ cuts on the imaginary $\ismmom$-axis.
The pole for the cut stretching to $\ismmom=\infty$ is fixed,
consequently, there are $N$ remaining poles $\ptdiv{\ismmom}_n$ on the imaginary axis.
Each one of these moves on one of the cuts
whose branch points $\ptsqrt{\ismmom}_n^\pm$
approach two consecutive soliton momenta $(\ismsing_{n-1},\ismsing_n)$
in the infinite-length limit.
It therefore makes sense to assume that the $N$ dynamical positions of the poles
encode the $N$ position variables $\ismres_n$ in some way,
however, not necessarily one-to-one.
In the following, let us gain some intuition how the poles of the eigenvector function $\laxcp_{\fldlen}$
are related to the soliton positions.

For elliptic solutions, we observed that the eigenvector function $\laxcp_{\fldlen}$
has a smooth infinite-length limit $\laxcp$
where the limit of the pole $\ptdiv{\ismmom}_{\fldlen}$ actually describes the pole $\ptdiv{\ismmom}$ for the asymptotic state.
We suppose that the same holds for the spectral curve approximations of a generic asymptotic state.
Concretely, we suggest that the poles at finite length $\fldlen$ all have a proper infinite-length limit,
and that this limit coincides with the poles of the asymptotic eigenvector function $\laxcp$.

For a better understanding of the situation,
it makes sense to inspect the poles of the asymptotic eigenvector function of pure soliton states.
We will do so in terms of the spectral parameter $\laxpar=\ismmom^2+\kdvfld_0$
where the resulting dependence of $\ptdiv{\laxpar}(x)$ on $x$ will be very illuminating.

For a single soliton, we find an $x$-dependence as described previously as
\[
\ptdiv{\laxpar}(x)= \kdvfld_0-\parscl^2\tanh\brk!{\parscl (x-\ptdiv{x})}^2.
\eqsep
\mpostusemath{KdVSolitonDivisorPlot}
\]
Note that the divisor pole starts and ends asymptotically at the
soliton point $\ptsol{\laxpar}=\kdvfld_0+\ismsing^2$,
and it switches to the other side of the branch cut
when it reaches the peak at $\kdvfld_0$.

For a two-soliton state, the divisor poles $\ptdiv{\laxpar}(x)$ are given by
the solutions to a quadratic equation (formulated in terms of $\ismmom$):
\[
\ismsing_1 (\ptdiv{\ismmom}^2-\ismsing_2^2)\coth\brk!{-\iunit\ismsing_1(x-\ptsol{x}_1)}
-\ismsing_2 (\ptdiv{\ismmom}^2-\ismsing_1^2)\tanh\brk!{-\iunit\ismsing_2(x-\ptsol{x}_2)}
-\ptdiv{\ismmom}(\ismsing_1^2-\ismsing_2^2)
=0.
\]
Here, $\ptsol{x}_{1,2}$ describe the positions of the two soliton centres.
\unskip\footnote{The asymmetric dependency on the soliton parameters
originates from an asymmetric yet convenient definition
of the soliton centres $\ptsol{x}_{1,2}$
using the assumption $\Im\ismsing_1>\Im\ismsing_2$.}
The qualitative dependence of $\ptdiv{\laxpar}$ on the position $x$ depends
largely on the distribution of the soliton centres.
When the two solitons are well-separated,
one finds that the resulting pair of functions is approximated
by the superposition of the functions for two individual solitons:
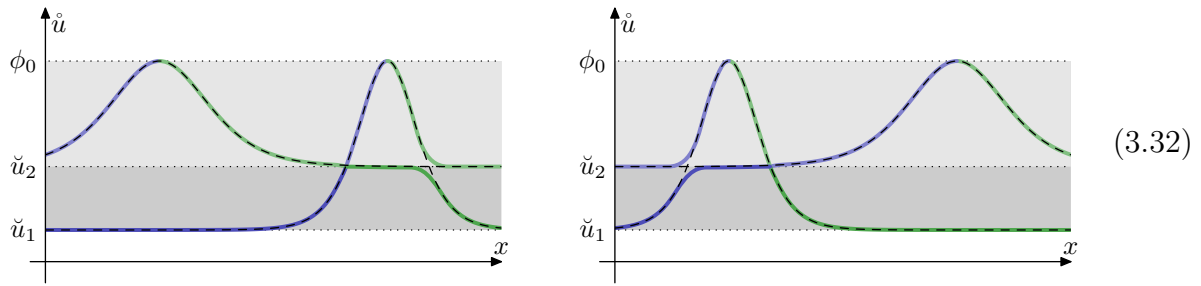
\[
\mpostusemath{KdVTwoSolitonDivisor1}
\eqsep
\mpostusemath{KdVTwoSolitonDivisor2}
\]
We note that there are two distinct cases based on whether the
slower and broader soliton is to the left or to the right of
the faster and more focussed soliton.
Conversely, when both solitons overlap significantly,
one finds an $x$-dependency which is less evidently
associated to the underlying state and its two contributing solitons:
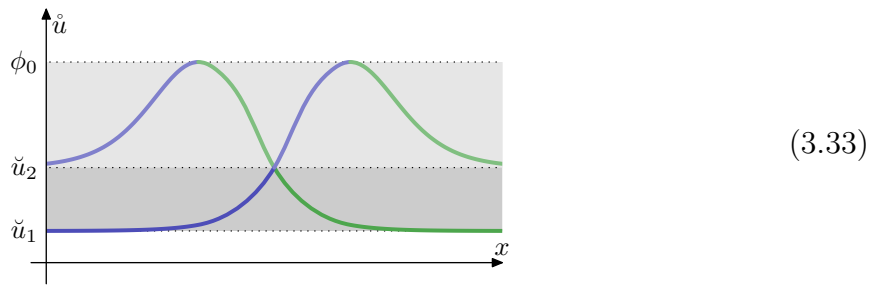
\[
\mpostusemath{KdVTwoSolitonDivisorC}
\]

Let us discuss the $x$-dependence of the divisor on the imaginary $\ismmom$-axis
and point out relevant features
supposing that the above picture generalises to arbitrary $N$.

Whenever the soliton centres are well-separated,
the composition of the divisor as the union of
the divisors of the individual solitons is well-expected
because every soliton has an exponentially weak residual influence
at the relevant regions for the other solitons.
The cases where some solitons overlap significantly
deforms smoothly to the cases where all solitons are well-separated

For all separations of soliton centres,
every interval between two consecutive soliton momenta $\ptsol{\laxpar}_n$ and $\ptsol{\laxpar}_{n+1}$
is populated by exactly one pole $\ptdiv{\laxpar}_n$ of the divisor for all positions $x$.

The pattern of crossings of the divisor poles as a function of the position $x$ is noteworthy:
There are $N(N-1)/2$ actual pairwise crossings
where the two poles $\ptdiv{\laxpar}_n$ and $\ptdiv{\laxpar}_{n+1}$
move in opposite directions and reside on opposite sides of the branch cuts.
Moreover, these crossings must occur exactly at the soliton momentum $\ptsol{\laxpar}_{n+1}$
that separates the intervals of the divisor poles.
The crossing can only occur at the prescribed point for the reason that
every interval must be populated by one divisor pole for all $x$.
This behaviour is compatible with spectral curves
where the divisor poles must remain on the same branch cut for all $x$.
The crossing at infinite length can thus be interpreted as an exchange of
roles of two consecutive divisor poles.

Furthermore, one observes one avoided crossing for all pairs of well-separated solitons.
Here, two consecutive poles come close while moving in the same direction,
i.e.\ while they are on the same side of the cuts.
Here, the crossing is avoided because such divisor poles repel each other.

It is interesting to observe that the divisor poles
become exponentially slow at the soliton points $\ptsol{\laxpar}_{n}$.
This feature reflects the asymptotic behaviour of solitons.
\unskip\footnote{In fact, this is how the position-dependence of the regularised monodromy
gets suppressed in the infinite-length limit: for $x\to\pm\infty$,
all soliton divisor poles $\ptdiv{\laxpar}_n$
are exponentially close to their starting and ending positions
at a soliton momentum $\ptsol{\laxpar}_n$.
Even though the full information on the soliton centres exists in the monodromy at finite but large $\fldlen$
in the form of exponentially small contributions,
it is impossible to extract it from the scattering matrix
and difficult to reimplement manually into an approximation for the monodromy.
Likewise, it is difficult to reconstruct the divisor at a position $x$
within the region of interest because the soliton divisor poles are very different
from their asymptotic values.}
However, either by means of an actual or an avoided crossing,
a pole can continue or pause and restart its journey around its branch cut.
At such a crossing, the divisor poles exchange their roles
in representing two different solitons.

Importantly, this pattern of crossings leads to the desired periodicity
of the $n$-th divisor pole $\ptdiv{\laxpar}_n$ encircling the $n$-th cut
exactly $n$ times as $x$ advances by one period or,
in the case of infinite length, along the whole real line.

In terms of physics,
the position of the highest soliton divisor pole $\ptdiv{\laxpar}_N$
is most immediately relevant:
whenever it moves around the branch point $\ptsol{\laxpar}_{N+1}=\kdvfld_0$,
the wave function displays a distinguished bump.
This happens $N$ times when all values of $x$ are traversed,
and these represent precisely the individual solitons.
The other $N-1$ divisor poles merely act auxiliary variables
that keep track of the positions of the other solitons.
In the case of well-separated solitons,
the other divisor poles are near their rest positions
without significant influence on the shape of the wave function.
However, when some solitons overlap, the subsequent divisor pole
takes on a non-trivial position which contributes to an effective joining
of the two soliton bumps into a single entity.

The most important conclusion to be drawn from the above discussion for our purposes
is that all allowable configurations of divisor poles are actually realised by
some configuration of the soliton centres relative to $x$.
Therefore, the only requirement for a proper infinite-length limit
is that the divisor poles associated to the cuts along the imaginary $\ismmom$-axis
possess a proper limit, whereas the limiting divisor can be in an arbitrary configuration.

%%%%%%%%%%%%%%%%%%%%%%%%%%%%%%%%%%%%%%%%%%%%%%%%%%%%%%%%%%%%%%%%%%%%%%%%%%%%%%%%
\subsection{Local Charges}

We have discussed a family of spectral curves that approximate a state on the infinite line.
The branch cuts of these spectral curves must behave in a specific way,
which also depends on the particular representation of the spectral curve,
e.g.\ using a particular spectral parameter.
Alternatively, the relevant moduli of spectral curves can be expressed in terms of action variables
which express the sizes of cuts in a more universal scheme.
Furthermore, action variables are the elementary phase space degrees of freedom
to express integrable systems in. For instance, they obey useful relations
with the local conserved charges and their corresponding wave numbers,
see e.g.\ \eqref{KdVEllChargeRel}.
Let us therefore investigate the infinite-length limit of
action variables and local charges based on the discussion above.

From the family of spectral curves, we can read off the action variables
as particular period integrals around the cuts.
For the small cuts due to the continuum,
we can locally assume as discussed in \eqref{KdVContF} that $\cos \laxqmom(\ismmom)$
oscillates with period $2\pi/\fldlen$ and amplitude $1/\abs{\ismtrans_n}$.
The action variable is determined by a period integral over the forbidden zone
of the wave at position $\ismmom_n$.
We find
\begin{align}
\label{KdVContAct}
\actvar_n = \frac{1}{\iunit\pi}\oint_{\contacyc_n}\laxpar \diff{\laxqmom}
&\approx
\frac{4\ismmom_n}{\pi \fldlen}
\int_{-\arccos\abs{\ismtrans_n}}^{+\arccos\abs{\ismtrans_n}}
\frac{\laxzpar\sin(\laxzpar)\diff{\laxzpar}}{\sqrt{\cos(\laxzpar)^2-\abs{\ismtrans_n}{}^2}}
\\
&=
-\frac{4\ismmom_n}{\fldlen} \log\abs{\ismtrans_n}
=
-\frac{2\pi n}{\fldlen^2} \log\brk!{1-\abs{\ismrefl_n}^2}.
\end{align}
We can convert the discrete action variable $\actvar_n$ to an action density function $\actvar(\ismmom)$
which remains finite in the infinite-length limit:
\[
\actvar_n \diff{n}
\approx
\actvar(\ismmom) \diff{\ismmom},
\eqsep
\actvar(\ismmom)
:=
-\frac{2\ismmom}{\pi} \log\brk!{1-\abs{\ismrefl(\ismmom)}^2}.
\]
Concerning the remaining branch cuts,
we have to bear in mind that solitons are associated to the interval between two branch cuts
rather than to a single branch cut.
Consequently, every branch cut connects two consecutive soliton singularities.
The action variable for the branch cut thus depends on the data for two solitons.
We can disentangle the two contributions by composing the cycle around the $n$-th branch cut
as the difference of two large cycles around all branch cuts from $n$ through $N$
and from $n+1$ through $N$.
For large $\fldlen$, the contributions from the other branch cuts on the period integral are suppressed
and we can reuse the result obtained from the elliptic curve. We find
\[
\label{KdVSolAct}
\actvar_n \approx \frac{4\fldlen}{3\pi}
\brk!{ (\Im\ismsing_n)^3-(\Im \ismsing_{n+1})^3},
\]
where we formally define $\ismsing_{N+1}=0$.
Note that here the action variable diverges linearly in the infinite-length limit.

Next, we would like to relate the action variables to the local charges
expressed through the scattering data for states on the infinite line.
These can be computed from the expansion of the quasi-momentum
$\laxqmom(\laxpar)$ at $\laxpar\to\infty$ and thus at $\ismmom\to\infty$.
Using that $\ismrefl(\ismmom)$ should be exponentially small at $\ismmom\to\infty$,
we have the relation $\laxqmom\approx\ismmom \fldlen+\arg\ismtrans$.
Furthermore, using the relation $\laxpar=\ismmom^2+\kdvfld_0$,
the expansion of $\laxqmom(\laxpar)$ at $\laxpar\to\infty$
translates to an expansion of $\arg\ismtrans(\ismmom)$ in terms
of the Galilei-invariant charges:
\[
\arg\ismtrans(\ismmom)
=
-\frac{\potshift-\half \kdvfld_0 \fldlen}{\ismmom}
-\frac{\tilde\totmom}{4\ismmom^3}
-\frac{\tilde\eng}{16\ismmom^5}
-\ldots.
\]
We have seen that the transmission coefficient $\ismtrans(\ismmom)$
is determined by a convolution integral of the other scattering data.
The expansion of $\arg\ismtrans(\ismmom)$ at $\ismmom\to\infty$
yields
\[
\arg\ismtrans(\ismmom)
=
\sum_{j=0}^\infty\frac{1}{\ismmom^{2j+1}}
\brk*{
\sum_{n=1}^N\frac{2(-1)^j}{2j+1}(\Im\ismsing_n)^{2j+1}
+\int_0^\infty \frac{\diff{\ismmom'}}{\pi} \ismmom'{}^{2j}\log\brk!{1-\abs{\ismrefl(\ismmom')}^2}
}
,
\]
where we have used the reality condition $\ismrefl(-\ismmom)=\ismrefl(\ismmom)^\conj$.
We can now directly read off the Galilei-invariant conserved charges
as
\begin{align}
\numberhere
\potshift &\approx
\half \kdvfld_0 \fldlen
-\sum\nolimits_n 2\Im\ismsing_n
-\int_0^\infty \frac{\diff{\ismmom}}{\pi} \log\brk!{1-\abs{\ismrefl(\ismmom)}^2},
\\
\tilde\totmom &\approx +\sum\nolimits_n\rfrac{8}{3}(\Im\ismsing_n)^3
- \int_0^\infty \frac{2\ismmom \diff{\ismmom}}{\pi} 2\ismmom \log\brk!{1-\abs{\ismrefl(\ismmom)}^2},
\\
\tilde\eng &\approx -\sum\nolimits_n\rfrac{32}{5}(\Im\ismsing_n)^5
- \int_0^\infty \frac{2\ismmom \diff{\ismmom}}{\pi} 8\ismmom^3 \log\brk!{1-\abs{\ismrefl(\ismmom)}^2}.
\end{align}
We point out that for a non-zero asymptotic field value $\kdvfld_0$,
the charge $\potshift$ (and thus the original charges $\totmom$, $\eng$)
contain contributions with a non-trivial asymptotic charge density.

The differentials of these conserved charges
can be expanded in the basis of differentials of action variables.
We find the generic form for a conserved charge $J=\potshift,\tilde\totmom,\tilde\eng$
(up to the contribution of $\der \kdvfld_0$ to $\potshift$)
as
\begin{align}
\numberhere
\der J &\approx \sum_{n=1}^N \brk*{\sum_{j=1}^n \frac{\pi f_J(\ismsing_j)}{\ismsing_j \fldlen}}\.\der \actvar_n
- \int_0^\infty \frac{2\ismmom \diff{\ismmom}}{\pi} f_J(\ismmom)\. \der\log\brk!{1-\abs{\ismrefl(\ismmom)}^2}
\\
&=
-\sum_{n=1}^N 4 \Im f_J(\ismsing_n)\. \der\brk!{\ismsing_n^2}
+\int_0^\infty \diff{\ismmom} f_J(\ismmom)\.  \der \actvar(\ismmom),
\end{align}
where the $f_J(\ismmom)$ are some applicable profile functions
for which we find
\[
f_{\potshift}(\ismmom)\approx\frac{1}{2\ismmom},
\eqsep
f_{\tilde\totmom}(\ismmom)= 2\ismmom,
\eqsep
f_{\tilde\eng}(\ismmom)\approx 8\ismmom^3.
\]
A fundamental relation of integrable systems
is that angle variables shift linearly under
transformations generated by conserved charges.
Therefore, the coefficients $f_J(\ismmom)$
represent periodicities for the corresponding angle variables.
We can thus verify that the above action variables
yield a consistent set of relations with the conserved charges and angle variables.
First, we note that the slopes corresponding to translations in $x$,
namely $f_{\tilde\totmom}(\ismmom_n)=2\pi n/\fldlen$
and the coefficient $2\pi n/\fldlen$ of $\der \actvar_n$ in $\der \tilde\totmom$,
take on the required discrete values $(2\pi/\fldlen) \Integer$
to ensure that the angle variables shift by integer multiples of $2\pi$ over one period.
At the level of spectral curves, this exact relation is ensured by the Riemann bilinear identity.
Furthermore, the slope $f_{\tilde\eng}\approx 8\ismmom^3$
is precisely the coefficient for time evolution of the argument of the reflection function $\ismrefl$.
Finally, all soliton slopes scale as $1/\fldlen$ indicating that the wave trains for solitons
lose their periodic behaviour in the infinite-length limit.

%%%%%%%%%%%%%%%%%%%%%%%%%%%%%%%%%%%%%%%%%%%%%%%%%%%%%%%%%%%%%%%%%%%%%%%%%%%%%%%%
\subsection{Asymptotic Limit}
\label{KdVAsympLim}

We have investigated how different structures of the inverse scattering method
are represented in spectral curves in the infinite-length limit.
We will now summarise the resulting asymptotic behaviour for spectral curves.
This describes the conditions for the existence of the infinite-length limit
at the level of spectral curves
and it provides expressions for the limiting scattering data.
We will also convert the expressions from the momentum variable $\ismmom$
to the spectral parameter $\laxpar=\kdvfld_0+\ismmom^2$ using \eqref{KdVLaxTrans}
which is a more natural parameter
for the spectral curve and all branch cuts reside on the real axis in this picture.

The distribution of the branch cuts for the spectral curve at large $\fldlen$
takes the following form:
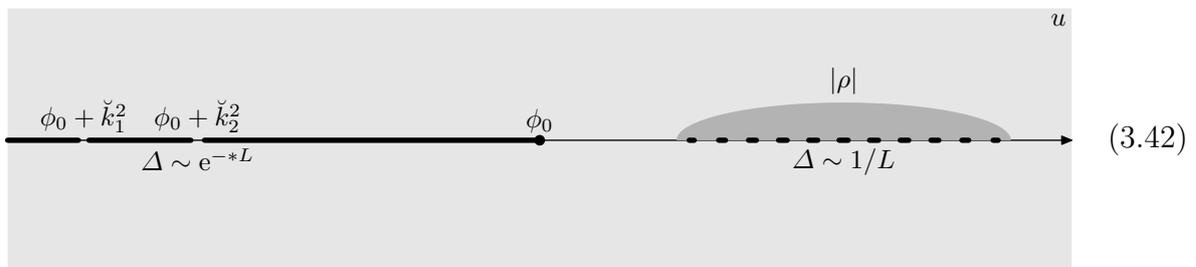
\[
\label{KdVsummary}
\mpostusemath{KdVLimitCuts}
\]
Let us sketch the expected distribution of branch cuts along the real line
for a family of spectral curves with an appropriate asymptotic limit:
The family of curves must consist of two distinct sets of branch cuts.
There must be a fixed number of $N+1$ branch cuts on the left.
All branch points must have a well-defined large-$\fldlen$ limit
such that the intervals between the cuts decrease exponentially with $\fldlen$.
Note that the first branch point is fixed to $-\infty$
while the last one limits to $\kdvfld_0$
that represents the asymptotic value of the KdV field.
The second set consists of an asymptotically increasing number of branch cuts to the right of $\kdvfld_0$.
\unskip\footnote{Importantly, the lengths of cuts must follow a certain limiting pattern.
This requirement yields some flexibility in assigning the individual cuts
so that their number will typically increase linearly with $\fldlen$,
but may actually fluctuate with $\fldlen$ or be infinite altogether.}
These are arranged in a regular pattern such that their lengths shrink to zero as $1/\fldlen$
while the relative filling along the real axis approaches a well-defined density in the large-$\fldlen$ limit.

The first set of $N+1$ branch cuts (labelled by $n=0,\ldots,N$) describes the configuration of solitons:
a pair of consecutive branch points must have a common limit 
\[
\ptsqrt{\laxpar}_n^- , \ptsqrt{\laxpar}_{n-1}^+ \to \ptsol{\laxpar}_n=\kdvfld_0+\ismsing_n^2,
\eqsep
\ismsing_n = \iunit \sqrt{\kdvfld_0-\ptsol{\laxpar}_n},
\]
and which describes the soliton pole $\ismsing_n$.
The remaining cuts (labelled by $n=N+1,\ldots$)
are delimited by the branch points \eqref{KdVContCutCentr}
which for $n\sim \fldlen$ must arrange in a regular pattern
\[
\ptsqrt{\laxpar}_n^\pm \sim \kdvfld_0
+\frac{\pi^2 n^2}{\fldlen^2}
\pm\frac{2\pi n}{\fldlen^2}\arcsin\abs!{\ismrefl(\pi n/\fldlen)}
+\ldots,
\]
which describes the magnitude of the reflection function $\ismrefl(\ismmom)$.
This constitutes the complete set of conserved data for the asymptotic state
in the inverse scattering method.

These data determine the transmission function $\ismtrans(\ismmom)$
according to the dispersion relation \eqref{faddeevformula}
which constrains the asymptotic behaviour of the spectral curve data further:
The separation of the branch points corresponding to solitons
shrinks exponentially according to the relation
\eqref{KdVSolSep}:
\[
\ptsqrt{\laxpar}_n^- - \ptsqrt{\laxpar}_{n-1}^+
\approx
8\Im(\ismsing_n) \Im \res \ismtrans(\ismsing_n) \exp(\iunit \ismsing_n \fldlen).
\]
The remaining branch points corresponding to the continuum
must take the limiting behaviour
\[
\ptsqrt{\laxpar}_n^\pm \sim \kdvfld_0+
\brk[s]*{\frac{\pi n-\arg\ismtrans(\pi n/\fldlen)\pm\arccos\abs{\ismtrans(\pi n/\fldlen)}}{\fldlen}}^2.
\]
In effect, this limiting relation describes the magnitude and phase
of the transmission function $\ismtrans(\ismmom)$ in terms of
the modulation pattern of the branch cuts which must behave as follows:
As the length $\fldlen$ increases, the individual branch cuts move towards $\kdvfld_0$
while they also shrink accordingly.
However, the relative filling of the cuts
remains fixed in place along the real axis
according to the function $\ismtrans(\pi n/\fldlen)$.
Note that a finite point $\laxpar$ resides near the $n$-th branch cut with $n\approx \fldlen\laxpar/\pi$,
therefore the number of branch cuts must asymptotically grow.
Note further the shift by $\arg\ismtrans(\pi n/\fldlen)$ has a relatively large effect
on the first few cuts with $n=N+1,\ldots$ where $\arg\ismtrans(0)\approx \pi N$
so that the first few cuts are positioned near the central point $\kdvfld_0$.

To avoid some of the above complications in fixing detailed relations for the branch point positions,
the spectral curve configuration can be described
in terms of action variables, see \eqref{KdVSolAct} and \eqref{KdVContAct}.
The relevant insight is that the action variables of the cuts on the left all scale as $\fldlen$.
The soliton poles $\ismsing_n$ on the imaginary axis are expressed in terms
of the limiting behaviour of the action variables as
\[
\sum_{m=n}^N \actvar_m\sim \frac{4\fldlen}{3\pi}(\Im\ismsing_n)^3.
\]
As discussed in \secref{KdVSolDiv}, the divisor poles for these cuts must all
have proper limiting points which can be in an arbitrary configuration.
The action variables corresponding to continuum cuts
were determined in \eqref{KdVContAct},
\begin{align}
\actvar_n \sim
-\frac{2\pi n}{\fldlen^2} \log\brk!{1-\abs!{\ismrefl(\pi n/\fldlen)}^2}
=
-\frac{4\pi n}{\fldlen^2} \log\abs!{\ismtrans(\pi n/\fldlen)}.
\end{align}
Note that their limiting behaviour directly specifies
the magnitude of the reflection and transmission functions. 

The limiting behaviour of the divisor is more intricate,
and it has been discussed in detail in \secref{KdVContDiv} and \secref{KdVSolDiv}.
Most noteworthy,
the analysis for the continuum cuts showed that the effective phases $\sigma_n$
for the position of a divisor pole on its respective cut
must be aligned properly for a coherent state to form in the infinite-length limit.
These phases must shift rapidly
by an angle of $2\pi/\fldlen$ from one cut to the next
according to \eqref{effectivephase}.
The effective phase then determines the argument of the reflection
function $\ismrefl$
\[
\frac{2\pi n x}{\fldlen}-\sigma_n\to \arg\ismrefl(\pi n/\fldlen).
\]

In this section, we have thus characterised how a family of spectral curves describing periodic KdV states
degenerates to the scattering data describing a KdV state on an infinite line with proper asymptotic boundary conditions.
We conclude that this family of spectral curves
necessarily displays two sectors of branch cuts
which behave very differently in the infinite-length limit
and which describe the two distinct structures of the spectral data,
namely the solitons and the continuum.

%%%%%%%%%%%%%%%%%%%%%%%%%%%%%%%%%%%%%%%%%%%%%%%%%%%%%%%%%%%%%%%%%%%%%%%%%%%%%%%%
%%%%%%%%%%%%%%%%%%%%%%%%%%%%%%%%%%%%%%%%%%%%%%%%%%%%%%%%%%%%%%%%%%%%%%%%%%%%%%%%
\section{Continuous Heisenberg Magnet}
\label{sec:CHM}

After having laid down the procedure for finite-length extrapolation
of a generic infinite-length KdV state in the previous section,
we will now apply it to the continuous Heisenberg magnet (CHM).
This further prototypical integrable model displays 
inverse scattering and spectral curve data with somewhat different features,
which thus makes it an interesting case to discuss and compare.
Its equation of motion is a version of the Landau--Lifshitz equation \cite{landau1935equation}
and it is equivalent to the nonlinear Schrödinger (NLS) equation.

The CHM model consists of a spin vector field $\spinvec\in\Sphere^2\subset\Real^3$
in one spatial dimension. The equation of motion is the isotropic Landau--Lifshitz equation
\[
\label{CHMeom0}
\dot{\spinvec} = \spinvec \vtimes \spinvec''.
\]
We will express the spin vector in terms of the complex stereographic projection $\spincp\in\ComplexComplete$
\[
\label{CHMstereovecrel}
\spinvec=\frac{1}{1+\spincp\bar\spincp}\begin{pmatrix}
\spincp+\bar\spincp\\
-\iunit\spincp+\iunit\bar\spincp\\
1-\spincp\bar\spincp
\end{pmatrix}.
\]
The equation of motion for $\spincp$ then takes the equivalent form
\[
\label{CHMeom}
\dot\spincp = \iunit\spincp''-\frac{2\iunit \bar\spincp\spincp'^2}{1+\spincp\bar\spincp}
.
\]
The integrable structure of the model can be expressed in terms of the auxiliary linear problem
\[
\laxvec' = \laxconn \laxvec,
\eqsep
\laxconn(\laxpar)=\frac{\iunit}{\laxpar}\paulivec\vdot\spinvec
\label{CHMALP}
\]
with the components
\[
\laxconn(\laxpar)
=\frac{\iunit}{\laxpar}\frac{1}{1+\spincp\bar\spincp}
\begin{pmatrix}
1-\spincp\bar\spincp & 2\bar\spincp
\\
2\spincp& \spincp\bar\spincp-1
\end{pmatrix},
\eqsep
\laxvec=
\begin{pmatrix}\laxvec_1 \\ \laxvec_2\end{pmatrix}.
\]

Let us start by introducing the inverse scattering method for infinite-length states
and the spectral curve method for finite-length periodic states
along with their sets of data.
We will then discuss some simple states and adjust our method
for obtaining the infinite-length limit of periodic states.
This allows us to extract the limiting structures
that give rise to the continuum and solitons.
Finally we summarise.

%%%%%%%%%%%%%%%%%%%%%%%%%%%%%%%%%%%%%%%%%%%%%%%%%%%%%%%%%%%%%%%%%%%%%%%%%%%%%%%%
\subsection{Scattering Data}

The inverse scattering method for the CHM model has been discussed in 
\cite{Takhtajan:1977rv,fogedby1980solitons,gelfand1951determination,marchenko1952,DEMONTIS201835,demontis2019continuous}.
For the infinite line, we assume the asymptotic conditions
\[
\lim_{x \to \pm \infty} \spinvec
=
\begin{pmatrix}0\\0\\1\end{pmatrix}
\eqjoin*{\iff}
\lim_{x \to \pm \infty} \spincp = 0.
\]
As usual, the solutions of the auxiliary problem can be characterised by their asymptotic properties
in the far left or in the far right
where the Lax connection simplifies to the diagonal form
\[
\lim_{x \to \pm \infty}  \laxconn(\laxpar)=
\begin{pmatrix}
-\iunit \ismmom& 0
\\
0 & \iunit \ismmom
\end{pmatrix}
\eqjoin{\text{with}} \ismmom := -\frac{1}{\laxpar}.
\]
Consequently, the states are characterised by the spin direction (up or down)
in combination with their momentum $\pm\ismmom$.
The two states with prescribed asymptotics on the left
and the two corresponding ones for the right are related by a
linear relation which is expressed through the scattering matrix
\[
\label{CHMtransportscattering}
\ismscat(\ismmom)
=
\lim_{\winbegend \to \mp \infty}
\begin{pmatrix}\eunit^{\iunit \ismmom \winend}&0\\0&\eunit^{-\iunit \ismmom \winend}\end{pmatrix}\.
\laxtransport(\laxpar;\winend,\winbeg)\.
\begin{pmatrix}\eunit^{-\iunit \ismmom \winbeg}&0\\0&\eunit^{\iunit \ismmom \winbeg}\end{pmatrix}.
\]
The underlying Lax parallel transport $\laxtransport(\laxpar;\winend,\winbeg)$
from $\winbeg$ to $\winend$ takes the same general form as before in \eqref{LaxParallelTransport}.
%
%\[
%\laxtransport(\winend,\winbeg):=\pathordleft \brk[s]*{\exp \int_{\winbeg}^{\winend} \diff{x}\laxconn(x)}.
%\]
%
We parametrise the auxiliary scattering matrix as
\[
\label{CHMscatteringmatrix}
\ismscat(\ismmom)
=
\begin{pmatrix}
1/\ismtrans(\ismmom) & -\ismrefl(\ismmom)/\ismtrans(\ismmom)
\\
\ismrefl(\ismmom^\conj)^\conj/\ismtrans(\ismmom^\conj)^\conj & 1/\ismtrans(\ismmom^\conj)^\conj
\end{pmatrix}
\]
in terms of the complex transmission and reflection functions $\ismtrans(\ismmom)$ and $\ismrefl(\ismmom)$, respectively.
The scattering matrix is unimodular by construction, $\det\ismscat=1$,
which implies the following relation between the scattering functions
\[
1+\ismrefl(\ismmom)\ismrefl(\ismmom^\conj)^\conj
=\ismtrans(\ismmom)\ismtrans(\ismmom^\conj)^\conj.
\]
Furthermore, the scattering matrix obeys the relation $\ismscat(\ismmom)^\hconj = \ismscat(\ismmom^\conj)^{-1}$
which reduces to plain unitarity for real momenta $\ismmom$.

The scattering data for the CHM model consists of the reflection function $\ismrefl(\ismmom)$
on the real axis $\ismmom\in\Real$
together with $N$ soliton momenta $\ptsol{\ismmom}_n \in \bar\Complex^+$
in the upper half complex plane
and corresponding dynamical coefficients $\ismres_n\in\Complex$:
\[
\mpostusemath{CHMScatteringData}
\]
The transmission function $\ismtrans(\ismmom)$
on the upper half of the complex plane including the real axis, $\Im \ismmom \geq 0$,
should be fixed by a dispersion relation.
In analogy to the formula \eqref{faddeevformula} for the KdV scattering problem,
we expect it to take the form:
\[
\ismtrans(\ismmom)=
\brk[s]*{\prod\nolimits_j\frac{\ismsing_j}{\ismsing_j^\conj}\frac{\ismmom-\ismsing_j^\conj}{\ismmom-\ismsing_j}}
\exp\brk[s]*
{
\frac{1}{2\pi\iunit}\int \diff{\ismmom'}
\brk*{\frac{1}{\ismmom'-\ismmom-\iunit 0}-\frac{1}{\ismmom'}}\log\brk!{1+\abs{\ismrefl(\ismmom')}^2}
}.
\label{CHMantiunitarydisp}
\]
By construction, this function features the following analytic behaviour:
It is meromorphic for $\Im k>0$ with poles at the prescribed positions $\ismsing_j$.
For $\ismrefl(\ismmom)=0$, it agrees with expectations for purely solitonic states,
see also \cite{DEMONTIS201835}.
For small $\ismrefl(\ismmom)$ we have confirmed that it agrees with
the leading-order result from the inverse scattering method.
Furthermore, the magnitude of $\ismtrans(\ismmom)$ for real $\ismmom$
is determined by the distributional identity in \eqref{distribution}
as
\[
\abs{\ismtrans(\ismmom)}=\sqrt{1+\abs{\ismrefl(\ismmom)}{}^2}.
\label{antidispersion}
\]
This implies the relation $\abs{\ismtrans}^2 = 1 + \abs{\ismrefl}^2$
which follows from hermiticity of the CHM model auxiliary linear problem on the real $\ismmom$-axis.
Finally, at $\ismmom=0$ and thus $\laxpar=\infty$, the Lax connection becomes trivial
implying $\ismtrans=1$.

Let us now collect the key differences to the scattering data for the KdV model.
First, the reflection and transmission functions are genuinely complex functions for the CHM model
whereas in the KdV model they obey the reality conditions $\ismtrans(\ismmom)^\conj=\ismtrans(-\ismmom)$
and $\ismrefl(\ismmom)^\conj=\ismrefl(-\ismmom)$.
Second, the reflection function $\ismrefl(\ismmom)$ is unconstrained for $\ismmom\in\Real$,
its magnitude is not bounded by $1$ as for the KdV model
while the magnitude of the transmission function $\ismtrans(\ismmom)$
is bounded from below by $1$ rather than from above.
The soliton momenta $\ismsing_j$ can reside anywhere on the upper complex half plane
rather than being restricted to the positive imaginary axis,
and the dynamical coefficients $\ismres_n$ are complex rather than real.
Finally, the soliton momenta do not need to be distinct for the CHM model,
there can be any number of coincident soliton momenta
(supposing the corresponding dynamical coefficients $\ismres_n$
are adjusted appropriately to this situation).

Finally, it makes sense to express some essential conserved charges
in terms of the scattering data.
The momentum $\totmom$ and energy $\eng$ can be expressed in terms of the expansion
of $\ismtrans(\ismmom)$ around the point $\ismmom=0$ as
\[
\arg\ismtrans(\ismmom)=\frac{\totmom}{2}+\frac{\eng}{4\ismmom}+\ldots.
\]
Likewise, the angular momentum is expressed by the expansion
around the point $\ismmom=\infty$:
\[
\arg\ismtrans(\ismmom)=\Delta\angmom\.\ismmom+\ldots.
\]
Here, $\Delta\angmom$ denotes the difference of the
angular momentum to the angular momentum of the ground state
which has a constant angular momentum density $1$
and thus carries an infinite amount of angular momentum on the infinite line.
The resulting charges for an asymptotic state specified through its scattering data thus read
\begin{align}
\label{CHMISMCharges}
\totmom&=
4\sum_j\arg\ismsing_j
+\frac{1}{\pi}\int \frac{\diff{\ismmom}}{\ismmom}\log\brk!{1+\abs{\ismrefl(\ismmom)}^2},
\\
\eng&=
8\sum_j\Im\ismsing_j
+\frac{2}{\pi}\int \diff{\ismmom}\log\brk!{1+\abs{\ismrefl(\ismmom)}^2},
\\
\Delta\angmom&=
-2\sum_j\frac{\Im\ismsing_j}{\abs{\ismsing_j}^2}
-\frac{1}{2\pi}\int \frac{\diff{\ismmom}}{\ismmom^2}\log\brk!{1+\abs{\ismrefl(\ismmom)}^2}.
\end{align}
The derivative of a charge $Q=\totmom,\eng,\Delta\angmom$ can be expressed universally as
\[
\label{CHMISMChargesDer}
\der Q=
\sum_j 2\Im\brk[s]*{f_Q(\ismsing_j)\.\der\brk*{-\frac{1}{\ismsing_j}}}
+\int f_Q(\ismmom) \frac{\diff{\ismmom}}{2\pi\ismmom^2}\der\log\brk!{1+\abs{\ismrefl(\ismmom)}^2},
\]
with the profile functions
\[
f_{\totmom}(\ismmom)=2\ismmom,
\eqsep
f_{\eng}(\ismmom)=4\ismmom^2,
\eqsep
f_{\Delta\angmom}(\ismmom)=-1.
\]

%%%%%%%%%%%%%%%%%%%%%%%%%%%%%%%%%%%%%%%%%%%%%%%%%%%%%%%%%%%%%%%%%%%%%%%%%%%%%%%%
\subsection{Spectral Curve and Divisor}

We now turn to the states for a finite interval with periodic boundary conditions
\[
\spinvec(x+\fldlen)=\spinvec(x).
\]
States with these boundary conditions are described by spectral curves
which are likewise based on the above auxiliary linear problem,
see \cite{kotlyarov1976periodic,its1976explicit}.
Here, the monodromy $\laxmono(\laxpar)$ is defined
as the parallel transport of the Lax connection $\laxconn(\laxpar;x)$ over one period
\[
\laxmono(\laxpar;x)=\laxtransport(\laxpar;x+\fldlen,x).
\]
The data of the underlying state are encoded by the eigensystem of the monodromy $\laxmono$ as follows.

The two eigenvalues are independent of the reference point $x$ and they are conserved in time.
As $\det \laxmono=1$ is fixed,
it suffices to denote the eigenvalue functions as $\laxval(\laxpar)$ and $1/\laxval(\laxpar)$.
The diagonalisation of $\laxmono(\laxpar)$ introduces $2K$ square-root branch points $\ptsqrt{\laxpar}_j$
for the eigenvalue function $\laxval(\laxpar)$ on the complex $\laxpar$-plane.
These are the points where the monodromy becomes non-diagonalisable
implying that both eigenvalues coincide, thus $\laxval(\ptsqrt{\laxpar}_j)=\pm 1$.
Due to reality of the physical system,
the branch points $\ptsqrt{\laxpar}_j$ are distributed symmetrically around the real axis.
Furthermore, the branch points cannot reside on the real axis
because the monodromy is unitary and thus diagonalisable for real $\laxpar$.
Therefore, it makes sense to pair up branch points as complex conjugates
$\ptsqrt{\laxpar}_{K+j}=\ptsqrt{\laxpar}_{j}^\conj$
such that a branch cut extends vertically between them in the complex plane.
\unskip\footnote{The position of branch cuts is not universally defined:
The simplest choice is to let all branch cuts extend purely vertically.
A more physically meaningful choice is to require the action density
to be real along the branch cut;
then the branch cuts curl slightly away from the origin of the $\laxpar$-plane \cite{Beisert:2003xu}.
Unfortunately, the resulting configuration of branch cuts can be rather intricate
even leaving some discrete freedom.}

Next to the branch points $\ptsqrt{\laxpar}_j$,
there are two distinguished values $\laxpar=\infty$ and $\laxpar=0$.
For the former, the Lax connection trivialises such that $\laxmono(\infty)=1$.
For the latter, the singularity of the Lax connection yields
an exponential singularity in $\laxval(\laxpar)$ at $\laxpar=0$.
To better control it, we introduce the quasi-momentum $\laxqmom(\laxpar)$
defined as the logarithm of the eigenvalue function,
$\laxqmom(\laxpar):=-\iunit \log\laxval(\laxpar)$.
This function has a single pole at $\laxpar=0$ with residue $\fldlen$.
As the logarithm introduces a $2\pi$-ambiguity in $\laxqmom$,
it makes sense to instead consider its derivative $\der\laxqmom/\der\laxpar$.

Altogether, the spectrum of the monodromy determines the spectral curve
as the set of permissible pairs $(\laxpar,\pm\der\laxqmom/\der\laxpar)$.
With the above properties of $\laxval(\laxpar)$,
the spectral curve is a Riemann surface of genus $K-1$ with punctures at the two points $\laxpar=0$.
The branch cuts split the spectral curve into two sheets corresponding
to the two eigenvalues. The distribution of branch cuts of can be depicted as follows:
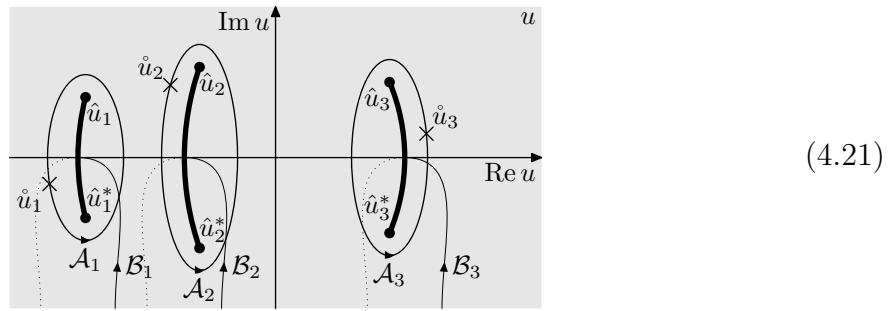
\[
\mpostusemath{CHMCurve}
\]

The configuration of branch points encodes the conserved charges of the state
through integrals on the spectral curve:
We define $\contacyc_j$ as a closed contour encircling the
complex conjugate pairs of branch points $\ptsqrt{\laxpar}_j,\ptsqrt{\laxpar}_j^\conj$
and $\contbcyc_j$ as an open contour connecting the two points with $\laxpar=\infty$
passing between $\ptsqrt{\laxpar}_j$ and $\ptsqrt{\laxpar}_j^\conj$.
\unskip\footnote{This assignment is not universal,
it depends on the choice of branch cut distribution.
Our choice is clearest if the branch cuts are short and well separated,
but there also exist situations where branch cuts collide, overlap or join.
To address the latter, we take a continuous deformation
to well-separated branch cuts to define a proper assignment.}
The A-periods of $\diff\laxqmom$ are all zero
while its B-periods
\unskip\footnote{Strictly speaking, only differences of B-periods
are period integrals for closed cycles.
Here, it is convenient to incorporate the fixed values $\laxval(\laxpar)=+1$ at $\laxpar=\infty$.}
evaluate to integer multiples of $2\pi$.
The remaining continuous data on conserved charges is expressed
by A-periods of $\laxpar\diff\laxqmom$.
Altogether we have the relations
\[
\label{CHMcurvequant}
\oint_{\contacyc_j} \diff\laxqmom=0,
\eqsep
\frac{1}{2\pi}
\int_{\contbcyc_j} \diff\laxqmom=\modenum_j,
\eqsep
\frac{1}{2\pi \iunit}
\oint_{\contacyc_j} \laxpar \diff\laxqmom=\actvar_j.
\]
In physical terms, the data of the spectral curve describe
a set of $K$ excitations with action variables $\actvar_j\in\Real$
and mode numbers $\modenum_j\in\Integer$ corresponding to the spatial period.
The mode numbers are equivalently obtained as the values of the quasi-momentum function
at the branch branch points, $\laxqmom(\ptsqrt{\laxpar}_j)=\laxqmom(\ptsqrt{\laxpar}_j^\conj)=\pi\modenum_j$.
All of the $\actvar_j$ and $\modenum_j$ represent independent quantities.
Furthermore, the expansions of the quasi-momentum at the special points
$\laxpar=\infty$ and $\laxpar=0$ yield the total angular momentum $\angmom$,
the momentum $\totmom$ and the energy $\eng$
(as well as the higher conserved charges):
\[
\laxqmom(\laxpar)=\frac{\angmom}{\laxpar} + \Order(1/\laxpar^2),
\eqsep
\laxqmom(\laxpar)=\frac{\fldlen}{\laxpar}-\half\totmom+\quarter\eng\laxpar + \Order(\laxpar^2).
\]
The conserved charges are related to the action variables by the
vanishing of the net residue of $\laxpar \diff\laxqmom$:
\[
\label{CHMactreal1}
\sum_{j=1}^K \actvar_j = \fldlen-\angmom.
\]
Furthermore, a Riemann bilinear identity implies the following relation
among the conserved charges
\[
\label{CHMactreal2}
-2\pi\sum_{j=1}^K \modenum_j\actvar_j = \fldlen\totmom.
\]
This concludes the description of conserved data.

The time-dependent data of the underlying state
are encoded by the eigenvectors of the monodromy $\laxmono(\laxpar)$.
It suffices to consider the points $\laxpar=\ptdiv{\laxpar}_j$
\footnote{More precisely, points on the spectral curve
because these also specify uniquely the corresponding eigenvalue or eigenvector.}
where the monodromy eigenvector points in a given reference direction.
There are precisely $K$ such points,
their set is called the dynamical divisor
and they encode the dynamical data for the state.

The divisor points depend on the reference point $x$ for the monodromy $\laxmono(\laxpar;x)$.
Towards reconstruction of a state $\spinvec(x)$,
it makes sense to consider the entire dependence of the divisor on $x$.
This dependency is well-prescribed by the Lax connection
and it which can be sketched qualitatively as follows:
Each divisor point $\laxpar=\ptdiv{\laxpar}_j$ is associated to one branch cut.
As $x$ increases by one period,
the divisor point moves $\modenum_j$ times
around the associated branch cut (loosely speaking along $\contacyc_j$).
\unskip\footnote{Unfortunately, this is merely an oversimplified picture
because the contours around the branch points may recombine
when they are too large (depending on the choice of reference direction)
or when branch cuts are too close or even unite.
Nevertheless, the simplified picture can be used for bookkeeping purposes.}

\medskip

To conclude, let us compare the spectral curve picture
with the one for the KdV model, and point out relevant differences.
The most immediate difference is that for the CHM model the branch cuts are oriented
in a vertical direction rather than horizontally along the real axis as for the KdV model.
There is no semi-infinite branch cut in the CHM model.
Also the behaviour of the divisor points is substantially more intricate in the CHM model
than in the KdV model.
In the KdV model, each divisor point resides on a unique branch point
as far as the reference direction for the monodromy eigenvectors is suitably chosen.
In the CHM model, there is no such choice, and the divisor points merely move around
a loosely associated branch cut.

%%%%%%%%%%%%%%%%%%%%%%%%%%%%%%%%%%%%%%%%%%%%%%%%%%%%%%%%%%%%%%%%%%%%%%%%%%%%%%%%
\subsection{Simple States}

In this section, we will gather some experience on the infinite-length limit
for the CHM model by considering three families of explicit states
and how they are related.
They all represent travelling wave solutions with an invariant shape
\cite{NAKAMURA1974321,LAKSHMANAN1976577,PhysRevB.15.3470,Takhtajan:1977rv,fogedby1980solitons}
travelling at a fixed velocity $v$ and rotating about the $z$-axis with an angular frequency $\parfreq$.
Using rotational symmetry, we always fix the overall angular momentum
to be aligned with the $z$-axis.

\medskip

The first family of states describes a single CHM soliton
which in stereographic projection takes the form
\[
\label{CHMsoliton}
\spincp(x)=
\exp\brk!{2\iunit \Re (\ismsing) (x-\parpos)+\iunit\parang}
\frac{\iunit \sin(\arg\ismsing)}{\cosh\brk!{2\Im(\ismsing)(x-\parpos)-\iunit \arg\ismsing}}.
\]
This state describes a localised travelling and rotating bump on the real line
with asymptotics $\spincp(x) \to 0$ as $x \to \pm \infty$.
The complex parameter $\ismsing$ with $\Im\ismsing>0$ describes the
singularity of the scattering matrix in the inverse scattering method.
It encodes the spatial velocity and angular frequency of the soliton wave function as
\[
\ismsing = -\rfrac{1}{4} \parvel + \rfrac{\iunit}{4} \sqrt{4\parfreq - \parvel^2}
\eqjoin{\iff}
\parvel=-4\Re(\ismsing),
\eqsep*
\parfreq=4|\ismsing|^2.
\]
An interesting consequence of these relations is that for a given velocity $\parvel$
there is a lower bound $\parfreq \geq \rfrac{1}{4}\parvel^2$ on the angular frequency.
The two dynamical variables $\parpos$ and $\parang$ describe the spatial position of the bump
and its angle about the $z$-axis.
The shape of the soliton can be depicted as follows:
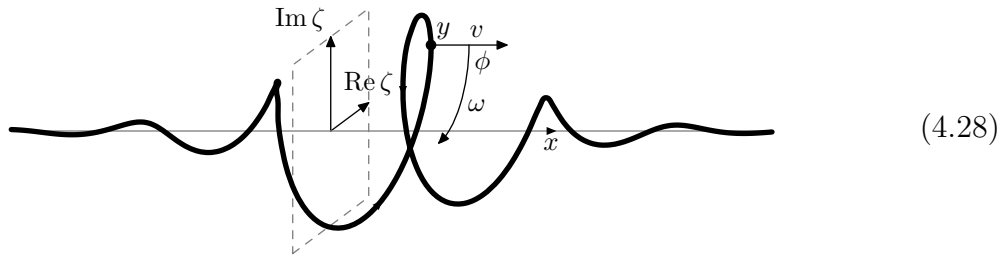
\[
\label{chmsolitonfig}
\mpostusemath{CHMSpinSoliton3D}
\]
The conserved charges of the soliton read:
\[
\totmom = 4\arg(\ismsing),
\eqsep
\eng = 8\Im(\ismsing),
\eqsep
\Delta\angmom = -2\frac{\Im(\ismsing)}{\abs{\ismsing}{}^2}.
\]
They are in perfect agreement with the general expressions
\eqref{CHMISMCharges} for the inverse scattering method.

\medskip

The second family of states describes spin waves with constant amplitude
\[\label{CHMspinwave}
\spincp(x) =
\tan(\half\vartheta)\exp\brk{-\iunit \parwnum x + \iunit \parang},
\eqsep
\parwnum:=\frac{2\pi \modenum}{\fldlen}.
\]
%
%and the corresponding spin vector field is given by
%%
%\[
%  \spinvec(x,t) =
%\brk*{\sqrt{1-\frac{\parvel^2}{\parwnum^2}}\cos\brk!{s\parwnum x + \parang},
%\sqrt{1-\frac{\parvel^2}{\parwnum^2}}\sin\brk!{s\parwnum x + \parang},
%\frac{\parvel}{s\parwnum}}.
%\]
%%
In this state, the spin vector rotates at a constant latitude $\vartheta$ around the $z$-axis.
The state is periodic with wave length $\fldlen>0$
and it has the periodicity mode number $\modenum\in\Integer$.
The wave number in the spatial direction is thus $\parwnum=2\pi \modenum/\fldlen$
so that the wave function actually repeats $\abs{\modenum}$ times within the periodicity interval $\fldlen$.
\unskip\footnote{One might effectively divide $\fldlen$ by $\modenum$
and consider only states with mode number $\pm 1$.
These have the same wave function,
but they belong to different sectors of the mechanical model
specified by the periodicity parameter $\fldlen$.
In particular, general states can be viewed as (non-linear) superpositions
of spin waves with different $\modenum$.}
The sign $s=\sign\modenum=\sign\parwnum=\pm 1$ determines the direction of motion.
As their amplitude is perfectly constant, the notion of initial position $\parpos$
is intertwined with the notion of initial angle $\parang$,
and the state is already specified by either quantity.
In terms of dynamics, the state travels along the spatial direction with velocity $\parvel=-\parwnum\cos\vartheta$
and rotates about the $z$-axis with angular frequency $\parfreq=-\parwnum\parvel=\parwnum^2\cos\vartheta$.
Again, the latter two processes are indistinguishable.

Let us briefly discuss the spectral curve data for this periodic class of states.
The quasi-momentum function is given by
\[
\laxqmom
= \pi\modenum +\frac{\pi\modenum}{\laxpar}\sqrt{\laxpar^2-\frac{4\laxpar}{\parwnum}\cos\vartheta+\frac{4}{\parwnum^2}}
= \pi\modenum +\frac{\pi\modenum}{\laxpar}\sqrt{\laxpar-\ptsqrt{\laxpar}}\sqrt{\laxpar-\ptsqrt{\laxpar}^\conj}
\]
with the complex branch point
\[
\ptsqrt{\laxpar}=\frac{\fldlen}{\pi\modenum}\exp(\iunit s\vartheta)
\eqjoin{\iff}
\fldlen
=\pi \abs{n\ptsqrt{\laxpar}},
\eqsep*
\vartheta=s\arg\brk*{s\ptsqrt{\laxpar}}.
\]
The spectral curve is of the rational kind
\[
\frac{\der\laxqmom}{\der\laxpar}
=
\pi \modenum\Re(\ptsqrt{\laxpar})
\frac{\laxpar-\ptcrit{\laxpar}}{\laxpar^2\sqrt{\laxpar-\ptsqrt{\laxpar}}\sqrt{\laxpar-\ptsqrt{\laxpar}^\conj}} ,
\eqsep
\ptcrit{\laxpar} = \frac{\fldlen}{\pi\modenum \cos\vartheta}= \frac{\abs{\ptsqrt{\laxpar}}^2}{\Re(\ptsqrt{\laxpar})}.
\]
The conserved charges of the state read:
\[
\totmom=-2\pi\modenum(1-\cos\vartheta),
\eqsep
\eng=\frac{2\pi^2\modenum^2}{\fldlen}\sin^2\vartheta,
\eqsep
\angmom=\fldlen \cos\vartheta.
\]
%
%\[
%\totmom=-2\pi\modenum+2\pi\modenum\frac{\Re(\ptsqrt{\laxpar})}{s\abs{\ptsqrt{\laxpar}}},
%\eqsep
%\eng=2\pi\frac{\Im(\ptsqrt{\laxpar})^2}{\abs{\ptsqrt{\laxpar}}^3},
%\eqsep
%\angmom=\pi\modenum\Re(\ptsqrt{\laxpar}).
%\]
%
Furthermore, the action variable integral yields:
\[
\label{CHMSpinWaveAct}
\actvar = \frac{1}{2\pi\iunit}\oint_{\contacyc} \laxpar\diff\laxqmom
=\fldlen(1-\cos\vartheta)
=\pi \abs{\modenum\ptsqrt{\laxpar}}-\pi\modenum\Re(\ptsqrt{\laxpar})
.
\]
We sketch a branch cut trivialisation of the spectral curve:
\[
\mpostusemath{CHMTrigonometricDivisor}
\]
We point out that states with the opposite angle $\vartheta'=\pi-\vartheta$
and the opposite mode number $\modenum'=-\modenum$
are related by a rotation of the spin vector by $\pi$.
Their spectral curves are therefore the same.
However, some of the above quantities are not identical,
in particular, $\totmom'=4\pi\modenum+\totmom$
and $\angmom'=-\angmom$ as well as $\actvar'=2\fldlen-\actvar$.
These differences arise from a different configuration of the branch cut:
In one case, the branch cut directly connects the two branch points,
whereas in the other, it also winds around the singular point at $\laxpar=0$.
The associated A,B-cycles differ in their direction and in additional contributions
from winding around $\laxpar=0$.

Let us finally discuss the dynamical divisor for the class of rational states.
From the explicit solution for $\laxvec$
of the auxiliary linear problem \eqref{CHMALP},
we obtain its direction function $\laxcp(\laxpar):=\laxvec_2/\laxvec_1$ as
\[
\laxcp(\laxpar) =
\frac{2\eunit^{\iunit\parwnum x-\iunit\parang}\sin(\vartheta)}
{2\cos(\vartheta)-\laxpar\parwnum - \parwnum \sqrt{\laxpar^2 - 4\laxpar\cos(\vartheta)/\parwnum + 4/\parwnum^2}}.
\]
This expression can be inverted to get the divisor location for a reference direction $\laxcp_0$ as
\[
\ptdiv{\laxpar} = \frac{2\cos(\vartheta)+\sin(\vartheta)
\brk*{\eunit^{\iunit\parwnum x-\iunit\parang}\laxcp_0 -(\eunit^{\iunit\parwnum x-\iunit\parang}\laxcp_0)^{-1} }}{\parwnum}.
\]
Effectively, the divisor as a function of $x$ encircles the branch cut as sketched in the above figure.
The initial angle $\parang$ of the spin wave essentially corresponds to the
phase of the around the branch cut.

\medskip

The third family of states comprises wave functions of elliptic kind
whose spectral curves therefore have genus $g=1$.
These states and corresponding curves are somewhat tractable in explicit form
but nevertheless they lead to lengthy expressions
which have not found much attention in the literature.
For our purposes they illustrate the link between periodic and asymptotic states,
and we work out the class of elliptic travelling wave solutions in \appref{EllipticCHMSolution}.
Here we suffice ourselves by a qualitative discussion.

The travelling wave profile resembles the soliton profile in \eqref{chmsolitonfig},
but where the central part is periodically repeated
(with properly smooth continuation).
In other words, the amplitude $\abs{\spincp(x)}$ does not asymptote to zero at $x\to\pm\infty$
as for the soliton, but rather periodically oscillates between a maximum and minimum value.
Next to the length of periodicity $\fldlen$ and their overall orientation,
travelling wave states are parametrised by two continuous static quantities
$\parvel$ (velocity) and $\parfreq$ (angular velocity),
two discrete static quantities $\modenum_{1,2}$ (mode numbers for amplitude and angle)
as well as two continuous dynamical quantities $\parpos$ (centre) and $\parang$ (angle at centre).

In the spectral curve picture, the four static quantities relate to
the four branch points $\ptsqrt{\laxpar}_{1,2,3,4}$ which are distributed
symmetrically in the complex plane around the real axis:
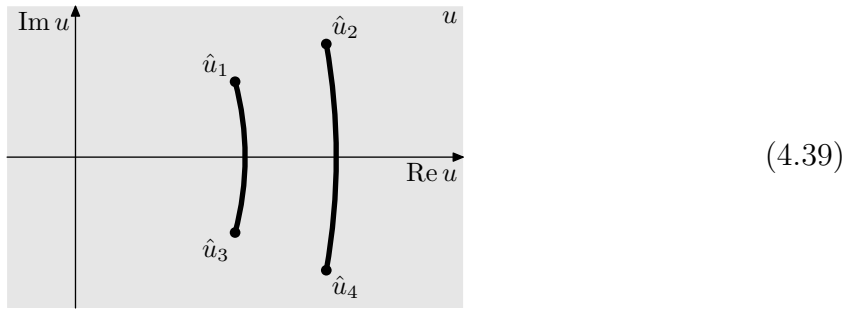
\[
\mpostusemath{CHMEllipticCuts}
\]
The dynamical divisor consists of two points $\ptdiv{\laxpar}_{1,2}$
which ultimately encode the two dynamical quantities of the state.
As discussed in \appref{CHMEllipticDivisor}, the functional shape of
the divisor is intricate and it strongly depends on the parameters
and additional choices that have to be made.
For example, a divisor point may encircle one or both cuts one time, multiple times or not at all.
We refrain from attempting to describe their shapes further.

\medskip

We conclude this section by considering limiting cases of the above families of states:
The soliton \eqref{CHMsoliton} degenerates into the ground state $\spincp(x)=0$
in the limit $\Re(\ismsing)\to 0$.
Similarly, the spin wave \eqref{CHMspinwave} degenerates into the ground state $\spincp(x)=0$
in the limit $\vartheta\to 0$ (or equivalently to $\spincp(x)=\infty$ at $\vartheta\to\pi$).
As the limit is approached, one can observe that the cut of
the spectral curve closes and the action variable $I$ vanishes.
The states are plain trigonometric waves with small amplitude
which in the KdV case correspondingly came to use for the finite-length approximation
of the continuum in the inverse scattering method (see \secref{smallamplitudelimit});
we will see the same phenomenon in the infinite-length limit for the CHM model.

For the elliptic travelling waves, there are two qualitatively different limiting cases:
We will discuss them in terms of the complex branch points $\ptsqrt{u}_{1,2}$
of its spectral curve.
When either of the branch points approaches the real axis, $\Re (\ptsqrt{u}_{1,2})\to 0$,
the state degenerates into a spin wave \eqref{CHMspinwave}
which is described by a spectral curve
with the remaining non-trivial branch point $\ptsqrt{u}_{2,1}$ and associated cut.
When approaching the limit, the state is composed from the spin wave
with a second small-amplitude spin wave superimposed on it.
In the limit, the fluctuation disappears completely,
and the parameter $\ptsqrt{u}_{1,2}$ has no impact.

The elliptic travelling wave can also degenerate to the soliton \eqref{CHMsoliton}
in the limit where the two branch cuts of the spectral curve are made to coincide,
i.e.\ in the limit $\ptsqrt{\laxpar}_2 \to \ptsqrt{\laxpar}_1$.
This degeneration results in a state of infinite length, $\fldlen \to \infty$,
while keeping $\parvel$, $\parfreq$ and $\modenum_{1,2}$ finite.
The common limit of the two branch points serves as the parameter $\ismsing$
for the soliton wave function \eqref{CHMsoliton} as $\ismsing = -1/\ptsol{\laxpar}$.
This is similar to how KdV elliptic states degenerate to solitons (see \secref{KdVInfLen})
with the main distinction that now branch points are off the real axis 
whereas in the KdV case they were restricted to it. 
Consequently, CHM solitons have one complex degree of freedom
rather than a real one for KdV solitons.
Similarly, it takes two branch cuts to coincide for the CHM spectral curve
whereas merely the end of one branch cut degenerates with the beginning of the next
in the KdV case.

%%%%%%%%%%%%%%%%%%%%%%%%%%%%%%%%%%%%%%%%%%%%%%%%%%%%%%%%%%%%%%%%%%%%%%%%%%%%%%%%
\subsection{Finite-Length Extrapolation}

In the following,
we will implement the finite-length extrapolation
introduced in \secref{sec:KdVFiniteLengthExtrapolation}
for the CHM model.

We start by constructing a family of states $\spinvec_{\fldlen}(x)$
approximating some fixed infinite-length state $\spinvec(x)$
with asymptotic boundary conditions.
We require the function $\spinvec_{\fldlen}(x)$
to periodically repeat a window of interest of $\spinvec(x)$ with the length $\fldlen$.
These periodic states are thus defined by
\[
\spinvec_{\fldlen}(x+\Integer\fldlen):=\spinvec(x)
\eqjoin*{\text{with}}
\winbegL<x<\winendL=\winbegL+\fldlen.
\]

In order to extract the Lax monodromy for $\spinvec_{\fldlen}$,
we again refer to the construction of the scattering matrix \eqref{CHMtransportscattering}.
To that end, the asymptotics of the Lax transport $\laxtransport(\winend,\winbeg)$
for small $\winbeg$ and large $\winend$
and for spectral parameter $\laxpar=-1/\ismmom$ must be of the form
\[
\laxtransport(\winend,\winbeg)
\sim
\diag(\eunit^{-\iunit \ismmom \winend},\eunit^{\iunit \ismmom \winend})\.
\ismscat\.
\diag(\eunit^{\iunit \ismmom \winbeg},\eunit^{-\iunit \ismmom \winbeg}).
\]
Here we substitute the form of the scattering matrix \eqref{CHMscatteringmatrix}
of the inverse scattering problem
in terms of transmission and reflection functions,
to find the approximation
\[
\laxtransport(\winend, \winbeg) \sim \begin{pmatrix}
\eunit^{-\iunit \ismmom (\winend-\winbeg)}/\ismtrans(\ismmom)
& -\eunit^{-\iunit \ismmom (\winend+\winbeg)}\ismrefl(\ismmom)/\ismtrans(\ismmom)
\\
\eunit^{\iunit \ismmom (\winend+\winbeg)}\ismrefl(\ismmom^\conj)^\conj/\ismtrans(\ismmom^\conj)^\conj
& \eunit^{\iunit \ismmom (\winend-\winbeg)}/\ismtrans(\ismmom^\conj)^\conj
\end{pmatrix}.
\]
The same qualifications as for the KdV model apply to this approximation.
As discussed above, the transmission coefficient $\ismtrans$
can be reconstructed from the magnitude of the reflection coefficient $\abs{\ismrefl}$
and the $N$ soliton momenta $\ptsol{\ismmom}_n$ using the dispersion relation.
%In particular, it is independent of the dynamical data $\arg \ismrefl$ and $\ismres_n$ and thus conserved.

To finally identify the branch points of the spectral curve,
we consider the quasi-momentum function $\laxqmom$ via the monodromy trace:
\[
\cos\brk!{\laxqmom(\ismmom)}
= \half\tr \laxmono(x)
= \half\tr \laxtransport(x+\fldlen,x)
\approx \frac{\eunit^{-\iunit \ismmom \fldlen}}{2\ismtrans(\ismmom)}
+ \frac{\eunit^{\iunit  \ismmom \fldlen}}{2\ismtrans(\ismmom^\conj)^\conj}.
\]
The branch points are located where $\laxqmom(\ismmom)\in\pi\Integer$ or, equivalently, $\cos\laxqmom(\ismmom)=\pm 1$.

%%%%%%%%%%%%%%%%%%%%%%%%%%%%%%%%%%%%%%%%%%%%%%%%%%%%%%%%%%%%%%%%%%%%%%%%%%%%%%%%
\subsection{Continuum Cuts}
\label{CHMContCutDiv}

First, we want to find the branch cuts associated to the continuum
due to the reflection function $\ismrefl(\ismmom)$.
These ought to be short branch cuts near the real axis,
so we can assume $\ismmom\in\Real$ such that the quasi-momentum can be expressed as
\[
  \label{CHMContF}
\cos\brk!{\laxqmom(\ismmom)}
\approx
\frac{1}{\abs{\ismtrans(\ismmom)}}\cos\brk!{\ismmom \fldlen+\arg\ismtrans(\ismmom) }.
\]
Recall that unitarity of the scattering problem for $\ismmom \in \Real$
implies that $\abs{\ismtrans(\ismmom)}\geq 1$,
which means $\abs{\cos \laxqmom(\ismmom)} \leq 1$.
The function $\cos \laxqmom(\ismmom)$ thus oscillates
with an amplitude less than $1$
and the values $\pm 1$ could be attained only exceptionally.
\unskip\footnote{This can happen only where $\ismrefl=0$,
and the resulting points are in fact no square-root branch points.}
This observation is in line with our earlier observation
that there are no branch cuts on the real axis for the CHM model.
However, if the argument to $\cos$ acquires an imaginary part,
also larger values absolute values are attained.
We assume the quasi-momentum $\laxqmom(\ismmom)$ to be analytic near the real axis,
and extrapolate $\cos\laxqmom(\ismmom)$ to the complex plane.
The solutions $\ptsqrt{\ismmom}$ to $\cos\laxqmom(\ismmom)=\pm1$ turn out to have
only a small imaginary part which is attenuated by the factor of $\fldlen$
in the argument of $\cos$.
We can approximate the solutions as
\[
\label{CHMContCutBranch}
\ptsqrt{\ismmom}_n^\pm
\approx
\frac{\pi n-\arg\ismtrans_n\pm \iunit \arcosh \abs{\ismtrans_n}}{\fldlen} ,
\]
leading to a distribution of branch cuts which we illustrate as follows:
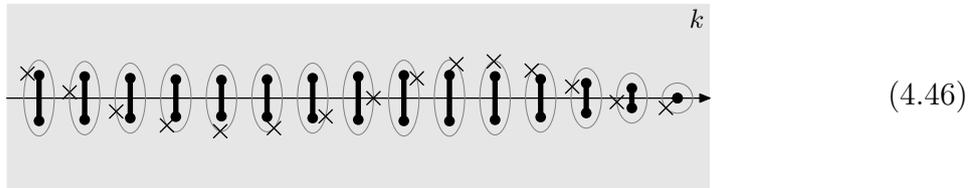
\[
\label{CHMContCutBranchFigure}
\mpostusemath{CHMContinuumCuts}
\]
This result is similar to the expression \eqref{KdVContCutBranch} for the KdV model,
merely the trigonometric functions have been traded in for hyperbolic functions
and the cuts are distributed vertically on the complex plane.

The distribution can be described as a deformed lattice of short cuts:
Each short cut represents a small fluctuation of the ground state.
The fluctuation with mode $n$ is naturally located at momentum
$\ismmom_n:=\pi n/\fldlen$.
The non-zero density of cuts along the real axis
leads to a distortion of the horizontal lattice positions $(\pi/\fldlen)\Integer$
by the phase of the transmission function $-\arg(\ismtrans_n)/\fldlen$:
\unskip\footnote{Note that the phase of the transmission function
should be understood as a continuous function on the real axis
with the asymptotics $\arg\ismtrans(\ismmom)\to 0$ as $\ismmom\to\infty$.}
\[
\label{CHMContCutCentr}
\half(\ismmom_n^++\ismmom_n^-)
\approx
\frac{\pi n-\arg\ismtrans_n}{\fldlen}
=
\ismmom_n-\frac{\arg\ismtrans_n}{\fldlen}.
\]
The lengths of the branch cuts are determined by the magnitude of the reflection function
\[
\label{CHMContCutLen}
\ptsqrt{\ismmom}_n^+-\ptsqrt{\ismmom}_n^-
\approx
\frac{2\iunit}{\fldlen}\arcosh\abs{\ismtrans_n}
=
\frac{2\iunit}{\fldlen}\arsinh\abs{\ismrefl_n}.
\]
For the second equality we have used the unitarity relation $\abs{\ismtrans}^2 = 1 + \abs{\ismrefl}^2$.
Therefore, the length of the cuts is proportional to $\abs{\ismrefl}$
for a small reflection coefficient $\ismrefl$,
and it scales as $1/\fldlen$ for large periodicity length.

Let us also discuss the action variables associated to the continuum cuts.
These relations will be useful in specifying the spectral curve
in terms of such abstract but universally defined data.
The action variable is given as a contour integral on the spectral curve.
According to \eqref{CHMContF}
the function $\cos\laxqmom(\ismmom)$ oscillates with amplitude $1/\abs{\ismtrans_n}$
and periodicity $2\pi/\fldlen$.
We then find
\begin{align}
\label{CHMContAct}
\actvar_n = \frac{1}{2\pi\iunit}\oint_{\contacyc_n}\laxpar \diff{\laxqmom}
&\approx
\frac{1}{\pi\fldlen \ismmom_n^2}
\int_{-\arcosh\abs{\ismtrans_n}}^{+\arcosh\abs{\ismtrans_n}}
\frac{\laxzpar\sinh(\laxzpar)\diff{\laxzpar}}{\sqrt{\abs{\ismtrans_n}{}^2-\cosh(\laxzpar)^2}}
\\
&=
\frac{1}{\fldlen\ismmom_n^2} \log\abs{\ismtrans_n}
=
\frac{1}{2\fldlen\ismmom_n^2} \log\brk!{1+\abs{\ismrefl_n}^2},
\end{align}
which implies the following density function for the infinite-length limit
\[
\actvar(\ismmom) =
\frac{1}{2\pi\ismmom^2} \log\brk!{1+\abs{\ismrefl(\ismmom)}^2}.
\label{CHMContDenFun}
\]
The expression $\actvar_n$ agrees with the action variable for
a rational state at leading order in small $\ismrefl_n$, see \eqref{CHMSpinWaveAct}
and the form $\actvar(\ismmom)$ matches with the integral kernel in \eqref{CHMISMChargesDer}.

The configuration of the dynamical divisor for asymptotic CHM states
is much harder to discuss analytically for various reasons.
For instance, the monodromy eigenvectors will change rapidly around the branch cuts
which become dense in the infinite-length limit.
Picking out a dedicated direction therefore is delicate
given that we have access to the reflection function only for real $\laxpar$.
Furthermore, see the comments on the divisor shape in \appref{CHMEllipticDivisor}.
Qualitatively, we expect a distribution of divisor points
which move around their associated branch cut in a coherent fashion
determined by the reflection function, see \eqref{CHMContCutBranchFigure},
and in analogy to the KdV case discussed in \secref{KdVContDiv}.

%%%%%%%%%%%%%%%%%%%%%%%%%%%%%%%%%%%%%%%%%%%%%%%%%%%%%%%%%%%%%%%%%%%%%%%%%%%%%%%%
\subsection{Soliton Cuts}
\label{CHMSolCutDiv}

To retrieve the branch points corresponding to solitons,
we approximate the quasi-\hspace{0pt}momentum
for $\ismmom$ in the upper half plane in analogy to \eqref{solitoncosq} as
\[
\cos\laxqmom \approx
\frac{\eunit^{-\iunit \ismmom \fldlen}}{2\ismtrans(\ismmom)}.
\label{CHMsolitoncosq}
\]
Branch points are determined by the condition $\cos\laxqmom=\pm1$.
Note that the exponent in $\cos\laxqmom$ becomes very large in the upper half plane,
so branch points must be exponentially close to the poles of the transmission function $\ismtrans$
so that the denominator can balance the numerator.
These points precisely describe the momenta of solitons and their generalisations.
Let us assume that there are $m$ coincident soliton momenta at $\ismmom=\ismsing$.
The transmission function then expands as
\[
\label{CHMSolTransExpd}
\ismtrans(\ismmom)=\alpha (\ismmom-\ismsing)^{-m}+\ldots.
\]
We thus find $2m$ solutions $\ptsqrt{\ismmom}_j$, $j=1,\ldots,2m$,
to the equation $\cos\laxqmom(\ismmom)=\pm 1$
given by
\unskip\footnote{There is always a even number of solutions,
and the sign of $\cos \laxqmom(\ptsqrt{\ismmom}_j)$ alternates with $j$.}
\[
\label{CHMSolSep}
\ptsqrt{\ismmom}_j = \ismsing + (2\alpha)^{1/m}
  \exp\brk*{\frac{\iunit\ismsing \fldlen+ \iunit \pi j}{m}}+\ldots.
\]
These branch points are symmetrically situated on a circle around the soliton momentum $\ismsing$.
\unskip\footnote{This feature provides a compelling reason why
configurations with $m>1$ coincident soliton momenta cannot exist in the KdV model:
In the spectral curve, configurations with $m>1$
would clearly imply branch points away from $\laxpar\in\Real$.
Only for $m=1$, the symmetric distributions yields (two) branch points aligned on a straight line.}
As for the KdV model, the soliton branch points are separated by an exponentially small distance
$\sim \exp(-\fldlen\Im \ismsing/m)$ at large length $\fldlen$.
In addition, the soliton branch points for CHM spiral towards the soliton momentum $\ismsing$
with the angle increasing as $\sim\fldlen\Re \ismsing/m$,
and thus the spiral is tilted with an angle of $\arg (-\iunit\ismsing)$.

\medskip

Let us first discuss the elementary configuration for a simple soliton, $m=1$,
which consists of $2m=2$ branch cuts.
This is different from the representation of solitons in the KdV model
where a soliton is encoded by the gap between two branch cuts
and thus requires only half as many branch cuts in total.
This is also related to the fact that the motion of a soliton in KdV
has merely a single real degree of freedom (speed),
whereas here we need two real degrees of freedom (speed and frequency)
or one complex degree of freedom ($\ismsing$) to describe the motion of CHM solitons.

The resulting structures for $m=1$ can be inferred
from the detailed discussion of the family of elliptic states in \appref{EllipticCHMSolution},
in particular from the discussion of the infinite-length limit in \appref{CHMSolitonLim}.
The configuration is described by two mode numbers $\modenum_{1,2}$ and two action variables $\actvar_{1,2}$.
The mode numbers are consecutive integers $\modenum_2=\modenum_1+1$.
\unskip\footnote{In fact, the labelling of cuts and mode numbers
is opposite in \appref{EllipticCHMSolution},
but it can be adjusted by switching labels or signs.}
The action variables have the leading-order behaviour
\[
\label{CHMEllLimitAct1}
\actvar_{1,2} = \pm\frac{2\fldlen}{\pi}\arg\ismsing+\ldots,
\]
which is determined by the complex angle of $\ismsing$ alone.
The magnitude of the variable $\ismsing$
appears only at the next-to-leading order
\[
\label{CHMEllLimitAct2}
\actvar_1+\actvar_2
= \frac{2\sin (\arg\ismsing)}{\abs{\ismsing}}+\ldots
= \frac{2\Im \ismsing}{\abs{\ismsing}^2}+\ldots.
\]

Curiously, the action variable $\actvar_2$ is negative:
This is related to the fact, that the soliton is in fact a large excitation
of the opposite vacuum state $\spincp=\infty$.
The action variable that describes the excitation
above the opposite vacuum is given by $\tilde\actvar_2=\actvar_2+2\angmom$
and it is positive.
Nevertheless, it is easier to work with action variables corresponding to the
$\spincp=0$ vacuum which can consequently attain negative values.

An important observation concerning the configurations
branch points and cuts was made in \cite{Bargheer:2008kj}:
As the continuous parameters of the family of elliptic states are varied,
the branch cuts move in the complex plane.
This can happen in such a way that the branch points circle around their centre,
and this is precisely what happens as the length $\fldlen$ increases, see \appref{CHMSolitonLim}:
Consider two branch cuts with adjacent mode numbers $\modenum$ and $\modenum+1$.
The branch cuts start out short and are then expanded in such a way that the branch points remain close.
One finds that at first the two cuts are separate (case I$_{n,n+1}$, see \cite{Bargheer:2008kj}).
At some point the growing cuts meet at their centres (case D$_{n,n+1}$)
and join by a condensate of two straight cuts (case II$_{n,n+1}$).
The branch points then circle around each other
until the one with mode number $n$ meets the condensate cut (case C$_{n+1}$).
By crossing the condensate cut the mode number of the branch cut changes from $n$ to $n+2$ (case II$_{n+1,n+2}$).
The circling of branch cut can continue indefinitely as the mode numbers increase one step for each half rotation.

A relevant operation in this sequence is to move branch points or singularities through a branch cut
which implies a reshuffling the A,B-cycles and thus of the mode numbers $\modenum_j$ and action variables $\actvar_j$.
For instance when branch cut $j$ including its branch points moves through branch cut $j+1$,
\[
\mpostusemath{BranchCutTransition1}
\eqjoin{\longrightarrow}
\mpostusemath{BranchCutTransition2}
\]
these numbers transform as follows:
\[
\label{CHMBranchCutTransformation}
\modenum_j'=\modenum_{j+1},
\eqsep
\modenum_{j+1}'=2\modenum_{j+1}-\modenum_{j},
\eqsep
\actvar_j'=2\actvar_j+\actvar_{j+1},
\eqsep
\actvar_{j+1}'=-\actvar_j.
\]
As the branch cuts exchange their relative position,
we also exchange the labels accordingly.
If we have two consecutive mode numbers, $\modenum_{j+1}=\modenum_j+1$,
the transformation effectively shifts both mode numbers by one unit,
$(\modenum'_j,\modenum'_{j+1})=(\modenum_j+1,\modenum_{j+1}+1)$.
This explains the transition from case I$_{n,n+1}$ to case I$_{n+1,n+2}$.
Conversely, the large-length assignments \eqref{CHMEllLimitAct1,CHMEllLimitAct2}
effectively remain unchanged.

\medskip

It remains to generalise the above limiting behaviour to the case of coincident solitons with poles of higher degree $m>1$,
but there are several difficulties in doing so:
We do not have explicit expressions to investigate the general case concretely.
For $2m$ branch cuts, we need to assign $2m$ mode numbers $\modenum_j$ and
$2m$ action variables $\actvar_j$. These are more values to be assigned
than the two degrees of freedom (speed and frequency),
hence there have to be some constraints or ambiguities among the values.
Even worse, the $2m$ branch cuts can be distributed in various entangled ways
to connect the branch points $\ptsqrt{\ismmom}_j$
to their respective complex conjugates $\ptsqrt{\ismmom}_j^\conj$.
Even though such different configurations are equivalent to each other,
they lead to different assignments of values for the mode numbers and action variables.
Nevertheless, let us explore how the above picture may generalise to a larger number of branch points.

We first consider some suitable (topological) configurations of branch cuts.
We highlight two useful choices for the configuration of branch cuts
and discuss their implications:
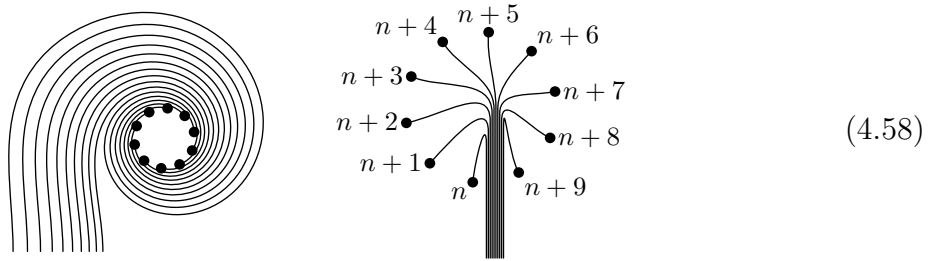
\[
\mpostusemath{CHMSolitonTwirl}
\eqsep
\mpostusemath{CHMSolitonFlower}
\]
The configurations essentially differ in that the cuts all extend either
outwards from the circle or inwards into it.

In the first configuration, the branch cuts extend on a spiralling path along
the large-length trajectories of the branch points described above.
One can obtain it starting with very short branch cuts
and letting the branch points circle around each other.
In doing so, we deform the branch cuts to avoid crossings with any of the branch points.
The cuts thus follow a spiralling path towards the branch points in the upper half of the complex plane.
Here, the configuration of mode numbers
\[
\modenum_j=\modenum+\half+\half(-1)^j,
\]
should alternate between two adjacent integers $\modenum$ and $\modenum+1$.
A basic argument in favour of this pattern is as follows:
The quasi-momentum function $\laxqmom$ is continuous on the disc centred at $\ismsing$
and it is approximated through \eqref{CHMSolTransExpd} there.
By construction the quasi-momentum alternates between even and odd integers on the perimeter.
If these were not adjacent integers, the quasi-momentum would attain further
integer values near $\ismsing$ which would correspond to further, undesirable branch points.

In the second configuration, all $2m$ branch cuts meet near $\ismmom$
before extending towards $\ismmom^\conj$ as a bundle and splitting up again.
Here, one naturally obtains a pattern of mode numbers
\[
\modenum_j=\modenum+j-1,
\]
which increases linearly along the circle from $\modenum$ to $\modenum+2m-1$.
The bundled branch cuts serve as the boundary which starts and ends the sequence.
Here, the branch cuts within the disk invalidate the argument
which constrained the mode number to be adjacent integers in the former configuration.
Note that the beginning and end of the sequence can be altered by moving the first
or last branch cut through all others.
The corresponding transformation \eqref{CHMBranchCutTransformation}
shifts all mode number effectively by $+1$ or $-1$, respectively,
but the overall increasing sequence remains intact.

The two configurations of branch cuts can be transformed into each other:
Here one has to revert the order of all branch cuts
which effectively flips the ends of branch cuts inside out.
It turns out that the corresponding sequence of transformations \eqref{CHMBranchCutTransformation}
maps between the alternating pattern $\modenum_j=\modenum+\half+\half(-1)^j$ for the first configuration
and the linearly increasing pattern $\modenum_j=\modenum+j-1$ for the second configuration.
Note that the base mode number $\modenum$ does not appear to play a role
in the infinite-length limit as the position $\ismsing$ is
fully determined by the action variables
in \eqref{CHMEllLimitAct1} and \eqref{CHMEllLimitAct2}.

\medskip

Based on the explicit results \eqref{CHMEllLimitAct1,CHMEllLimitAct2}
for the elementary soliton with $m=1$,
we conjecture that the action variables have the following large-length
behaviour
\[
\actvar_j = (-1)^{j+1}\frac{2\fldlen}{\pi m}\arg\ismsing+\ldots,
\eqsep
\sum_{j=1}^{2m}\actvar_j
= \frac{2\Im \ismsing}{\abs{\ismsing}^2}+\ldots.
\]
The second relation simply reflects the soliton contribution to $\Delta\angmom=\angmom-\fldlen$ in \eqref{CHMISMCharges}
to the identity \eqref{CHMactreal1} $\sum_j\actvar_j = \fldlen-\angmom = -\Delta\angmom$.
The pattern of leading-order contributions to the action
variables merely alters between negative and positive sign.
This pattern is distinguished by being stable
under the transformation \eqref{CHMBranchCutTransformation}
for any pair of adjacent branch cuts,
and in particular for the cyclic shift
that shifts the mode numbers by one unit in the second configuration for the branch cuts.
The magnitude of the leading-order contributions to the action variables $\actvar_j$
can then be inferred from the identity \eqref{CHMactreal2}
such that
\[
-2\pi\sum_{j=1}^{2m}\modenum_j\actvar_j
=
4\fldlen \arg\ismsing
+\ldots
=\fldlen \totmom.
\]

%%%%%%%%%%%%%%%%%%%%%%%%%%%%%%%%%%%%%%%%%%%%%%%%%%%%%%%%%%%%%%%%%%%%%%%%%%%%%%%%
\subsection{Asymptotic Limit}
\label{CHMAsympLim}

Now we summarise the resulting asymptotic behaviour for a family of spectral curves in parallel to \secref{KdVAsympLim}
in the $\laxpar$-plane.
We recall from the previous sections
that the branch cuts have two different types of limiting behaviour
for large length.
The distribution of the branch cuts for the spectral curve at large $\fldlen$
can be sketched as follows:
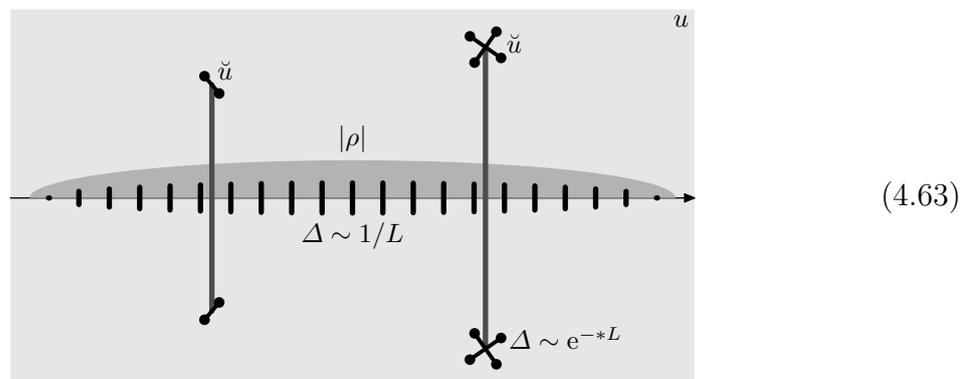
\[
\mpostusemath{CHMLimitCuts}
\]
We point out that some of the features of the infinite-length asymptotics
of spectral curve data have been described in \cite{ForestMcL1982},
albeit for the case of the sine-Gordon model whose spectral curves share several aspects
with those of the CHM model.
In particular, one can recognise the lattice of short cuts
which is described as a sequence of spines protruding from the real axis.

\medskip

The first admissible configuration
is a bundle of long branch cuts
which is identified with (multiple coincident) solitons:
A general soliton configuration consists of $N$
pole singularities $\ismsing_j$ of degree $m_j$
in the upper half of the complex plane (with their conjugates in the lower half).
For each singularity of degree $m$,
there are $2m$ corresponding branch points $\ptsqrt{\laxpar}_j$, $j=1,\ldots,2m$
(as well as their complex conjugates),
which approach a common value
\[
\ptsqrt{\laxpar}_j \to \ptsol{\laxpar} = -\frac{1}{\ismsing}.
\]
This limiting behaviour determines the momentum $\ismsing$ and multiplicity $m$
of the associated soliton.

The second set of short branch cuts located symmetrically around the real $\laxpar$-axis
are identified with the continuum.
They are delimited by pairs of complex conjugate branch points
which for $n\sim\fldlen$ are approximated by
\[
\ptsqrt{\laxpar}_n,
\ptsqrt{\laxpar}_n^\conj
=
-\frac{\fldlen}{\pi n}
\pm \frac{\fldlen}{\pi^2 n^2}
\iunit\arsinh \abs!{\ismrefl(\pi n/\fldlen)}
+\ldots
.
\]
Here, the length determines the magnitude of the reflection function for the continuum.
We thus recover all invariant data required for the inverse scattering method.

\medskip

The transmission function $\ismtrans(\ismmom)$ is then fixed by 
the dispersion relation \eqref{CHMantiunitarydisp}.
As for the KdV model, it constrains the limiting behaviour further.
In particular, the $2m$ branch points $\ptsqrt{\laxpar}_j$ corresponding to an $m$-fold soliton
must be located symmetrically around $\ptsol{\laxpar}$.
As $\fldlen\to\infty$, their distance shrinks exponentially
and they spin around the centre
according to the relation \eqref{CHMSolSep}:
\[
\ptsqrt{\laxpar}_j \approx \ptsol{\laxpar} + (2\alpha)^{1/m}\ptsol{\laxpar}^2
\exp\brk*{
-\frac{\fldlen}{m}\frac{\Im \ptsol{\laxpar}}{\abs{\ptsol{\laxpar}}^2}
}
\exp\brk*{
-\iunit\frac{\fldlen}{m}\frac{\Re \ptsol{\laxpar}}{\abs{\ptsol{\laxpar}}^2}
+\frac{\iunit \pi j}{m}}.
\]
Here, $\alpha$ is the coefficient of the $m$-fold pole
in the transmission function $\ismtrans(\ismmom)$
at $\ismmom=\ismsing$ defined by the expansion
\[
\ismtrans(\ismmom) = \frac{\alpha}{(\ismmom-\ismsing)^m}+\ldots.
\]

The branch points corresponding to the continuum
must obey the limiting behaviour
\[
\ptsqrt{\laxpar}_n
=
-\frac{\fldlen}{\pi n}
+\frac{\fldlen}{\pi^2 n^2}\brk[s]2{-\arg\ismtrans(\pi n/\fldlen)
+\iunit \arcosh \abs!{\ismtrans(\pi n/\fldlen)}}
+\ldots
.
\]
Note that the mode number $n$ is assumed to be of the order of $\fldlen$
such that the above relation consists of a leading-order term $-\fldlen/\pi n$
and a sub-leading correction.
This limiting relation describes the magnitude and phase of the transmission function $\ismtrans(\ismmom)$
in terms of the modulation pattern of the branch cuts which must behave as follows:
As the length $\fldlen$ increases, the individual branch cuts grow but also move towards $\infty$ in a linear fashion.
However, the relative filling of the cuts remains fixed
in place along the real axis according to the function $\ismtrans(\pi n/\fldlen)$.
Since a finite point $\laxpar$ resides near the $n$-th branch cut with $n \approx \fldlen u/\pi$,
the number of branch cuts must asymptotically grow and only those modes contribute whose $n$ grows with the same rate as $\fldlen$.

%%%%%%%%%%%%%%%%%%%%%%%%%%%%%%%%%%%%%%%%%%%%%%%%%%%%%%%%%%%%%%%%%%%%%%%%%%%%%%%%
%%%%%%%%%%%%%%%%%%%%%%%%%%%%%%%%%%%%%%%%%%%%%%%%%%%%%%%%%%%%%%%%%%%%%%%%%%%%%%%%
\section{Conclusions and Outlook}
\label{sec:conclusions}

In this paper we have explored the infinite-length limit
of states of the spectral curve method in detail. 
For two prototypical integrable models of one-dimensional fields,
the Korteweg-de Vries equation and the Continuous Heisenberg Magnet, 
we have investigated the features of families of states
that provide them with a proper infinite-length limit
which is consequently described by the inverse scattering method.
In brief our results can be summarised as follows:

There are two distinct classes of branch cut configurations for spectral curves
with proper infinite-length limits.
One class consists of a collection of a growing number of branch cuts whose size
and separation shrinks correspondingly. This configuration gives rise to
continuum states in the inverse scattering method
such that the density of branch cuts relates to the magnitude of the reflection and transmission functions.
The other class consists of a fixed number of branch cuts whose 
ends approach each other exponentially fast. Such a configuration
gives rise to soliton states where the limit of branch points
describes the soliton momenta.
The qualitative features of the branch cut configurations
depend somewhat on the underlying model.
In particular there is a major distinction between
a positive and negative signature of excitations
which is related to the compactness of the fields,
signature of unitarity and
orientation of branch cuts in the complex plane
which also has an impact on the statistics for soliton excitations.

Next to the conserved data for states,
we have also addressed the time-dependent data
which is embodied by the dynamical divisor for spectral curves
as well as by phase information for the inverse scattering method.
The configuration of dynamical divisors turn out to be rather complex
and they depend strongly on the situation and on various implementation details.
Therefore we have merely discussed them at a qualitative level.
Roughly the divisor points orbit around the branch cuts
and the phases of the orbits must be correlated strongly for a proper infinite-length limit.
The phases of the orbits describe the complex phases of the infinite scattering method data.

\medskip

The scheme for the infinite-length limit of integrability methods
has several generalisation and applications that may be explored further:

As emphasised above, our understanding of the time-dependent data is incomplete
and could be described in more detail. This would require a better understanding
of the dynamical divisor which likely requires a mapping to the Jacobian
of the spectral curve by means of the Abel map and theta-functions.
This may come close to a formal functional treatment of the spectral curve solutions,
see the earlier work \cite{its1976problems,its1983problems,bbeim1994,Osborne:1985,Osborne:1986ua,ForestMcL1982}.
Moreover, we have have only addressed the limit of the data.
It would be interesting to derive the relevant inverse scattering equations,
in particular the Gel'fand--Levitan--Marchenko integral equation,
from the equations of the spectral curve method.

The scheme ought to be applicable to wider classes of integrable one-dimensional models.
These could be deformations of the above models with twisted closed boundary conditions
such that their infinite-length limit can result in different boundary conditions 
in the left and right asymptotic region.
These could as well be different kinds of models, such as the sine-Gordon model,
where it will be interesting to understand and interpret the the relevant adjustments.
Furthermore, the methods of integrability extend to one-dimensional integrable lattice models
with minor adjustments due to discreteness. We expect that the scheme 
extends to chain models with minor adjustments due to discreteness.

Finally, in terms of applications, one might use states of the inverse 
scattering method as a first approximation to states of the spectral curve method.
Such states can be used as a suitable starting point 
in an approximation method for reasonably long states,
which avoids having to deal with elaborate functional analysis for 
higher-genus Riemann surfaces.

%%%%%%%%%%%%%%%%%%%%%%%%%%%%%%%%%%%%%%%%%%%%%%%%%%%%%%%%%%%%%%%%%%%%%%%%%%%%%%%%
\pdfbookmark[1]{Acknowledgements}{ack}
\section*{Acknowledgements}

We thank
Jacqueline Angersbach,
Patrick Dorey,
Egor Im,
Alexander Its,
Fedor Smirnov,
Charles Thull
and
Anders Wallberg
for discussions related to the project
and previous work on the subject.
The work of NB is partially supported
by the Swiss National Science Foundation through the NCCR SwissMAP. 
The work of KG is supported by the Inlaks Shivdasani Foundation Scholarship, Knut and Alice Wallenberg Foundation Grant KAW 2021.0170, and by the Olle Engkvists Stiftelse Grant 2180108.

%%%%%%%%%%%%%%%%%%%%%%%%%%%%%%%%%%%%%%%%%%%%%%%%%%%%%%%%%%%%%%%%%%%%%%%%%%%%%%%%
%%%%%%%%%%%%%%%%%%%%%%%%%%%%%%%%%%%%%%%%%%%%%%%%%%%%%%%%%%%%%%%%%%%%%%%%%%%%%%%%
\appendix

%%%%%%%%%%%%%%%%%%%%%%%%%%%%%%%%%%%%%%%%%%%%%%%%%%%%%%%%%%%%%%%%%%%%%%%%%%%%%%%%
%%%%%%%%%%%%%%%%%%%%%%%%%%%%%%%%%%%%%%%%%%%%%%%%%%%%%%%%%%%%%%%%%%%%%%%%%%%%%%%%
\section{Elliptic Solution in CHM}
\label{EllipticCHMSolution}

In this appendix we derive the class of elliptic solutions
to the continuous Heisenberg model.
We extend the early work \cite{NAKAMURA1974321,LAKSHMANAN1976577,PhysRevB.15.3470}
by analysing and discussing some of the relevant properties 
of these solutions in detail.

We will proceed in analogy with the derivation of the KdV elliptic solution in \secref{sec:KdVElliptic}
by assuming a travelling wave with a constant shape.
This will turn out to produce a general elliptic solution
which is quasi-periodic in space and which can be specialised
to properly periodic or asymptotic solutions by suitable choices of parameters.

%%%%%%%%%%%%%%%%%%%%%%%%%%%%%%%%%%%%%%%%%%%%%%%%%%%%%%%%%%%%%%%%%%%%%%%%%%%%%%%%
\subsection{Equation of Motion}

The continuous Heisenberg model is based on a field $\vect{S}(x,t)\in \Real^3$
which is restricted to the unit sphere $\Sphere^2$ by the constraint $\norm{\vect{S}}=1$.
The equation of motion with normalised parameters takes the form
\[
\dot{\spinvec}=\spinvec\vtimes\spinvec''.
\]
We parametrise points $\spinvec\in \Sphere^2$
in terms of a complex variable $\spincp\in\ComplexComplete$
by means of the stereographic projection
\[
\spinvec=\frac{1}{1+\spincp\bar\spincp}\begin{pmatrix}
\spincp+\bar\spincp\\
-\iunit\spincp+\iunit\bar\spincp\\
1-\spincp\bar\spincp
\end{pmatrix}.
\]
The equations of motion in these variables take the form
\[
\dot\spincp = \iunit\spincp''-\frac{2\iunit \bar\spincp\spincp'^2}{1+\spincp\bar\spincp},
\eqsep
\dot{\bar\spincp} = -\iunit\bar\spincp''+\frac{2\iunit \spincp\bar\spincp'^2}{1+\spincp\bar\spincp}.
\]
Here, it will be convenient to assume $\bar\spincp\in\ComplexComplete$
to be another complex variable independent of $\spincp$,
and only eventually identify it with the complex conjugate $\spincp^\conj$
by means of the reality condition $\bar\spincp=\spincp^\conj$.
With that assumption, the first equation is linear in $\bar\spincp$
and can be solved exactly in this variable:
\[
\label{zetabar}
\bar\spincp = - \frac{\iunit\dot\spincp +\spincp''}{\iunit\dot\spincp\spincp +\spincp''\spincp - 2 \spincp'^2}.
\]
By substituting this solution back into the second equation
yields a second-order differential equation for $\spincp$:
\[
\frac{\ddot\spincp}{\spincp'}
-\frac{4\dot\spincp\dot\spincp'}{\spincp'^2}
+\frac{3\dot\spincp^2\spincp''}{\spincp'^3}
+\frac{3\spincp''^3}{\spincp'^3}
-\frac{4\spincp''\spincp'''}{\spincp'^2}
+\frac{\spincp''''}{\spincp'}
=0.
\]
As an aside, we note that this equation can be integrated once to a conservation equation:
\[
\frac{\partial}{\partial t} \frac{\dot\spincp}{\spincp'}
+\frac{\partial}{\partial x}\brk[s]*{
-\frac{3\dot\spincp^2}{2\spincp'^2}
+\frac{\spincp'''}{\spincp'}
-\frac{3\spincp''^2}{2\spincp'^2}
}=0.
\]
Here, we observe the Schwarzian derivative in the latter two terms:
The model has an $\grp{SO}(3)$ rotational symmetry
which amounts to a Möbius transformation for $\spincp$.
The Schwarzian derivative is invariant under this transformation
and also the terms involving time derivatives happen to be invariant.

%%%%%%%%%%%%%%%%%%%%%%%%%%%%%%%%%%%%%%%%%%%%%%%%%%%%%%%%%%%%%%%%%%%%%%%%%%%%%%%%
\subsection{Travelling Wave Ansatz}

We now make an ansatz of a wave function with constant shape
analogously to the ansatz for the KdV equation that led to the class of elliptic solutions.
Such an ansatz links the spatial translation symmetry with time translation symmetry
such that the wave function takes on a constant velocity $\parvel$.
In addition, here we can make use of rotational symmetry and link it to time translations as well
such that the wave function still preserves shape but rotates with a constant angular velocity $\parfreq$.
Out of the $\grp{SO}(3)$ symmetries, without loss of generality, we choose rotations about the $z$-axis.
The travelling wave ansatz reads
\[
\spincp(x,t)=\exp(\iunit\parfreq t)\.\spincp(x-x_0-\parvel t).
\]
The equation of motion translates into a fourth-order differential equation for $\spincp(x)$
which also ensures that the equation of motion for $\spincp(x,t)$ holds for all $x$ and all $t$
using the above symmetry considerations.
Taking into account that the residual 2D rotational symmetry acts by multiplication on $\spincp(x)$,
it makes sense to formulate the wave equation in terms of the logarithmic derivative $G(x)$:
\[
G(x):=-\iunit\frac{\zeta'(x)}{\zeta(x)},
\eqsep
\frac{G'''}{G}-\frac{4G'G''}{G^2}+\frac{3G'^3}{G^3}+GG'-\frac{2\parfreq \parvel G'}{G^2}+\frac{3\parfreq G'}{G^3}=0.
\]
It is straight-forward to integrate this differential equation once:
\[
\frac{G''}{G}
-\frac{3G'^2}{2G^2}
+\half G^2
+\frac{2\parfreq\parvel}{G}
-\frac{3\parfreq^2}{2G^2}
+\kappa
=0.
\]
However, we may as well divide the original equation by $G$
and integrate to obtain an alternate second-order differential equation:
\[
\frac{G''}{G^2}
-\frac{G'^2}{G^3}
+G
+\frac{\parfreq \parvel}{G^2}
-\frac{\parfreq^2}{G^3}
+\chi
=0.
\]
%
\iffalse
define $\zeta=\exp(\iunit F)$, $\bar\zeta=\exp(-\iunit\bar F)$
translates to differential equation in $x$
%
\[
\frac{\zeta''''}{\zeta'}
-\frac{4\zeta''\zeta'''}{\zeta'^2}
+\frac{3\zeta''^3}{\zeta'^3}
-2\iunit\parfreq \parvel\brk[s]*{\frac{\zeta\zeta''}{\zeta'^2}-1}
-3\parfreq^2\brk[s]*{\frac{\zeta^2\zeta''}{\zeta'^3}-\frac{\zeta}{\zeta'}}
=0
\]
%\[
%\frac{F''''}{F'}
%-\frac{4F''F'''}{F'^2}
%+\frac{3F''^3}{F'^3}
%+F'F''
%-\frac{2\parfreq \parvel F''}{F'^2}
%+\frac{3\parfreq^2 F''}{F'^3}
%=0
%\]
%
which solves the differential equation for $\zeta$ for all $x$ and all $t$.

differential equation for $F$ integrates to
note that up to $\parfreq$ this is a Schwarzian derivative
%
\[
\frac{\zeta'''}{\zeta'}
-\frac{3\zeta''^2}{2\zeta'^2}
+\frac{2\iunit\parfreq \parvel \zeta}{\zeta'}
+\frac{3\parfreq^2\zeta^2}{2\zeta'^2}
+\kappa
=0
\]
%\[
%\frac{F'''}{F'}
%-\frac{3F''^2}{2F'^2}
%+\half F'^2
%+\frac{2\parfreq \parvel}{F'}
%-\frac{3\parfreq^2}{2F'^2}
%+\kappa
%=0
%\]
%
but can also divide by $F'$ before integration to get
%
\[
\frac{\iunit\zeta\zeta'''}{\zeta'^2}
-\frac{\iunit\zeta\zeta''^2}{\zeta'^3}
-\frac{\iunit\zeta''}{\zeta'}
-\frac{\parfreq \parvel\zeta^2}{\zeta'^2}
+\frac{\iunit\parfreq^2\zeta^3}{\zeta'^3}
+\chi
=0
\]
%\[
%\frac{F'''}{F'^2}
%-\frac{F''^2}{F'^3}
%+F'
%+\frac{\parfreq \parvel}{F'^2}
%-\frac{\parfreq^2}{F'^3}
%+\chi
%=0
%\]
%
two new integration constants $\kappa$ and $\chi$.
these are in fact both differential equation for $G:=F'$.
\fi
By combining the two equations, we
eliminate the second-order term $G''$
to obtain a first-order differential equation:
\[
G'^2=
-P(G),
\eqsep
P(g):=
g^4
+2\chi g^3
-2\kappa g^2
-2\parfreq \parvel g
+\parfreq^2.
\]
The polynomial $P(g)$ on the right-hand side is quartic
with coefficients freely adjustable by choosing
the parameters $\parvel$ and $\parfreq$ as well as the integration constants $\kappa$ and $\chi$.
We can cast the polynomial in factorised form:
\[
P(g)=(g-\ptsqrt{g}_1)(g-\ptsqrt{g}_2)(g-\ptsqrt{g}_3)(g-\ptsqrt{g}_4),
\]
where the roots $\ptsqrt{g}_k$ are related to the parameters as follows:
\unskip\footnote{For any given $P(g)$, there are two choices for the sign of $\parfreq$.
These correspond to two distinct solutions $\spincp(x)$ and $1/\spincp(-x)$.}
\begin{align}
\parfreq&=\sqrt{\ptsqrt{g}_1\ptsqrt{g}_2\ptsqrt{g}_3\ptsqrt{g}_4},
\\
\parvel&=\half\sqrt{\ptsqrt{g}_1\ptsqrt{g}_2\ptsqrt{g}_3\ptsqrt{g}_4}
\brk!{\ptsqrt{g}_1^{-1}+\ptsqrt{g}_2^{-1}+\ptsqrt{g}_3^{-1}+\ptsqrt{g}_4^{-1}},
\\
\kappa&=-\half(\ptsqrt{g}_1\ptsqrt{g}_2+\ptsqrt{g}_1\ptsqrt{g}_2+\ptsqrt{g}_1\ptsqrt{g}_4+\ptsqrt{g}_2\ptsqrt{g}_3+\ptsqrt{g}_2\ptsqrt{g}_4+\ptsqrt{g}_3\ptsqrt{g}_4),
\\
\chi&=-\half(\ptsqrt{g}_1+\ptsqrt{g}_2+\ptsqrt{g}_3+\ptsqrt{g}_4).
\end{align}

The differential equation is solved by the elliptic function
%%
%\[
%\frac{(\ptsqrt{g}_3-\ptsqrt{g}_2)(G-\ptsqrt{g}_1)}{(\ptsqrt{g}_1-\ptsqrt{g}_2)(G-\ptsqrt{g}_3)}=\ellSN(\parscl x+\parsolb,\ellmods)^2,
%\]
%%
%
\[
G(x)=\frac{(\ptsqrt{g}_4-\ptsqrt{g}_1)\ptsqrt{g}_3-(\ptsqrt{g}_3-\ptsqrt{g}_1)\ptsqrt{g}_4\ellSN(\parscl x+\parsolb,\ellmods)^2}
{(\ptsqrt{g}_4-\ptsqrt{g}_1)-(\ptsqrt{g}_3-\ptsqrt{g}_1)\ellSN(\parscl x+\parsolb,\ellmods)^2},
\]
where the elliptic modulus $\ellmods$ and the scaling parameter $\parscl$ are given by
\[
\ellmods=\frac{(\ptsqrt{g}_3-\ptsqrt{g}_1)(\ptsqrt{g}_4-\ptsqrt{g}_2)}{(\ptsqrt{g}_3-\ptsqrt{g}_2)(\ptsqrt{g}_4-\ptsqrt{g}_1)},
\eqsep
\parscl=\half\sqrt{(\ptsqrt{g}_3-\ptsqrt{g}_2)(\ptsqrt{g}_1-\ptsqrt{g}_4)},
\]
and where $\parsolb$ is an integration constant incorporating the offset $x_0$.

In order to integrate the logarithmic derivative function $G(x)$,
let us discuss its analytic structure:
It is a meromorphic function with poles at $\parscl x+\parsolb=\pm \parsolc$
with residues
\[
G\brk!{(-\parsolb\pm \parsolc)/\parscl+\epsilon}
=\mp\frac{\iunit}{\epsilon}+\Order(\epsilon),
\]
where the constant $\parsolc$ is defined by the inverse elliptic relations
\unskip\footnote{Note that it takes two independent evaluations of elliptic functions
to uniquely specify a point on the elliptic curve
(up to shifts by the periods $2\ellKval$ and $2\iunit\ellKvalc$).}
\[
\ellSN(\parsolc)^2=\frac{\ptsqrt{g}_1-\ptsqrt{g}_3}{\ptsqrt{g}_3-\ptsqrt{g}_1},
\eqsep
\frac{\ellSN(\parsolc)\ellCN(\parsolc)}{\ellDN(\parsolc)}
=\frac{2\iunit \parscl}{\ptsqrt{g}_3-\ptsqrt{g}_1}.
\]
The above poles imply zeros and poles at the respective positions.
With this information, we can integrate $G(x)$ and exponentiate
to ultimately obtain the wave function $\spincp(x)$
in terms of the Neville elliptic theta function $\ellTS(z,\ellmods)$:
\[
\spincp(x)=
\parsoleta\exp\brk!{\iunit \parqper(\parscl x+\parsolb)/2\ellKval}
\frac{\ellTS(\parscl x+\parsolb-\parsolc)}{\ellTS(\parscl x+\parsolb+\parsolc)}.
\]
The constant $\parqper$ of the exponent is defined
in terms of the elliptic zeta function $\ellZN(z,\ellmods)$:
\[
\parqper:=\brk[s]*{\frac{\ptsqrt{g}_4}{\parscl}- 2\iunit\ellZN(\parsolc)}2\ellKval.
\]

%%%%%%%%%%%%%%%%%%%%%%%%%%%%%%%%%%%%%%%%%%%%%%%%%%%%%%%%%%%%%%%%%%%%%%%%%%%%%%%%
\subsection{Conjugate Field}

We should now address the conjugate field $\bar\spincp(x,t)$.
We first introduce a corresponding travelling wave ansatz
and define the conjugate logarithmic derivative $\bar G$:
\[
\bar\spincp(x,t)=\exp(-\iunit\parfreq t)\.\bar\spincp(x-x_0-\parvel t),
\eqsep
\bar G(x):=\iunit\frac{\bar\spincp'(x)}{\spincp(x)}.
\]
As before, this reduces the equation of motion to a wave equation for $\bar G$,
which is precisely the same as for $G$.
\unskip\footnote{Note that the polynomial $P(g)$ is real and hence it equally
applies to $G$ as for its complex conjugate $\bar G$.}
Note that the wave equation is homogeneous, therefore the solution
for $\bar G$ can differ from the one for $G$ by a shift in the argument;
in anticipation of the reality conditions,
we choose the shift to be $\bar\parsolb+\ellKval$ rather than $\parsolb$
with some parameter $\bar\parsolb$ so that
\[
\bar G(x)=G\brk!{x-(\parsolb-\bar\parsolb-\ellKval)/\parscl}
%=\frac{(\ptsqrt{g}_3-\ptsqrt{g}_2)\ptsqrt{g}_1-(\ptsqrt{g}_1-\ptsqrt{g}_2)\ptsqrt{g}_3\ellSN(\parscl x+\bar\parsolb+\ellKval)^2}
%{(\ptsqrt{g}_3-\ptsqrt{g}_2)-(\ptsqrt{g}_1-\ptsqrt{g}_2)\ellSN(\parscl x+\bar b+\ellKval)^2}
.
\]
The integrated wave function $\bar\spincp$ then takes the form
\[
\bar\spincp(x)=
\bar\parsoleta\exp\brk!{-\iunit\parqper(\parscl x+\bar\parsolb)/2\ellKval}
\frac{\ellTS(\parscl x+\bar\parsolb-\parsolc+\ellKval)}{\ellTS(\parscl x+\bar\parsolb+\parsolc-\ellKval)}.
\]
As such, $\bar\spincp(x)$ merely describes a solution to the conjugate wave equation.
We still need to satisfy the compatibility relation \eqref{zetabar} between $\bar\spincp(x)$ and $\spincp(x)$:
\[\label{eq:spincpbar}
\bar\spincp = -\frac{1}{\spincp}\frac{\parfreq-\parvel G-\iunit G'+G^2}{\parfreq-\parvel G-\iunit G'-G^2}.
\]
Here, it is sufficient to consider selected points $x$
to determine how the parameters of the two solutions are related.
We find a relation between $\parsolb$ and $\bar\parsolb$
as well as between $\parsoleta$ and $\bar\parsoleta$:
\[
\label{CHMEllConjPar}
\parsolb-\bar\parsolb=2\parsold+\ellKval,
\eqsep
\parsoleta \bar\parsoleta
=
\exp\brk!{-\iunit \parqper(2\parsold+\ellKval)/2\ellKval}
\frac{\ellTN(\parsold+\parsolc)^2}{\ellTN(\parsold-\parsolc)^2},
%\frac{\parfreq-\ptsqrt{g}_1v+\ptsqrt{g}_1^2}{\parfreq-\ptsqrt{g}_1v-\ptsqrt{g}_1^2}
%\frac{\ellTS(\parsolb-\bar\parsolb-\parsolc-\ellKval)}{\ellTS(\parsolb-\bar\parsolb-\parsolc+\ellKval)}.
\]
where the parameter $\parsold$ is defined by
\[
\ellSN(\parsold)^2=\frac{\ptsqrt{g}_1(\ptsqrt{g}_2-\ptsqrt{g}_2)}{\ptsqrt{g}_3(\ptsqrt{g}_4-\ptsqrt{g}_2)},
\eqsep
\frac{\ellSN(\parsold)\ellCN(\parsold)}{\ellDN(\parsold)}=\frac{-2\iunit \parscl \ptsqrt{g}_4\ptsqrt{g}_2}
{\parfreq(\ptsqrt{g}_4-\ptsqrt{g}_2)}.
\]
We have now satisfied all of the defining complexified equations.

In order to further restrict to physical solutions with $\bar\spincp(x)=\spincp(x)^\conj$,
it makes sense to consider the combination $\spincp\bar\spincp$
given by \eqref{eq:spincpbar}.
%
%\[
%\spincp\bar\spincp
%%=- \frac{\iunit\dot\spincp\spincp +\spincp''\spincp}{\iunit\dot\spincp\spincp +\spincp''\spincp - 2 \spincp'^2}
%=-\frac{\parfreq-\parvel G-\iunit G'+G^2}{\parfreq-\parvel G-\iunit G'-G^2}
%=
%-\frac{\ellSN(\parsolc-\parsold)^2-\ellSN(\parscl x+\paroff)^2}
%{\ellSN(\parsolc+\parsold)^2-\ellSN(\parscl x+\paroff)^2}
%.
%\]
%
Evidently, this function represents an absolute value squared and it needs to be positive real for all real values of $x$.
As a consequence it turns out that the set of points $\set{\ptsqrt{g}_k}$ must be self-conjugate.
Without loss of generality and for concreteness, we make the assignments
\[
\ptsqrt{g}_3=\ptsqrt{g}_1^\conj,
\eqsep
\ptsqrt{g}_4=\ptsqrt{g}_2^\conj,
\eqsep
(\Im \ptsqrt{g}_3)(\Im \ptsqrt{g}_4)\geq 0.
\]
Furthermore, the dynamical parameters $\parsolb$ and $\parsoleta$ need to be chosen suitably.
For that, we consider the extrema of the function $\spincp\bar\spincp$
which are located at $\parscl x+\parsolb=\parsold$ and $\parscl x+\parsolb=\parsold+\ellKval$.
As the function $\spincp\bar\spincp$ is real for real $x$,
the extrema need to reside on the real axis
thus requiring $\Im \parsolb=\Im \parsold$.
Together with the relation \eqref{CHMEllConjPar}
we can solve for $\parsolb$ and $\bar\parsolb$:
\[
\parsolb = \parsold+\paroff,
\eqsep
\bar\parsolb = -\ellKval-\parsold+\paroff,
\]
where the remaining real constant $\paroff$ incorporates the spatial offset $x_0$.
The relation between the pre-factors is resolved by
\[
\parsoleta
=
\exp\brk!{-\iunit \parqper\parsold/2\ellKval+\iunit \parang}
\frac{\ellTN(\parsold+\parsolc)}{\ellTN(\parsold-\parsolc)},
\eqsep
\bar\parsoleta
=
\exp\brk!{-\iunit \parqper(\parsold+\ellKval)/2\ellKval-\iunit \parang}
\frac{\ellTN(\parsold+\parsolc)}{\ellTN(\parsold-\parsolc)}
,
\]
where $\parang$ describes the initial angle around the $z$-axis.

%\[
%\sqrt{\frac{2\parfreq-\ptsqrt{g}_3\ptsqrt{g}_4-\ptsqrt{g}_1\ptsqrt{g}_2}{2\parfreq+\ptsqrt{g}_3\ptsqrt{g}_4+\ptsqrt{g}_1\ptsqrt{g}_2}}
%=
%\frac{\Im \sqrt{\ptsqrt{g}_3\ptsqrt{g}_4}}{\Re\sqrt{\ptsqrt{g}_3\ptsqrt{g}_4}}
%\]

This completes the derivation of the physical solution; let us summarise:
The wave function takes the form
\unskip\footnote{In this form, the function is invariant
under shifts of the parameters $\parsolc$ and $\parsold$ by $2\ellKval$ and $2\iunit\ellKvalc$.
Note that the shift of $\parsolc$ by $2\iunit\ellKvalc$ also induces a shift of $\parqper$ by $-4\pi$.}
\[
\spincp(x)=
\exp\brk!{\iunit \parqper (\parscl x+\paroff)/2\ellKval+\iunit \parang}
\frac{\ellTN(\parsold+\parsolc)}{\ellTN(\parsold-\parsolc)}
\frac{\ellTS(\parscl x+\paroff+\parsold-\parsolc)}{\ellTS(\parscl x+\paroff+\parsold+\parsolc)}.
\label{CHMEllipticState}
\]
Furthermore, the combination $\spincp\bar\spincp$ can be written as the elliptic function
\[
\spincp\bar\spincp
=
-\frac{\ellSN(\parsolc-\parsold)^2-\ellSN(\parscl x+\paroff)^2}
{\ellSN(\parsolc+\parsold)^2-\ellSN(\parscl x+\paroff)^2}.
\]
The various contributing constants take the following reality conditions
\[
\parvel,\parfreq,\parscl,\paroff,\parqper,\parang,\ellmods\in\Real,
\eqsep
\parsolc,\parsold\in(\half+\Integer)\ellKval+\iunit\Real,
\]
as well as
\[
\ptsqrt{g}_3=\ptsqrt{g}_1^\conj,
\eqsep
\ptsqrt{g}_4=\ptsqrt{g}_2^\conj,
\eqsep
(\Im \ptsqrt{g}_3)(\Im \ptsqrt{g}_4)\geq 0,
\eqsep
\bar\parsolb=\parsolb^\conj,
\eqsep
\bar \parsoleta=\parsoleta^\conj.
\]
This choice implies that the elliptic modulus
is in the principal interval:
\[
0\leq \ellmods\leq 1.
\]
Qualitatively, the wave function $\spincp(x)$ takes on a maximum magnitude at $\parscl x+\paroff=0$
and a minimum magnitude at $\parscl x+\paroff=\ellKval$.
The magnitude further oscillates periodically between these two values,
while the complex phase develops linearly with some superimposed oscillation
of the same periodicity.

For the further discussion, it makes sense to express
the various constants in terms of the independent parameters $\parscl,\ellmods,\parsold,\parsolc$.
We find
\unskip\footnote{For what it is worth, the following derivative relations hold:
$\partial \parqper/\partial \parsold = -\iunit \parfreq \fldlen/\parscl$
and $\partial \parfreq/\partial \parsold = -\iunit\parfreq \parvel/\parscl$.}
\begin{align}
\ptsqrt{g}_1 &=
\iunit\parscl\ellDC(\parsold)\ellDC(\parsolc)
\brk!{\ellSN(\parsold-\parsolc)-\ellSN(\parsold+\parsolc)},
\\
\ptsqrt{g}_2 &=
\iunit\parscl\ellmods\ellCD(\parsold)\ellCD(\parsolc)
\brk!{\ellSN(\parsold-\parsolc)-\ellSN(\parsold+\parsolc)},
\\
\ptsqrt{g}_3 &=
\iunit\parscl\ellNS(\parsold)\ellNS(\parsolc)
\brk!{\ellSN(\parsold-\parsolc)+\ellSN(\parsold+\parsolc)},
\\
\ptsqrt{g}_4 &=
\iunit\parscl\ellmods\ellSN(\parsold)\ellSN(\parsolc)
\brk!{\ellSN(\parsold-\parsolc)+\ellSN(\parsold+\parsolc)},
\\
\parfreq &= \parscl^2 \ellmods \brk!{\ellSN(\parsold-\parsolc)^2-\ellSN(\parsold+\parsolc)^2},
\\
\parvel&=2\iunit \parscl \frac{\ellSN(\parsold-\parsolc)\ellCN(\parsold-\parsolc)\ellDN(\parsold-\parsolc)
  -\ellSN(\parsold+\parsolc)\ellCN(\parsold+\parsolc)\ellDN(\parsold+\parsolc)}
{\ellSN(\parsold-\parsolc)^2-\ellSN(\parsold+\parsolc)^2},
\\
\parqper&=2\iunit\brk!{\ellZN(\parsold-\parsolc)- \ellZN(\parsold+\parsolc)}\ellKval.
\end{align}
The parameters $\parscl,\ellmods,\parsold,\parsolc$
take the following qualitative roles for the points $\set{\ptsqrt{g}_k}$:
The shape of the configuration of $\set{\ptsqrt{g}_k}$ is determined by
$\parsolc$ and $\ellmods$,
while changing $\parsold$ and $\parscl$ merely shifts and scales all $\set{\ptsqrt{g}_k}$, respectively.

%%%%%%%%%%%%%%%%%%%%%%%%%%%%%%%%%%%%%%%%%%%%%%%%%%%%%%%%%%%%%%%%%%%%%%%%%%%%%%%%
\subsection{Auxiliary Linear Problem}
\label{CHMellipticALP}

Let us now construct the solution to the auxiliary linear problem
for the CHM elliptic state:
\[
\laxvec'=\laxconn\laxvec,
\eqsep
\laxconn=\frac{\iunit}{\laxpar}\paulivec\vdot\spinvec
=\frac{\iunit}{\laxpar}\frac{1}{1+\spincp\bar\spincp}
\begin{pmatrix}
1-\spincp\bar\spincp & 2\bar\spincp
\\
2\spincp& \spincp\bar\spincp-1
\end{pmatrix},
\eqsep
\laxvec=
\begin{pmatrix}
\laxvec_1
\\
\laxvec_2
\end{pmatrix}.
\label{CHMALPeqn}
\]
As above, we can algebraically solve for the second component of the vector $\laxvec$
in terms of the first:
\[
\laxvec_2=-\frac{\iunit(1+\spincp\bar\spincp)\laxpar}{2\bar\spincp}\laxvec_1'
- \frac{(1-\spincp\bar\spincp)}{2\bar\spincp}\laxvec_1.
\]
Upon substitution of $\laxvec_2$ into the second component of the auxiliary linear problem
we obtain a second-order differential equation for $\laxvec_1$
\[
\laxvec_1''
+\brk[s]*{\iunit\frac{\bar G+G\spincp\bar\spincp}{1+\spincp\bar\spincp}}\laxvec_1'
+\brk[s]*{\frac{\bar G-G\spincp\bar\spincp}{1+\spincp\bar\spincp}\frac{1}{\laxpar}+\frac{1}{\laxpar^2}}\laxvec_1
=
0.
\]

By careful analysis of the coefficient function properties and singularities,
the form of the solution can be deduced:
\begin{align}
\laxvec_1(x)&=
\exp\brk[s]!{\iunit\laxqmom(\laxzpar)(\parscl x+\paroff)/2\ellKval}
\frac{\ellTN(\parscl x+\paroff+\parsolc-\laxzpar)}
{\ellTN(\parscl x+\paroff)\ellTS(\laxzpar-\parsolc)},
\\
\laxvec_2(x)&=
\exp\brk[s]!{\iunit \brk!{\laxqmom(\laxzpar)+\parqper} (\parscl x+\paroff)/2\ellKval+\iunit\parang}
\frac{\ellTN(\parscl x+\paroff-\parsolc-\laxzpar)}
{\ellTN(\parscl x+\paroff)\ellTS(\laxzpar+\parsolc)}.
\end{align}
The spectral parameter $\laxpar=\laxpar(\laxzpar)$ in the above differential equation
is expressed in terms of the uniformised parameter $\laxzpar$ by the function
\[
\laxpar(\laxzpar)=
\frac{\iunit}{\parscl\ellmods }\frac{1-\ellmods \ellSN(\parsold)^2\ellSN(\parsolc)^2}{\ellSN(\parsold)\ellCN(\parsold)\ellDN(\parsold) }
\frac{1-\ellmods \ellSN(\parsold)^2\ellSN(\laxzpar)^2}{\ellSN(\parsolc)^2-\ellSN(\laxzpar)^2}.
\]
%
%\[
%\laxpar(\laxzpar)=
%-\frac{2}{\parfreq}\frac{\ptsqrt{g}_1(\ptsqrt{g}_3-\ptsqrt{g}_2)-(\ptsqrt{g}_1-\ptsqrt{g}_2)\ptsqrt{g}_3\ellSN(\laxzpar)^2}
%{(\ptsqrt{g}_3-\ptsqrt{g}_2)-(\ptsqrt{g}_1-\ptsqrt{g}_2)\ellSN(\laxzpar)^2}.
%\]
%
The function $\laxqmom(\laxzpar)$ takes the form
\unskip\footnote{The elliptic zeta functions can be combined into $-2\ellZN(\laxzpar-\parsolc)$
plus an elliptic function by means of addition theorems.
However, doing so may not be convenient.}
\[
\laxqmom(\laxzpar)=\iunit\brk!{
\ellZN(\parsold-\laxzpar)
-\ellZN(\parsold-\parsolc)
-\ellZN(\parsold+\laxzpar)
+\ellZN(\parsold+\parsolc)
}\ellKval.
\]
Note that the quasi-momentum function and spectral curve for the elliptic state
of CHM have been discussed in detail in \cite{Bargheer:2008kj}
in the case of periodic boundary conditions.

Let us discuss some properties of the above quantities:
First of all, we note that the above differential equation
is of second order and therefore has a basis of two solutions.
Correspondingly, the map $\laxzpar\mapsto\laxpar$ has two pre-images $\pm\laxzpar$ for every $\laxpar$
(on the fundamental domain of the elliptic curve),
and each one corresponds to one of the two elements of the basis.
Flipping the sign of $\laxzpar$
acts on the above functions $\laxpar(\laxzpar)$ and $\laxqmom(\laxzpar)$ as follows:
\[
\laxpar(-\laxzpar)=\laxpar(\laxzpar),
\eqsep
\laxqmom(-\laxzpar)=-\parqper-\laxqmom(\laxzpar).
\]
Second, we note that the functions $\laxpar(\laxzpar)$ and $G(x)$ are closely related:
\[
\laxpar(\parscl x+\paroff+\parsold)=-\frac{2}{\parfreq}G(x).
\]
Furthermore, the function $\laxqmom(\laxzpar)$ serves as the quasi-momentum
defining the spectral curve:
\[
\brk[s]*{\frac{\der\laxqmom}{\der\laxpar}}^2
=
\brk[s]*{\frac{\laxqmom'(\laxzpar)}{\laxpar'(\laxzpar)}}^2
=
\frac{4\ellKval^2 \ptsqrt{\laxpar}_1\ptsqrt{\laxpar}_2\ptsqrt{\laxpar}_3\ptsqrt{\laxpar}_4}{\parscl^2}
\frac{\brk!{1+\half \parvel\laxpar+\rfrac{1}{2}(\parfreq^2\ptsqrt{\laxpar}_1\ptsqrt{\laxpar}_2+\parfreq^2\ptsqrt{\laxpar}_3\ptsqrt{\laxpar}_4-\parscl^2\ellEval/\ellKval)\laxpar^2}^2}
{\laxpar^4(\laxpar-\ptsqrt{\laxpar}_1)(\laxpar-\ptsqrt{\laxpar}_2)(\laxpar-\ptsqrt{\laxpar}_3)(\laxpar-\ptsqrt{\laxpar}_4)}.
\]
Here, the branch points $\ptsqrt{\laxpar}_k$
coincide with the distinguished points $\ptsqrt{g}_k$ up to a common factor:
\[
\ptsqrt{\laxpar}_k=-\frac{2}{\parfreq}\ptsqrt{g}_k.
\]
Hence, they take the form
\begin{align}
\ptsqrt{\laxpar}_1 &=
-\frac{2\iunit}{\parscl\ellmods} \frac{\ellDC(\parsold)\ellDC(\parsolc)}{\ellSN(\parsold-\parsolc)+\ellSN(\parsold+\parsolc)},
&
\ptsqrt{\laxpar}_2 &=
-\frac{2\iunit}{\parscl} \frac{\ellCD(\parsold)\ellCD(\parsolc)}{\ellSN(\parsold-\parsolc)+\ellSN(\parsold+\parsolc)},
\\
\ptsqrt{\laxpar}_3 &=
-\frac{2\iunit}{\parscl\ellmods} \frac{\ellNS(\parsold)\ellNS(\parsolc)}{\ellSN(\parsold-\parsolc)-\ellSN(\parsold+\parsolc)},
&
\ptsqrt{\laxpar}_4 &=
-\frac{2\iunit}{\parscl} \frac{\ellSN(\parsold)\ellSN(\parsolc)}{\ellSN(\parsold-\parsolc)-\ellSN(\parsold+\parsolc)}.
\end{align}
In fact, these branch points are given by the following points on the spectral curve:
\begin{align}
\label{CHMellSqrts}
\laxzpar&=\ptsqrt{\laxzpar}_1=\ellKval ,& \laxpar(\laxzpar) &= \ptsqrt{\laxpar}_1 ,& \laxqmom(\laxzpar) &= -\half\parqper,
\\
\laxzpar&=\ptsqrt{\laxzpar}_2=\ellKval+\iunit\ellKvalc ,& \laxpar(\laxzpar) &= \ptsqrt{\laxpar}_2 ,& \laxqmom(\laxzpar) &= -\pi-\half\parqper.
\\
\laxzpar&=\ptsqrt{\laxzpar}_3=0 ,& \laxpar(\laxzpar) &= \ptsqrt{\laxpar}_3 ,& \laxqmom(\laxzpar) &= -\half\parqper,
\\
\laxzpar&=\ptsqrt{\laxzpar}_4=\iunit\ellKvalc ,& \laxpar(\laxzpar) &= \ptsqrt{\laxpar}_4 ,& \laxqmom(\laxzpar) &= -\pi-\half\parqper.
\end{align}
\[
\mpostusemath{CHMEllipticCuts2}
\eqsep
\mpostusemath{CHMEllipticTorus}
\]
Note that the branch points are located where the periodicity factor $\exp(\iunit\laxqmom)$
coincides with the periodicity factor $\exp(-\iunit\laxqmom-\iunit\parqper)$ for the
second solution of the auxiliary linear problem with the same value of $\laxpar$.
Finally, we state the representation of the points $\laxpar=\infty$ and $\laxpar=0$
which are related to rotational symmetries and local quantities, respectively:
\begin{align}
\laxzpar&=+\parsolc ,& \laxpar(\laxzpar) &= \infty ,& \laxqmom(\laxzpar) &= 0,
\\
\laxzpar&=-\parsolc ,& \laxpar(\laxzpar) &= \infty ,& \laxqmom(\laxzpar) &= -\parqper,
\\
\laxzpar&=+\parsold-\iunit\ellKvalc ,& \laxpar(\laxzpar) &= 0 ,& \laxqmom(\laxzpar) &= \infty,
\\
\laxzpar&=-\parsold+\iunit\ellKvalc ,& \laxpar(\laxzpar) &= 0 ,& \laxqmom(\laxzpar) &= \infty.
\end{align}

%%%%%%%%%%%%%%%%%%%%%%%%%%%%%%%%%%%%%%%%%%%%%%%%%%%%%%%%%%%%%%%%%%%%%%%%%%%%%%%%
\subsection{Periodicity and Moduli}

Let us now consider the periodicities of the travelling wave state.
The logarithmic derivative $G(x)$ is given by an elliptic function,
and as such it has two periods in the complex plane.
These are given by the lengths
\[
\fldlen=\frac{2\ellKval}{\parscl},
\eqsep
\fldlen'=\frac{2\iunit\ellKvalc}{\parscl}.
\]
The actual state function $\spincp(x)$ is quasi-periodic as follows:
\[
\spincp(x+\fldlen)
=\exp(\iunit \parqper)\.\spincp(x),
\eqsep
\spincp(x+\fldlen')
=
\exp\brk*{\frac{2\pi\iunit\parsolc-\ellKvalc\parqper}{\ellKval}}
\spincp(x),
\]
Quasi-periodicity means that upon a shift of $x$ by $\fldlen$
the wave function $\vect{S}(x)$ returns to its original shape
but rotated by an angle of $\parqper$ about the $z$-axis.
We will thus call $\parqper$ the angle of quasi-periodicity.
In particular, the solution is perfectly periodic
if and only if $\parqper\in 2\pi\Integer$.
The imaginary period $\fldlen'$ leads to a complex factor,
but it bears no physical significance
because it shifts off the physical real $x$-axis.

The quasi-periodicity is linked to the mode numbers of the two elementary modes
contributing to the elliptic state. According to \eqref{CHMellSqrts}
with $\pi\modenum=\laxqmom(\ptsqrt{\laxzpar})-\laxqmom(\parsolc)$
these mode numbers are $\modenum_1=-\lambda/2\pi$ and $\modenum_2=-(\lambda+2\pi)/2\pi$.
As such, the mode numbers of this state are always separated by one unit.
A more general state with two arbitrary integer mode numbers $\modenum_1,\modenum_2$ is obtained
by regarding $\modenum_1-\modenum_2$ copies of the elementary interval $\fldlen$
as the effective periodicity $\fldlen_\text{eff}=(\modenum_1-\modenum_2)\fldlen$.
The quasi-periodicity for the elementary interval then takes
a rational value $\parqper=-2\pi \modenum_1/(\modenum_1-\modenum_2)$.

There are different ways to deal with the quasi-periodicity of the state.
On the one hand, the periodicity of the field is determined by
the boundary conditions.
Closed boundaries can be specified as $\spincp(x+\fldlen)=\spincp(x)$ or as
$\spincp(x+\fldlen)=\exp(\iunit \parqper)\spincp(x)$, both are equally valid,
however, they define distinct physical models a priori.
For instance, the twist parameter $\exp(\iunit \parqper)$
does not belong to phase space
(otherwise it would have to be accompanied by a canonically conjugate variable).
On the other hand, the function and equation of motion could be
transformed by the local rotation $\tilde\spincp(x)=\exp(-\iunit\parqper x/\fldlen)\spincp(x)$.
In this case, the equation of motion as well as the Lax connection will receive
few additional local terms from the relationship
$\tilde\spincp'=\exp(-\iunit\parqper x/\fldlen)(\spincp'-\iunit\parqper x \spincp/\fldlen)$.
Here, we will mostly disregard the implications of quasi-periodicity on the mechanical system
by allowing an arbitrary quasi-periodicity angle $\parqper$.
Where needed, we may fix $\parqper\in 2\pi\Integer$ for perfect periodicity of the state $\spincp(x)$.

The solution $\laxvec(x)$ to the auxiliary linear problem obeys the following quasi-periodicity relations:
\[
\laxvec_j(x+\fldlen) = \exp\brk{\iunit\laxqmom_j}\laxvec_j(x),
\eqsep
\laxvec_j(x+\fldlen') = \exp\brk{\iunit\tilde\laxqmom_j}\laxvec_j(x)
\]
with
\[
\laxqmom_1=\laxqmom,
\eqsep
\laxqmom_2=\laxqmom+\parqper,
\eqsep
\tilde\laxqmom_1=\frac{\iunit\ellKvalc\laxqmom+\pi(\laxzpar-\parsolc)}{\ellKval},
\eqsep
\tilde\laxqmom_2=\frac{\iunit\ellKvalc(\laxqmom+\parqper)+\pi(\laxzpar+\parsolc)}{\ellKval}.
\]
The solution $\laxvec_1(x)$ to the auxiliary linear problem is quasi-periodic by construction.
Interestingly, the corresponding second component $\laxvec_2(x)$ to the auxiliary linear problem
is quasi-periodic as well, yet with a different phase.
Essentially, the additional phase $\parqper$ is needed to compensate for the quasi-periodicity of the state $\spincp(x)$.
The two different quasi-periodicity phases applicable to the components $\laxvec_1(x)$ and $\laxvec_2(x)$
render the combined vector a twisted eigenvector of the Lax parallel transport by $\fldlen$.
Note further, that the second solution to the auxiliary linear problem has the eigenvalues
$\exp(-\iunit\laxqmom-\iunit\parqper)$ and $\exp(-\iunit\laxqmom)$,
so that again the quasi-periodicity of $\spincp(x)$ is respected.

The following periodicity relations hold for the uniformised spectral parameter $\laxzpar$:
The original spectral parameter $\laxpar(\laxzpar)$ is properly periodic.
The solutions to the auxiliary linear problem display some anti-periodicity:
\[
\laxvec_{1,2}(\laxzpar+2\ellKval) = -\laxvec_{1,2}(\laxzpar),
\eqsep
\laxvec_{1,2}(\laxzpar+2\iunit\ellKvalc) = +\laxvec_{1,2}(\laxzpar).
\]
The quasi-momentum shifts by a constant:
\[
\laxqmom(\laxzpar+2\ellKval) = \laxqmom(\laxzpar),
\eqsep
\laxqmom(\laxzpar+2\iunit\ellKvalc) = \laxqmom(\laxzpar)-2\pi.
\]
In fact, the shifts represent the integrals of $\der\laxqmom$ around the two
cycles of the elliptic curve:
\[
\oint_{\contacyc}\diff\laxqmom = 0,
\eqsep
\oint_{\contbcyc}\diff\laxqmom = -2\pi.
\]
As these are shifts by multiples of $2\pi$,
they leave the relevant exponent $\exp(\iunit\laxqmom)$ invariant.

\medskip

Next, it makes sense to consider transformations of the parameters $\parsolc$ and $\parsold$:
All functions and quantities are properly periodic under shifts of $\parsolc$ and $\parsold$
by $2\ellKval$ and $2\iunit\ellKvalc$.
The only exception is the quasi-periodicity angle $\parqper$
which shifts as $\parqper\to\parqper -4\pi$ upon the periodicity $\parsolc\to\parsolc+2\iunit\ellKvalc$.
Furthermore, it is possible to simultaneously shift $\parsolc$ and $\parsold$ by $\pm\ellKval$ or by $\pm\iunit\ellKvalc$.
These generate inessential permutations of the branch points as follows:
\begin{align}
\parsold\to\parsold+\ellKval,
\quad
\parsolc\to\parsolc+\ellKval:&
\qquad
\ptsqrt{\laxpar}_{3,4}\leftrightarrow\ptsqrt{\laxpar}_{1,2},
\quad
\parang\to\parang+\pi;
\\
\parsold\to\parsold\pm\iunit\ellKvalc,
\quad
\parsolc\to\parsolc+\iunit\ellKvalc:&
\qquad
\ptsqrt{\laxpar}_{1,3}\leftrightarrow\ptsqrt{\laxpar}_{2,4},
\quad
\parqper\to\parqper - 2\pi.
\end{align}

Apart from these, there are a couple of discrete transformations.
The following transformation effectively generates a parity operation:
\[
(\parsolc,\parsold,\paroff)\to-(\parsolc,\parsold,\paroff):
\qquad
\spincp(x)\to \spincp(-x),\quad
(\parvel,\parqper,\laxqmom,\laxpar,\laxzpar)\to-(\parvel,\parqper,\laxqmom,\laxpar,\laxzpar).
\]
Here, we note that the elliptic state has the following
self-conjugation property:
\[
\bar\spincp(x)=-\exp(-2\iunit\parang)\.\spincp(-x-2\paroff/\parscl).
\]
Finally, there is a transformation
which effectively rotates the spin field by $\pi$ around the $y$-axis:
\[
\parsolc\to-\parsolc,\quad
\parang\to-\parang:
\qquad
\spincp(x)\to 1/\spincp(x),\quad
(\parfreq,\parqper,\laxzpar,\laxqmom)\to-(\parfreq,\parqper,\laxzpar,\laxqmom).
\]
Curiously, this transformation preserves all the branch points.
Moreover, it flips the sign of $\parfreq$ which cannot be achieved by a continuous transformation of the parameters.
In fact, the sign of $\parfreq$ is linked to the real parts of $\parsolc$ and $\parsold$ as follows:
\begin{align}
\Re(\parsold+\parsolc)\equiv 0,\quad
\Re(\parsold-\parsolc)\equiv \ellKval
:&\qquad \parfreq>0;
\\
\Re(\parsold+\parsolc)\equiv \ellKval,\quad
\Re(\parsold-\parsolc)\equiv 0:&\qquad \parfreq<0.
\end{align}
Furthermore, the former case includes the vacuum state $\spincp(x)=\infty$ with $\angmom=-\fldlen$
while the latter case includes the opposite vacuum $\spincp(x)=0$ with $\angmom=+\fldlen$.

%%%%%%%%%%%%%%%%%%%%%%%%%%%%%%%%%%%%%%%%%%%%%%%%%%%%%%%%%%%%%%%%%%%%%%%%%%%%%%%%
\subsection{Dynamical Divisor}
\label{CHMEllipticDivisor}

The dynamical divisor for the state consists of the points $\laxzpar=\ptdiv{\laxzpar}(x)$
where the solution $\laxvec(x)$ to the auxiliary linear problem
aligns with a particular reference direction.
To that end, we consider the directional function $\laxcp(x)\in\ComplexComplete$
for the solution $\laxvec(x)$:
\[
\laxcp(x)
:=\frac{\laxvec_2(x)}{\laxvec_1(x)}
=
\exp\brk[s]!{\iunit\parqper (\parscl x+\paroff)/2\ellKval+\iunit\parang}
\frac{\ellTS(\laxzpar-\parsolc)\ellTN(\parscl x+\paroff-\parsolc-\laxzpar)}
{\ellTS(\laxzpar+\parsolc)\ellTN(\parscl x+\paroff+\parsolc-\laxzpar)}
.
\]
This is a perfectly periodic function in $\laxzpar$,
but by construction it is quasi-periodic in $x$ as
$\laxcp(x+\fldlen)=\exp(\iunit\parqper)\laxcp(x)$.
In particular, this function directly reproduces the corresponding state at the point $\laxpar=0$ as
\begin{align}
\laxzpar&=-\parsold+\iunit \ellKvalc:
&
\laxcp(x)&=\spincp(x),
\\
\laxzpar&=+\parsold-\iunit \ellKvalc:
&
\laxcp(x)&=-1/\bar\spincp(x).
\end{align}
For a given reference direction $\laxcp_0$,
we define the dynamical divisor $\set{\ptdiv{\laxzpar}(x)}$
as the set of solutions to the equation
\[
\laxcp(x)=\laxcp_0
\eqjoin{\text{for}}
\laxzpar=\ptdiv{\laxzpar}(x).
\]
We note that since $\laxcp(x)$ is an elliptic function in $z$ of lowest degree,
it can be solved in terms of inverse elliptic functions
which yields two solutions $\ptdiv{\laxzpar}_{1,2}(x)$.

The choice of reference direction $\laxcp_0$
strongly influences the qualitative $x$-dependency of the divisor points $\ptdiv{\laxzpar}_{1,2}(x)$.
Two special reference directions are given by $\laxcp_0=0$ or $\laxcp_0=\infty$.
Here, one of the two divisor points is evidently fixed to $\ptdiv{\laxzpar}_1=\pm\parsolc$,
respectively, while the other moves along a linear trajectory
$\ptdiv{\laxzpar}_2(x)=\parscl x+\paroff\mp\parsolc+\iunit\ellKvalc$ as $x$ increases.
In terms of $\laxpar$, one divisor points is fixed to $\ptdiv{\laxpar}_1=\infty$,
the other one $\ptdiv{\laxpar}_2$ moves around one of the two branch cuts as $x$ increases.
In particular, this is a perfectly closed path winding once around the cut for one period in $x$.
Unfortunately, this divisor does not capture the information on the angle $\parang$.

Two further distinguished reference directions are given by
\[
\laxcp_0=\pm \exp(\iunit\parang).
\]
Here, the divisor points meet two of the branch points at the maximum $\parscl x+\paroff=0$,
\[
\set{\ptdiv{\laxpar}(-\paroff/\parscl)}=
\set{\ptsqrt{\laxpar}_3,\ptsqrt{\laxpar}_4}
\eqjoin{\text{or}}
\set{\ptdiv{\laxpar}(-\paroff/\parscl)}=
\set{\ptsqrt{\laxpar}_1,\ptsqrt{\laxpar}_2}.
\]
As $x$ increases by one period,
the divisor points $\ptdiv{\laxpar}_{1,2}(x)$ qualitatively oscillate
along and transversely to the branch cut.
There will be $\parqper/2\pi$ and $(\parqper+2\pi)/2\pi$ oscillations along the cuts for one period.
Note that for generic values of the quasi-periodicity angle $\parqper$, the path of the divisor poles does not close.
The paths are closed only if $\parqper\in 2\pi \Integer$.
In the special case $\parqper\in \pi+2\pi \Integer$,
the path connects both ends of the branch cuts during one period.

For intermediate values of the magnitude $\abs{\laxcp_0}$,
one finds that the divisor points $\ptdiv{\laxpar}_{1,2}(x)$
take a highly complex path in the complex plane.
Their path can be understood as some interpolation between the trivial
fixed and circular behaviour at $\laxcp_0=0,\infty$
and the mostly longitudinal oscillations at $\abs{\laxcp_0}=1$.
The qualitative description of the paths is complicated by the observation
that tuning $\laxcp_0$ continuously may change the topology of the paths
including their winding around the branch points
(such changes are necessary to connect the two topologically distinct situations at both ends).
We observe two effects: First, a path is strongly deflected by nearby branch points.
For discrete values of $\laxcp_0$ a path will directly hit a branch point.
Here, the winding of the path around this branch point can change.
Second, the two paths $\ptdiv{\laxpar}_{1,2}(x)$ of the divisor points
may come arbitrarily close. For discrete values of $\laxcp_0$ they perfectly align and can recombine.

A further obstacle in describing the divisor paths is their quasi-periodicity.
Only for $\parqper\in 2\pi\Integer$ the paths close after one period in $x$.
For $\parqper\not\in 2\pi\Integer$, we have two alternatives for our consideration:
We can either accept that the paths are merely quasi-periodic.
There is nothing wrong with this point of view, especially since the
state itself is merely quasi-periodic in these cases.
However, topology will provide a less clear description of the open paths.
Otherwise, we may generalise the divisor condition
by picking an $x$-dependent reference direction $\laxcp_0(x)$.
A simple but particularly useful choice is
\[
\laxcp_0(x)\sim \exp(-\iunit \parqper x/\fldlen).
\]
Here, the divisor condition becomes perfectly periodic
such that the paths of the divisor points $\ptdiv{\laxzpar}(x)$ close.
We might also consider more elaborate choices for $\laxcp_0(x)$,
and certainly, the qualitative behaviour of the paths will depend on them.
Nevertheless, even for $\parqper \not\in 2\pi\Integer$,
we can achieve close paths.

Finally, we point out that even for a perfectly periodic divisor,
there are two qualitatively distinct cases:
The divisor consists of two points which are perfectly equivalent,
and therefore the divisor takes the role of an unordered set rather than an ordered pair.
Hence, a periodic divisor merely implies that the set agrees for $x$ and $x+\fldlen$.
For the paths of the individual points $\ptdiv{\laxzpar}_{1,2}(x)$
we can have one of the two periodicity situations
\[
\ptdiv{\laxzpar}_{1,2}(x+\fldlen)=\ptdiv{\laxzpar}_{1,2}(x)
\eqjoin{\text{or}}
\ptdiv{\laxzpar}_{1,2}(x+\fldlen)=\ptdiv{\laxzpar}_{2,1}(x).
\]
Both cases do arise, but it is hard to predict for which values of the reference direction $\laxcp_0$
one obtains which case.

%%%%%%%%%%%%%%%%%%%%%%%%%%%%%%%%%%%%%%%%%%%%%%%%%%%%%%%%%%%%%%%%%%%%%%%%%%%%%%%%
\subsection{Charges}

We can read off conserved charges from the
expansion of the quasi-momentum at $\laxpar=0$, $\laxzpar=\parsold-\iunit\ellKval$
and $\laxpar=\infty$, $\laxzpar=\parsolc$ as
\[
\laxqmom = \frac{\fldlen}{\laxpar} - \half \totmom + \quarter \eng \laxpar + \Order(\laxpar^2),
\eqsep
\laxqmom = \frac{\angmom}{\laxpar} + \Order(\laxpar^{-2}).
\]
These are the momentum $\totmom$, energy $\eng$ and total angular momentum $\angmom$:
\unskip\footnote{Incidentally,
$\partial \parqper/\partial \parsolc = -\iunit \parfreq \angmom/\parscl$,
as well as $\partial \totmom/\partial \parsolc = \iunit \parfreq (\fldlen-\angmom)/\parscl$
and $\partial \totmom/\partial \parsold +\partial \totmom/\partial \parsolc = 2\iunit \eng/\parscl$.}
\begin{align}
\totmom &= \int \frac{\iunit\spincp\bar\spincp'-\iunit\spincp'\bar\spincp}{1+\spincp\bar\spincp}\diff x
=
\fldlen\brk!{-\parvel-2\iunit\parscl\ellZN(\parsold+\parsolc)}
+2\pi,
\\
\eng &=
\int \diff x\half \spinvec'^2
=\int \diff x \frac{2\spincp'\bar\spincp'}{(1+\spincp\bar\spincp)^2}
=
\frac{\fldlen\parscl^2}{8}
\brk[s]*{
4\frac{\ellEval}{\ellKval}
+
\brk*{
\frac{\ellCN(\parsold)}{\ellSN(\parsold)\ellDN(\parsold)}
-\frac{\ellSN(\parsold)\ellDN(\parsold)}{\ellCN(\parsold)}
}^2
},
\\
\angmom &=
\int \diff x\spinvar_z
=
\int \diff x\frac{1-\spincp\bar\spincp}{1+\spincp\bar\spincp}
=\frac{\fldlen\parscl^2}{\parfreq}
\brk*{
-2\frac{\ellEval}{\ellKval}
+\ellDN(\parsold+\parsolc)^2+\ellDN(\parsold-\parsolc)^2
}.
\end{align}
%
%\[
%-\angmom=\frac{\fldlen}{2\parfreq}
%\brk*{
%4\parscl^2\frac{\ellEval}{\ellKval}
%-\ptsqrt{g}_1\ptsqrt{g}_2-\ptsqrt{g}_3\ptsqrt{g}_4
%}
%\]
%
%\[
%\eng=\frac{\fldlen}{8}
%\brk*{
%4\parscl^2\frac{\ellEval}{\ellKval}
%-\parvel^2
%+(\ptsqrt{g}_1+\ptsqrt{g}_2)(\ptsqrt{g}_3+\ptsqrt{g}_4)
%}
%\]
%
The spectral curve also defines appropriate action variables via the
contour integral
\[
\actvar=\frac{1}{2\pi\iunit}\oint_{\contacyc} \laxpar\diff\laxqmom.
\]
The indefinite integral yields
\begin{align}
\frac{1}{2\pi\iunit}\int \laxpar\diff\laxqmom
&=
\frac{\laxzpar}{4\pi\ellKval}
\brk!{\fldlen(\totmom-2\pi-\parqper)+\parqper\angmom}
\\
&\alignrel
+\frac{\angmom}{2\pi\iunit}\log\brk[s]*{-\frac{\ellTS(\laxzpar-\parsolc)}{\ellTS(\laxzpar+\parsolc)}}
+\frac{\fldlen}{2\pi\iunit}\log\brk[s]*{\frac{\ellTN(\laxzpar-\parsold)}{\ellTN(\laxzpar+\parsold)}}.
\end{align}
The elliptic curve has two branch cuts, and the contour integrals can be evaluated as
\[
\actvar_1=
\frac{-\fldlen\totmom+(\fldlen-\angmom)(\parqper+2\pi)}{2\pi},
\eqsep
\actvar_2=
\frac{\fldlen\totmom-(\fldlen-\angmom)\parqper}{2\pi}.
\]
Here, the cycles around the two cuts can be deformed into each other
while crossing pole singularities at $\laxpar=0$ and $\laxpar=\infty$
whose residues are $\angmom$ and $\fldlen$, respectively.
Altogether this implies a relationship with the total angular momentum $\angmom$:
\[
\actvar_1+\actvar_2=\fldlen-\angmom.
\]
Furthermore, the Riemann bilinear identity implies a second
exact relationship with the momentum $\totmom$:
\[
\parqper \actvar_1+(\parqper+2\pi)\actvar_2=\fldlen\totmom.
\]

Let us consider the action variables for small branch cuts of the spectral curve.
Suppose a small branch cut of length $\Delta\ptsqrt{\laxpar}$ is located near
the point $\ptsqrt{\laxpar}$ on the real axis.
Then the action variable can be approximated as
\[
\actvar\approx\frac{\fldlen}{8} \frac{(\Im\Delta\ptsqrt\laxpar)^2}{\ptsqrt\laxpar^2}.
\]

We have to point out that the above discussion fits well the one half of phase space
with $\parfreq<0$ (for $\Re\parsold\equiv +\Re\parsolc$)
while the charges are not ideally represented on the other half with $\parfreq>0$
(for $\Re\parsold\equiv -\Re\parsolc$).
The contours defining the action variables $\actvar_{1,2}$ have been chosen
such that the contour integrals are zero, $\actvar_1=\actvar_2=0$, at the vacuum state
\[
\spincp(x)=0:\qquad
\ellmods=0,\quad \parsold=\parsolc=\half\ellKval+\ihalf\ellKvalc.
\]
The situation for the opposite vacuum is not as fortunate:
\[
\spincp(x)=\infty:\qquad
\ellmods=0,\quad \parsold=-\parsolc=\half\ellKval+\ihalf\ellKvalc.
\]
For instance, the definition of the total momentum $\totmom$ as an integral of a local density
is singular at $\spincp=\infty$ and thus ill-defined.
Instead, we can evaluate relevant changes via expansion of the quasi-momentum.
Here, the momentum $\totmom=2\parqper$ and action variables $\actvar_1=-2\angmom$ and $\actvar_2=0$
take peculiar non-trivial values. It thus makes sense to redefine these quantities as follows
\[
\tilde\totmom=\totmom-2\parqper,
\eqsep
\tilde\actvar_1=\actvar_1+2\angmom,
\eqsep
\tilde\actvar_2=\actvar_2.
\]
The alternative action variable can be achieved by picking a different contour.
The alternative total momentum $\tilde\totmom$ is obtained from expansion
of the quasi-momentum at the alternate point $\laxzpar=-\parsold+\iunit\ellKval$ corresponding to $\laxpar=0$
\[
\laxqmom = -\frac{\fldlen}{\laxpar} + \half \tilde\totmom - \quarter \eng \laxpar + \Order(\laxpar^2).
\]
The alternative momentum also has as an expression as an integral of a local density
\[
\tilde\totmom = \totmom-2\parqper
=\int \frac{\iunit\spincp'/\spincp-\iunit\bar\spincp'/\bar\spincp}{1+\spincp\bar\spincp}\diff x
=
\fldlen\brk!{-\parvel-2\iunit\parscl\ellZN(\parsold-\parsolc)}
+2\pi,
\]
where the quasi-periodicity angle $\parqper$ can be defined via the integrals
\[
\parqper=
-\iunit\int\diff x
\frac{\spincp'}{\spincp}
=
\iunit\int\diff x
\frac{\bar\spincp'}{\bar\spincp}.
\]
In fact, the alternative total momentum properly vanishes for the vacuum $\spincp(x)=\infty$,
but it is ill-defined for $\spincp(x)=0$.
The alternative action variables satisfy the relations
\[
\tilde\actvar_1+\tilde\actvar_2=\fldlen+\angmom,
\eqsep
-\parqper \tilde\actvar_1+(-\parqper+2\pi)\tilde\actvar_2=\fldlen\tilde\totmom.
\]

The action variables display an interesting behaviour under shifts
of the variables $\parsolc,\parsold$.
For the combined discrete shift introduced above, we find
\[
\parsold\to\parsold\pm\iunit\ellKvalc,\quad
\parsolc\to\parsolc+\iunit\ellKvalc:\qquad
\actvar_1\to \actvar_1 +(\angmom\pm \fldlen),\quad
\actvar_2\to \actvar_2 -(\angmom\pm \fldlen).
\]
Note that here the variable $\parsold$ fails to be strictly periodic
under a shift by $2\iunit\ellKvalc$ even though all other quantities remain invariant.
This potentially puzzling behaviour can be understood as follows:
Shifting the imaginary parts of $\parsold$ and $\parsolc$
by equal or opposite continuous amounts, effectively makes the
branch points circle around each other.
For every shift of $\ellKvalc$, the two branch points are exchanged
while a shift of $2\ellKvalc$ brings the branch points back to their original positions.
The resulting situation appears to be the same as before. However, this is not so
because the action variables $\actvar_{1,2}$ are constructed as contour integrals around the branch cuts.
If the branch cuts or their surrounding contours are continuously deformed
in such a way that no other branch cuts or structures on the elliptic curve are crossed,
they will have to twist around the pair of encircling branch points before they can reach them.
Each half-turn will lead to a different situation, and therefore it is justified that
$\actvar_{1,2}$ are merely quasi-periodic.
In fact, the curling of branch cuts can be unwound via the remainder of the elliptic curve.
By doing so, they necessarily cross the poles at $\laxpar=0$ and $\laxpar=\infty$
from which they pick up the respective contributions.
In fact, the above discussion is reminiscent and related to the detailed discussion of branch cuts in \cite{Bargheer:2008kj}.
There, it was observed that branch points may move through other branch cuts
resulting in modified mode numbers.
This phenomenon occurs especially in the case of two adjacent mode numbers,
which is equivalent to the present case where $\parqper/2\pi$ and $(\parqper+2\pi)/2\pi$
serve as the mode numbers.

We further point out that the moduli space is
subdivided by singular configurations of the branch points
at the locations $\parsold\pm\parsolc\equiv \iunit\ellKvalc$.
For continuous deformations of the moduli,
the parameters should remain within the strips defined by these boundaries.
To this end, we note that the vacuum states with $\actvar_1=\actvar_2=0$
or with $\tilde\actvar_1=\tilde\actvar_2=0$
reside within the following fundamental intervals:
\[
\Re(\parsold)\equiv \pm\Re(\parsolc):\qquad
-\ellKvalc<\Im(\parsold\mp\parsolc)<+\ellKvalc.
\]
%

%%%%%%%%%%%%%%%%%%%%%%%%%%%%%%%%%%%%%%%%%%%%%%%%%%%%%%%%%%%%%%%%%%%%%%%%%%%%%%%%
\subsection{Soliton Limit}
\label{CHMSolitonLim}

In the following, we will derive the CHM soliton as a limit of the elliptic state.
We will proceed analogously to the KdV elliptic solution
where we saw that two branch points degenerate in the limit $\ellmods\to 1$.
Here, the branch points are organised in complex conjugate pairs, hence,
the limit will consist in degenerating two pairs of branch points.
As the branch points can move freely on the complex plane,
the approach now takes place in two dimensions and thus with additional degrees of freedom.

We choose the two branch points $\ptsqrt{\laxpar}_{2,4}$
to approach a common point $\ptsol{\laxpar}$ while their distance
shrinks exponentially with the length $\fldlen$.
In addition, in this two-dimensional setting
we can assume a linear rotation of the separation vector.
The target point $\ptsol{\laxpar}$ will describe the momentum $\ismsing$
of the bound state pole corresponding to the soliton according to the map
\[
\ismsing=-\frac{1}{\ptsol{\laxpar}}.
\]
Altogether this results in a spiralling motion ansatz for the branch points
\[
\ptsqrt{\laxpar}_{2,4}
=
\ptsol{\laxpar}
\mp 4\Im(\ptsol{\laxpar})\exp\brk!{-\half(1+\iunit \delta)\parscl \fldlen-\ihalf \pi-\iunit\gamma}+\ldots.
\]
The coefficients $\gamma$ and $\delta$ describe the angle and angular velocity of the separation.
Their values specify certain aspects of the limit,
and we shall tune them to achieve specific desirable effects.
The limit described by the above branch points corresponds to the following parameters of the elliptic curve
\begin{align}
\ellmods &= 1-16\exp(-\parscl \fldlen)+\ldots,
\\
\parscl &= \frac{2\Im \ptsol{\laxpar}}{\abs{\ptsol{\laxpar}}^2}=2\Im \ismsing,
\\
\parsold &= -\half\ellKval + \iunit \brk!{\quarter\delta \parscl \fldlen+\half\gamma+\half\pi-\arg \ismsing}+\ldots,
\\
\parsolc &= +\half\ellKval - \iunit \brk!{\quarter\delta \parscl \fldlen+\half\gamma}+\ldots.
\end{align}
Note that these parameters correspond to a state based on the vacuum $\spincp(x)=\infty$.
However, for large $\fldlen$ the state will be highly excited so that most parts
reside at $\spincp(x)\approx 0$.

By careful analysis we find the large-length limit of the wave function
\[
\spincp(x)=
\exp\brk!{\iunit \cot(\arg\ismsing) (\parscl x+\paroff)+\iunit\parang}
\frac{\iunit \sin(\arg\ismsing)}{\cosh\brk!{\parscl x+\paroff-\iunit \arg\ismsing}}
+\ldots,
\]
as well as of the various characteristic quantities of the state
\begin{align}
\parvel&= -4\Re(\ismsing)+\ldots,
&
\parfreq &= 4\abs{\ismsing}^2+\ldots,
\\
\parqper &= 2\fldlen\brk!{\delta \Im(\ismsing)+\Re(\ismsing)}+2\gamma+\ldots,
&
\totmom &= 4\arg\ismsing+\ldots,
\\
\angmom &= \fldlen-2\frac{\Im(\ismsing)}{\abs{\ismsing}{}^2}+\ldots,
&
\eng &= 8\Im(\ismsing)+\ldots,
\\
\tilde\actvar_1 &=
\frac{2 \fldlen (\pi-\arg\ismsing)}{\pi}
+\frac{(\parqper-2\pi) \Im(\ismsing)}{\pi\abs{\ismsing}{}^2}+\ldots,
&
\tilde\actvar_2 &=
\frac{2 \fldlen \arg\ismsing}{\pi}
-\frac{\parqper \Im(\ismsing)}{\pi\abs{\ismsing}{}^2}+\ldots.
\end{align}
The wave function and associated quantities agree, where applicable,
with the CHM soliton.
We note that this soliton approaches the vacuum $\spincp(x)=0$ when
$\abs{\parscl x+\paroff}$ is large. However, it also has $\parfreq>0$,
so it belongs to the family of solutions which are connected to the other vacuum $\spincp(x)=\infty$.
In that sense, all of the soliton wave function can be considered highly excited.

In the limit, the spectral curve splits into two disconnected components
near $\laxzpar=\pm\parsolc$ corresponding to the two independent solutions of the auxiliary linear problem
for common spectral parameter $\laxpar$.
In the first case, we set $\laxzpar=\parsolc+\laxzpar_\infty$.
The limit of the uniformisation of the spectral parameter reads
\[
\laxpar
=\abs{\ptsol{\laxpar}}\frac{\sinh(\laxzpar_\infty+\iunit\arg \ptsol{\laxpar})}{\sinh(\laxzpar_\infty)},
\eqsep
\ismmom
:=-\frac{1}{\laxpar}
=\frac{\abs{\ismsing}\sinh(\laxzpar_\infty)}{\sinh(\laxzpar_\infty-\iunit\arg \ismsing)},
\]
which has the following inverse relation
\[
\laxzpar_\infty
= \frac{1}{2} \log \frac{\ismsing(\ismmom-\ismsing^\conj)}{\ismsing^\conj(\ismmom-\ismsing)}
= \frac{1}{2} \log \frac{\laxpar-\ptsol{\laxpar}^\conj}{\laxpar-\ptsol{\laxpar}}.
\]
The solution to the auxiliary linear problem has the following limit
\begin{align}
\laxvec_1(x) &= \exp\brk[s]*{\frac{-\iunit\sinh(\laxzpar_\infty)(\parscl x+\paroff)}{2\sin(\arg\ismsing)\sinh(\laxzpar_\infty-\iunit\arg \ismsing)}}
\frac{\cosh (\parscl x+\paroff-\laxzpar_\infty)}{\sinh(\laxzpar_\infty)\cosh(\parscl x+\paroff)}+\ldots,
\\
\laxvec_2(x) &= \exp\brk[s]*{
\frac{\iunit\sinh(\laxzpar_\infty-2\iunit\arg\ismsing)(\parscl x+\paroff)}{2\sin(\arg\ismsing)\sinh(\laxzpar_\infty-\iunit\arg\ismsing)}+\iunit\parang}
\frac{1}{\cosh(\parscl x+\paroff)}+\ldots,
\end{align}
and the corresponding quasi-momentum expands as
\[
\laxqmom
=
-\fldlen \ismmom
+\iunit\log \brk[s]*{\frac{\ismsing}{\ismsing^\conj}\frac{\ismmom -\ismsing^\conj}{\ismmom -\ismsing}} +\ldots
=
\frac{\fldlen}{\laxpar}
+\iunit\log \frac{\laxpar-\ptsol{\laxpar}^\conj}{\laxpar-\ptsol{\laxpar}} +\ldots.
\]
These functions and dependencies describe the integrable structure for the CHM soliton.
In particular, the regularised quasi-momentum $\laxqmom+\fldlen\ismmom$
describes the transfer coefficient of the auxiliary scattering matrix.
In the second case, we set $\laxzpar=-\parsolc-\laxzpar_\infty$
in order to obtain obtain agreement with the above uniformisation relations.
The limit then takes the form
\begin{align}
\laxvec_1(x) &= -\exp\brk[s]*{
\frac{-\iunit\sinh(\laxzpar_\infty-2\iunit\arg\ismsing)(\parscl x+\paroff)}{2\sin(\arg\ismsing)\sinh(\laxzpar_\infty-\iunit\arg\ismsing)}}
\frac{1}{\cosh(\parscl x+\paroff)}+\ldots,
\\
\laxvec_2(x) &= -\exp\brk[s]*{\frac{\iunit\sinh(\laxzpar_\infty)(\parscl x+\paroff)}{2\sin(\arg\ismsing)\sinh(\laxzpar_\infty-\iunit\arg \ismsing)}+\iunit\parang}
\frac{\cosh (\parscl x+\paroff+\laxzpar_\infty)}{\sinh(\laxzpar_\infty)\cosh(\parscl x+\paroff)}+\ldots,
\\
\laxqmom
&=
-\parqper+\fldlen \ismmom
-\iunit\log \brk[s]*{\frac{\ismsing}{\ismsing^\conj}\frac{\ismmom -\ismsing^\conj}{\ismmom -\ismsing}} +\ldots.
=
-\parqper-\frac{\fldlen}{\laxpar}
-\iunit\log \frac{\laxpar-\ptsol{\laxpar}^\conj}{\laxpar-\ptsol{\laxpar}} +\ldots.
\label{CHMSolitonLimqexp}
\end{align}
This describes the other solution to the auxiliary linear problem
and the other transfer coefficient.

We note that in order to perform the limit $\ellmods\to 1$
it was necessary to carefully expand the elliptic theta and zeta functions
with arguments of order $\ellKval\sim \log (1-\ellmods)$.
For instance, for an argument $z$ with a large imaginary part, we have to use the non-standard relations
\begin{align}
\ellTN(z)&=\exp(-z^2/2\ellKval)\cosh(z)+\ldots,
\\
\ellZN(z)&=\tanh(z) - z/\ellKval+\ldots.
\end{align}
In particular, these relations are consistent with the exact imaginary periodicity relation
for these elliptic functions.

Let us now discuss the significance and effect of the parameters $\gamma,\delta$
which relate to the orientation and rotation of the branch point configuration.
First, we observe that the resulting expressions are largely independent of them.
Merely the quasi-periodicity angle $\parqper$ and the action variables depend on them,
but these are necessarily finite-length concepts.

There are several useful options to adust $\gamma,\delta$:
We can aim to fix the quasi-periodicity angle $\parqper$ for all lengths $\fldlen$.
This is achieved by the assignment
\[
\delta=\frac{\Re(\ptsol{\laxpar})}{\Im(\ptsol{\laxpar})}
=-\frac{\Re(\ismsing)}{\Im(\ismsing)} = -\cot(\arg\ismsing),
\eqsep
\gamma=-\half\parqper.
\]
In this case, the branch points perform a spiralling exponential motion
towards their limiting position
\[
\ptsqrt{\laxpar}_{2,4}
=
\ptsol{\laxpar}
\mp 4\Im(\ptsol{\laxpar})\exp\brk!{-\iunit \fldlen/\ptsol{\laxpar}-\ihalf \pi+\ihalf\parqper}+\ldots.
\]
In particular, the driving term $-\iunit \fldlen/\ptsol{\laxpar}=\iunit\ismsing \fldlen$ of the twirling motion
takes a very natural form here.
With this choice of parameters, we can adjust the length $\fldlen$ continuously
while preserving the desired quasi-periodicity angle $\parqper$
which can be considered an essential parameter of the underlying mechanical model.

Another viable option is to fix the orientation of the branch points as $\fldlen$ increases
by setting $\delta=0$. In this case, the quasi-periodicity angle
has a linear asymptotic dependency on the length
\[
\parqper(\fldlen) = 2\fldlen\Re(\ismsing)+2\gamma+\ldots.
\]
We might also choose $\delta$ differently or, even more generally,
assume $\gamma$ to be some more elaborate function of $\fldlen$.
In all of these cases, $\parqper$ will depend non-trivially on $\fldlen$.
We can then either accept a continuously changing $\parqper$.
A freely adjustable parameter $\parqper$ can be dealt with well at the level
of the differential equations.
However, it may represent a conflict with the role of $\parqper$
as specifying the periodic boundary conditions within the model of Hamiltonian mechanics.

Alternatively, we can deal with a varying quasi-periodicity angle $\parqper$
by making use of the fact that the angle is defined modulo $2\pi$:
Supposing that the function $\parqper(\fldlen)$ keeps crossing the fixed lattice $2\pi\Integer+\parqper_0$ as $\fldlen\to\infty$,
we can focus on states at specific lengths $\fldlen_k$
which are chosen such that
\[
\parqper(\fldlen_k)\in 2\pi\Integer+\parqper_0.
\]
We can thus reduce the family of states with continuous $\fldlen$
to a sequence of states at specific lengths $\fldlen_k$
which define the large-$\fldlen$ limit.
Here, the value of $\parqper(\fldlen)$ changes between the states of the sequence,
but the quasi-periodicity angle $\parqper(\fldlen)$ is equivalent to a fixed target value $\parqper_0$.
While this approach appears to resolve the technical issue of a variable $\parqper(\fldlen)$,
we point out that it is equivalent to assuming a fixed $\parqper$:
To that end, we note that $\parqper$ may only shift by a multiple of $2\pi$
between elements of the sequence. Such a shift is equivalent
to the rotation of the pair of branch points by a multiple of $\pi$.
Such a rotation merely exchanges the branch points
(as well as a deformation of the branch cuts on the spectral curve),
and has no physical significance.
In conclusion, we may either take $\parqper$ to be fixed
or to be continuously adjustable within an extended notion of boundary conditions.

Let us finally discuss how to abstractly describe the configuration of branch cuts
that limit to a CHM soliton configuration.
We shall assume $\parqper$ to be fixed.
Since $\abs{\parqper}\ll \fldlen$, the two branch cuts that make up the soliton
correspond to small and fixed mode numbers $\modenum_1=-\parqper/2\pi$ and $\modenum_2=-(\parqper+2\pi)/2\pi$.
Furthermore, these two branch cuts are adjacent.
As a consequence we have for the quasi-momentum at the branch points
\[
\exp\brk!{\iunit \laxqmom(\ptsqrt{\laxzpar}_1)}
=
-
\exp\brk!{\iunit \laxqmom(\ptsqrt{\laxzpar}_3)}
=
\exp\brk!{-\ihalf\parqper}.
\]
In particular, $\exp\brk{\iunit \laxqmom(\ptsqrt{\laxzpar}_{1,3})}$
have opposite signs.
The position $\ismsing$ of the corresponding bound state pole can be identified
through the action variables $\actvar_{1,2}$ or their alternative versions
$\tilde\actvar_{1,2}$.
The large-$\fldlen$ behaviour specifies the angle $\arg\ismsing$
\[
\label{CHMEllActLimitLarge}
\frac{\tilde\actvar_1}{\fldlen}
=\frac{\actvar_1+2\angmom}{\fldlen}
=\frac{2}{\pi}\brk{\pi-\arg\ismsing}
+\ldots,
\eqsep
\frac{\tilde\actvar_2}{\fldlen}
=\frac{\actvar_2}{\fldlen}
=\frac{2}{\pi}\arg\ismsing
+\ldots.
\]
Furthermore, the finite remainder in the sum $\actvar_1+\actvar_2$
determines the magnitude $\abs{\ismsing}$
\[
\label{CHMEllActLimitSum}
\actvar_1+\actvar_2
=
\tilde\actvar_1+\tilde\actvar_2-2\angmom
=
\frac{2\sin(\arg\ismsing)}{\abs{\ismsing}}
+\ldots.
\]
%

%%%%%%%%%%%%%%%%%%%%%%%%%%%%%%%%%%%%%%%%%%%%%%%%%%%%%%%%%%%%%%%%%%%%%%%%%%%%%%%%
%%%%%%%%%%%%%%%%%%%%%%%%%%%%%%%%%%%%%%%%%%%%%%%%%%%%%%%%%%%%%%%%%%%%%%%%%%%%%%%%
\section{Scattering Unitarity on the Double Space}
\label{leftrightinout}

In this appendix we consider how
space-like and time-like scattering problems in two dimensions
and their unitarity properties are related.
Here, space-like scattering relates left and right asymptotic states
(which is specific to two dimensions)
while time-like scattering relates in (past) and out (future) asymptotic states.
The curious observation is that unitarity of one scattering problem
translates to unitarity of the other, albeit with regard to a different signature.
Here we treat the transformation between the two pictures
as an abstract problem of linear algebra.

Suppose we have a matrix which identifies vectors $L,R$
from two equivalent spaces via
\[
L=\laxscat R.
\]
Suppose further the matrix satisfies a unitarity relation
\[
\laxscat^{-1} = H \laxscat H^{-1}
\]
with $H$ some hermitian form we take to be a diagonal matrix
with entries $\pm 1$.

One can now exchange a subset of directions between the two subspaces,
\unskip\footnote{In the scattering picture,
the left and the right states with prescribed asymptotics $\eunit^{\pm\iunit kx}$
typically split evenly
into incoming and outgoing states with prescribed asymptotics $\eunit^{\pm\iunit \omega t}$
such that $k=n/2$.
Here we consider a more general permutation of elements between two abstract spaces.}
\begin{align}
I&=(L_1,\ldots,L_k,R_{k+1},\ldots,R_{n})^\trans,
\\
O&=(R_1,\ldots,R_k,L_{k+1},\ldots,L_{n})^\trans,
\end{align}
and again solve one subspace in terms of the other.
To that end, we have to invert the above relation for the subset
of directions that is to be exchanged.
The resulting relation takes the same form
\[
I=\tilde\laxscat O
\]
with a new matrix $\tilde\laxscat$.
The new matrix turns out to satisfy a unitarity relation as well
\[
\tilde\laxscat^{-1} = \tilde H \tilde\laxscat \tilde H^{-1} .
\]
Curiously, the new hermitian form $\tilde H$ is the same as the old one
up to a change of signs for all the directions that were exchanged between the spaces
\[
\tilde H=\diag(H_{11},\ldots,H_{kk},-H_{k+1,k+1},\ldots,-H_{n,n}).
\]
For two left/right or in/out states ($n=2$, $k=1$), 
this implies that unitarity with signature $++$ translates 
to unitarity with signature $+-$ and vice versa.

Let us try to understand better how the unitarity relations are related
by going to the double space.
The original relation between the two spaces can be formulated on the double space
which is the direct sum of the two spaces
\[
\begin{pmatrix} 1 & -\laxscat \end{pmatrix}
\begin{pmatrix} L \\ R \end{pmatrix}
=0.
\]
Here $1$ denotes the unit matrix. The solution to the relation takes the form of a kernel
of a rectangular matrix.
The alternative relation can be formulated on the same space by
introducing a permutation matrix $P$
which interchanges some rows between the subspaces
\[
\begin{pmatrix} L \\ R \end{pmatrix}
=P\begin{pmatrix} I \\ O \end{pmatrix}.
\]
Then we can transform the rectangular matrix by an appropriate matrix $U$
\[
\begin{pmatrix} 1 & -\laxscat \end{pmatrix} P
= U \begin{pmatrix} 1 & -\tilde\laxscat \end{pmatrix},
\]
such that the left block returns to an identity matrix.
This relation implicitly defines both $\tilde\laxscat$ and $U$.
With this we find the alternative relation right away
\[
\begin{pmatrix} 1 & -\tilde\laxscat \end{pmatrix}
\begin{pmatrix} I \\ O \end{pmatrix}
=0.
\]
Now let us also double the original relation by introducing a square matrix on the double space
\[
W:=\begin{pmatrix} 1 & -\laxscat \\ \laxscat^{-1} & -1\end{pmatrix},
\eqsep
V:=\begin{pmatrix} L \\ R \end{pmatrix},
\eqsep
WV=0.
\]
This matrix is nilpotent and hermitian
\[
W^2=0,
\eqsep
MW^\hconj M^{-1} = W,
\eqsep
M:=\begin{pmatrix} H & 0 \\ 0 & -H\end{pmatrix},
\]
where $M$ is a hermitian form on the double space.
It also has half rank such that the kernel of $W$ equals the image of $W$.
The nilpotent matrix for the alternative picture reads
\[
\tilde W:=\begin{pmatrix} 1 & -\tilde\laxscat \\ \tilde\laxscat^{-1} & -1\end{pmatrix},
\eqsep
\tilde W^2=0,
\eqsep
\tilde V:=\begin{pmatrix} I \\ O \end{pmatrix},
\eqsep
\tilde W\tilde V=0.
\]
The unitarity relation for $\tilde \laxscat$ is equivalent to hermiticity of $\tilde W$
\[
\tilde M\tilde W^\hconj \tilde M^{-1} = \tilde W,
\eqsep
\tilde M:=\begin{pmatrix} \tilde H & 0 \\ 0 & -\tilde H\end{pmatrix}.
\]
It turns out that it $\tilde \laxscat$ is unitary if and only if
\[
P\tilde M P^\hconj = M.
\]
This would manifestly hold if $P\tilde W P^{-1}=W$,
but the two nilpotent matrices are in fact more immediately related by the relation
\[
\begin{pmatrix} U & 0 \\ 0 & \laxscat^{-1}U\tilde\laxscat\end{pmatrix} \tilde W P^{-1} = W.
\]
However, note that the two matrices are singular, hence the above form of the relation is not unique.
In order to obtain a conclusive answer,
we assume that $H$ is diagonal and the permutation matrix takes the simple form
\[
P=\begin{pmatrix} 1-X & X \\ X & 1-X \end{pmatrix}
\]
with $X$ a diagonal matrix of elements $0$ and $1$.
The relation between $M$ and $\tilde M$ then
flips a sign between $H$ and $\tilde H$ precisely where $X$ has a non-zero diagonal element
which is achieved by permuting elements between the blocks $H$ and $-H$.

%%%%%%%%%%%%%%%%%%%%%%%%%%%%%%%%%%%%%%%%%%%%%%%%%%%%%%%%%%%%%%%%%%%%%%%%%%%%%%%%
%%%%%%%%%%%%%%%%%%%%%%%%%%%%%%%%%%%%%%%%%%%%%%%%%%%%%%%%%%%%%%%%%%%%%%%%%%%%%%%%
\ifarxiv\else
\begin{bibtex}[\jobname]

@article{marchenko1952,
  title={Some questions of the theory of one-dimensional linear differential operators of the second order. I},
  author={Marchenko, V. A.},
  journal={Trudy Mosk. Mat. Obs.},
  volume={1},
  pages={327-420},
  year={1952},
  url={https://www.mathnet.ru/eng/mmo/v1/p327}
}

@article{Marchenko1955695,
  author = {Marchenko, V. A.},
  title = {On reconstruction of the potential energy from phases of the scattered waves},
  year = {1955},
  journal = {Dokl. Akad. Nauk SSSR},
  volume = {104},
  number = {5},
  pages = {695-698},
}

@article{Akhiezer1961,
  author = {Akhiezer, N. I.},
  title = {Continuous analogues of orthogonal polynomials on a system of intervals},
  year = {1961},
  journal = {Dokl. Akad. Nauk SSSR},
  volume = {141},
  number = {2},
  pages = {263-266},
  url = {https://www.mathnet.ru/eng/dan/v141/i2/p263}
}

@article{marchenko1974dokl,
  title={A periodic Korteweg–de Vries problem},
  author={Marchenko, V. A.},
  journal={Dokl. Akad. Nauk SSSR},
  volume={217},
  pages={276-279},
  year={1974},
  url={https://www.mathnet.ru/eng/dan/v217/i2/p276}
}

@book{marchenko2011sturm,
  title={Sturm-Liouville operators and applications},
  author={Marchenko, Vladimir Aleksandrovich},
  volume={373},
  year={2011},
  publisher={American Mathematical Soc.}
}

@article{Lax:1968fm,
    author = "Lax, P. D.",
    title = {Integrals of Nonlinear Equations of Evolution and Solitary Waves},
    journal = "Commun. Pure Appl. Math.",
    volume = "21",
    pages = "467-490",
    year = "1968",
    doi = "10.1002/cpa.3160210503"
}

@book{russell1845report,
  title={Report on Waves: Made to the Meetings of the British Association in 1842--43},
  author={Russell, John Scott},
  year={1845}
}

@article{marchenko1974periodic,
  title={Periodic problem of Korteweg de Vries equation},
  author={Marchenko, V. A.},
  journal={Matem. Sbornik},
  volume={95},
  pages={331-356},
  year={1974}
}

@article{faddeev1964properties,
  title={Properties of the S-matrix of the one-dimensional Schrödinger equation},
  author={Faddeev, Lyudvig Dmitrievich},
  journal={Amer. Math. Soc. Transl. Ser. 2},
  volume={65},
  pages={4},
  year={1967},
  doi={10.1090/trans2/065/04}
}

@book{boussinesq1877essai,
  title={Essai sur la théorie des eaux courantes},
  author={Boussinesq, Joseph},
  year={1877},
  publisher={Académie des sciences de l'Institut de France, mémoires présentés par divers savants 23}
}

@article{kordeweg1895change,
  title={On the change of form of long waves advancing in a rectangular channel, and a new type of long stationary wave},
  author={Korteweg, D. J. and de Vries, G},
  journal={Phil. Mag.},
  volume={39},
  pages={422-443},
  year={1895}
}

@article{zabusky1965interaction,
  title={Interaction of solitons in a collisionless plasma and the recurrence of initial states},
  author={Zabusky, Norman J and Kruskal, Martin D},
  journal={Phys. Rev. Lett.},
  volume={15},
  number={6},
  pages={240},
  year={1965},
  publisher={APS}
}

@article{gelfand1951determination,
  title={On the determination of a differential equation from its spectral function},
  author={Gel'fand, Izrail Moiseevich and Levitan, Boris Moiseevich},
  journal={Amer. Math. Soc. Transl. Ser. 2},
  volume={1},
  pages={253-304},
  year={1955},
  doi={10.1090/trans2/001/11}
}

@article{miura1968korteweg1,
  title={Korteweg-de Vries equation and generalizations I: A remarkable explicit nonlinear transformation},
  author={Miura, Robert M},
  journal={J. Math. Phys.},
  volume={9},
  number={8},
  pages={1202-1204},
  year={1968},
  publisher={American Institute of Physics}
}

@article{gardner1971,
  title={Korteweg‐de Vries Equation and Generalizations. IV. The Korteweg‐de Vries Equation as a Hamiltonian System},
  author={Gardner, Clifford S},
  journal={J. Math. Phys.},
  volume={12},
  pages={1548–1551},
  year={1971},
  doi={10.1063/1.1665772}
}

@article{miura1968korteweg2,
  title={Korteweg-de Vries equation and generalizations II: Existence of conservation laws and constants of motion},
  author={Miura, Robert M and Gardner, Clifford S and Kruskal, Martin D},
  journal={J. Math. Phys.},
  volume={9},
  number={8},
  pages={1204-1209},
  year={1968},
  doi={10.1063/1.1664701}
}

@article{its1975schrodinger,
  title={Schrödinger operators with finite-gap spectrum and N-soliton solutions of the Korteweg--de Vries equation},
  author={Its, Alexander Rudol'fovich and Matveev, Vladimir Borisovich},
  journal={Teoret. Mat. Fiz.},
  volume={23},
  number={1},
  pages={51-68},
  year={1975},
  publisher={Russian Academy of Sciences, Steklov Mathematical Institute of Russian}
}

@article{lax1975periodic,
  title={Periodic solutions of the KdV equation},
  author={Lax, Peter D},
  journal={Commun. Pure Appl. Math.},
  volume={28},
  number={1},
  pages={141-188},
  year={1975},
  doi={10.1002/cpa.3160280105},
  publisher={Wiley Online Library}
}

@article{dubrovin1976finite,
  title={Finite-zone linear operators and Abelian manifolds},
  author={Dubrovin, B. A.},
  journal={Uspekhi Mat. Nauk},
  volume={31},
  number={4},
  pages={259-260},
  year={1976}
}

@article{levitan1981almost,
  title={Almost periodicity of infinite-zone potentials},
  author={Levitan, Boris Moiseevich},
  journal={Izvestiya Rossiiskoi Akademii Nauk. Seriya Matematicheskaya},
  volume={45},
  number={2},
  pages={291-320},
  year={1981},
  publisher={Russian Academy of Sciences, Steklov Mathematical Institute of Russian~…}
}

@article{Gardner:1967wc,
    author = "Gardner, Clifford S. and Greene, Johm M. and Kruskal, Martin D. and Miura, Robert M.",
    title = {Method for solving the Korteweg-deVries equation},
    doi = "10.1103/PhysRevLett.19.1095",
    journal = "Phys. Rev. Lett.",
    volume = "19",
    pages = "1095-1097",
    year = "1967"
}

@article{aktosun1985non,
  title={Non-uniqueness in the one-dimensional inverse scattering problem},
  author={Aktosun, Tuncay and Newton, Roger G},
  journal={Inverse Problems},
  volume={1},
  number={4},
  pages={291},
  year={1985},
  publisher={IOP Publishing}
}

@article{novikov1974periodic,
  title={The periodic problem for the Korteweg-de Vries equation},
  author={Novikov, S. P.},
  journal={Funct. Anal. Appl.},
  volume={8},
  pages={236-246},
  year={1974},
  doi={10.1007/BF01075697}
}

@article{dubrovin1975InvPerShort,
  title={Inverse problem of scattering theory for periodic finite-zone potentials},
  author={Dubrovin, B. A.},
  journal={Funct. Anal. Appl.},
  volume={9},
  pages={61-62},
  year={1975},
  doi={10.1007/BF01078183}
}

@article{dubrovin1975InvPerLong,
  title={Periodic problems for the Korteweg-de Vries equation in the class of finite band potentials},
  author={Dubrovin, B. A.},
  journal={Funct. Anal. Appl.},
  volume={9},
  pages={215-223},
  year={1975},
  doi={10.1007/BF01075598}
}

@article{its1976problems,
  title={On a Class of Solutions of the Korteweg-deVries Equation},
  author={Its, A. R. and Matveev, V. B.},
  journal={Problemy Mat. Fiz.},
  volume={8},
  pages={70-92},
  year={1976},
  publisher={Izdat. Leningrad University}
}

@article{its1983problems,
  title={Connection between Soliton and Finite-Gap Solutions to the NS equations},
  author={Its, A. R.},
  journal={Problemy Mat. Fiz.},
  volume={10},
  pages={118-136},
  year={1983},
  publisher={Izdat. Leningrad University}
}

@book{bbeim1994,
  title={Algebro-Geometric Approach to Nonlinear Integrable Equations},
  author={Belokolos, E. D. and Bobenko, Alexander I. and Enol'skii, V. Z. and Its, Alexander R. and Matveev, V. B.},
  pages={337},
  series={Springer Series in Nonlinear Dynamics},
  year={1994},
  publisher={Springer},
  address = "Berlin, Germany",
}

@article{its1975Hills,
  title={Hill's operator with finitely many gaps},
  author={Its, A. R. and Matveev, V. B.},
  journal={Funct. Anal. Appl.},
  volume={9},
  pages={65-66},
  year={1975},
  doi={10.1007/BF01078185}
}

@article{ZakharovFaddeev1971,
  title={Korteweg-de Vries equation: A completely integrable Hamiltonian system},
  author={Zakharov, V. E. and Faddeev, L. D.},
  journal={Funct. Anal. Appl.},
  volume={5},
  pages={280-287},
  year={1971},
  doi={10.1007/BF01086739}
}

@article{dubrovin1976survey,
  title={Nonlinear equations of Korteweg-deVries type, finite zoned linear operators, and Abelian varieties},
  author={Dubrovin, B. A. and Matveev, V. B. and Novikov, S. P.},
  journal={Russ. Math. Surveys},
  volume={31},
  pages={59-146},
  year={1976},
  doi={10.1070/RM1976v031n01ABEH001446}
}

@article{novikov1995periodic,
  title={The Periodic Problem for the Korteweg-de Vries Equation},
  author={Novikov, S. P.},
  journal={World Scientific series in 20th century physics},
  volume={11},
  pages={326-336},
  year={1995},
  publisher={World Scientific Publishing}
}

@article{levitan1964determination,
  title={Determination of a differential equation by two of its spectra},
  author={Levitan, Boris Moiseevich and Gasymov, Mirabbas Geogdzhaevich},
  journal={Russian Mathematical Surveys},
  volume={19},
  number={2},
  pages={1},
  year={1964},
  publisher={IOP Publishing}
}

@article{aktosun2001small,
  title={Small-energy asymptotics for the Schrödinger equation on the line},
  author={Aktosun, Tuncay and Klaus, Martin},
  journal={Inverse problems},
  volume={17},
  number={4},
  pages={619},
  year={2001},
  publisher={IOP Publishing}
}

@article{aktosun2004inverse,
  title={Inverse scattering transform, KdV, and solitons},
  author={Aktosun, Tuncay},
  journal={Current trends in operator theory and its applications},
  pages={1-22},
  year={2004},
  publisher={Springer}
}

@book{belokolos1994algebro,
  title={Algebro-geometric approach to nonlinear integrable equations},
  author={Belokolos, Eugene D and Bobenko, Alexander I and Enolskii, Viktor Z and Its, Alexander R and Matveev, Vladimir B},
  volume={550},
  year={1994},
  publisher={Springer}
}

@article{novikov1988multidimensional,
  title={Multidimensional inverse spectral problem for the equation—$\Delta$ $\psi$+(v (x)—Eu (x)) $\psi$= 0},
  author={Novikov, Roman G},
  journal={Functional Analysis and Its Applications},
  volume={22},
  number={4},
  pages={263-272},
  year={1988},
  publisher={Springer}
}

@article{dubrovin1976non,
  title={Non-linear equations of Korteweg-de Vries type, finite-zone linear operators, and Abelian varieties},
  author={Dubrovin, Boris Anatol'evich and Matveev, Vladimir Borisovich and Novikov, Sergei Petrovich},
  journal={Russian Mathematical Surveys},
  volume={31},
  number={1},
  pages={59},
  year={1976},
  publisher={IOP Publishing}
}

@book{newton2013scattering,
  title={Scattering theory of waves and particles},
  author={Newton, Roger G},
  year={2013},
  publisher={Springer Science \& Business Media}
}

@article{NAKAMURA1974321,
title = {Solitons and wave trains in ferromagnets},
journal = {Phys. Lett. A},
volume = {48},
number = {5},
pages = {321-322},
year = {1974},
issn = {0375-9601},
doi = {10.1016/0375-9601(74)90447-2},
author = {K. Nakamura and T. Sasada},
}

@article{PhysRevB.15.3470,
  title = {Solitons in the continuous Heisenberg spin chain},
  author = {Tjon, J. and Wright, Jon},
  journal = {Phys. Rev. B},
  volume = {15},
  issue = {7},
  pages = {3470-3476},
  numpages = {0},
  year = {1977},
  month = {Apr},
  publisher = {American Physical Society},
  doi = {10.1103/PhysRevB.15.3470},
}

@article{LAKSHMANAN1976577,
title = {On the dynamics of a continuum spin system},
journal = {Physica A},
volume = {84},
number = {3},
pages = {577-590},
year = {1976},
issn = {0378-4371},
doi = {10.1016/0378-4371(76)90106-0},
author = {M. Lakshmanan and Th. W. Ruijgrok and C. J. Thompson},
}

@article{LAKSHMANAN197753,
title = {Continuum spin system as an exactly solvable dynamical system},
journal = {Phys. Lett. A},
volume = {61},
number = {1},
pages = {53-54},
year = {1977},
issn = {0375-9601},
doi = {10.1016/0375-9601(77)90262-6},
author = {M. Lakshmanan},
}

@article{shabat1972exact,
  title={Exact theory of two-dimensional self-focusing and one-dimensional self-modulation of waves in nonlinear media},
  author={Shabat, Aleksei and Zakharov, Vladimir},
  journal={Sov. Phys. JETP},
  volume={34},
  number={1},
  pages={62},
  year={1972}
}

@article{Takhtajan:1977rv,
    author = "Takhtajan, L. A.",
    title = {Integration of the Continuous Heisenberg Spin Chain Through the Inverse Scattering Method},
    doi = "10.1016/0375-9601(77)90727-7",
    journal = "Phys. Lett. A",
    volume = "64",
    pages = "235-237",
    year = "1977"
}

@article{dubrovin1976schrodinger,
  title={The Schrödinger equation in a periodic field and Riemann surfaces},
  author={Dubrovin, Boris Anatol'evich and Krichever, Igor Moiseevich and Novikov, Sergei Petrovich},
  journal={Dokl. Akad. Nauk SSSR},
  volume={229},
  number={1},
  pages={15--18},
  year={1976}
}

@book{levitan1987inverse,
  title={Inverse Sturm-Liouville Problems},
  author={Levitan, Boris Moiseevich},
  year={1987},
  publisher={VSP}
}

@article{DEMONTIS201835,
  author = {F. Demontis and S. Lombardo and M. Sommacal and C. van der Mee and F. Vargiu},
  title = {Effective generation of closed-form soliton solutions of the continuous classical Heisenberg ferromagnet equation},
  journal = {Commun. Nonlinear Sci. Numer. Simul.},
  volume = {64},
  pages = {35-65},
  year = {2018},
  issn = {1007-5704},
  doi = {10.1016/j.cnsns.2018.03.020},
}

@article{bian2015high,
  author={Bian, Dongfen and Guo, Boling and Ling, Liming},
  title={High-Order Soliton Solution of Landau--Lifshitz Equation},
  journal={Stud. Appl. Math.},
  volume={134},
  number={2},
  pages={181--214},
  year={2015},
  doi={10.1111/sapm.12051}
}

@article{zakharov1979equivalence,
  author={Zakharov, Vladimir Evgen'evich and Takhtajan, Leon Armenovich},
  title={Equivalence of the nonlinear Schrödinger equation and the equation of a Heisenberg ferromagnet},
  journal={Theor. Math. Phys.},
  volume={38},
  number={1},
  pages={17--23},
  year={1979},
  doi={10.1007/BF01030253}
}

@article{Bargheer:2008kj,
    author = "Bargheer, Till and Beisert, Niklas and Gromov, Nikolay",
    title = {Quantum Stability for the Heisenberg Ferromagnet},
    eprint = "0804.0324",
    archivePrefix = "arXiv",
    primaryClass = "hep-th",
    reportNumber = "AEI-2008-010, LPTENS-08-19, SPHT-T08-051",
    doi = "10.1088/1367-2630/10/10/103023",
    journal = "New J. Phys.",
    volume = "10",
    pages = "103023",
    year = "2008"
}

@article{Osborne:1986ua,
    author = "Osborne, A. R. and Bergamasco, L.",
    title = {The solitons of Zabusky and Kruskal revisited: Perspective in terms of the periodic spectral transform},
    journal = "Physica D",
    volume = "18",
    pages = "26--46",
    year = "1986",
    doi = {10.1016/0167-2789(86)90160-0}
}

@article{Osborne:1985,
    author = "Osborne, A. R. and Bergamasco, L.",
    title = {The small-amplitude limit of the spectral transform for the periodic Korteweg-de Vries equation},
    journal = "Nuovo Cim. B",
    volume = "85",
    pages = "229-243",
    year = "1985",
    doi = {10.1007/BF02721563}
}

@article{ForestMcL1982,
    author = {Forest, M. Gregory and McLaughlin, David W.},
    title = {Spectral theory for the periodic sine‐Gordon equation: A concrete viewpoint},
    journal = {J. Math. Phys.},
    volume = {23},
    pages = {1248-1277},
    year = {1982},
    doi = {10.1063/1.525509}
}

@article{PhysRevB.16.1763,
  title = {Solitons as nonlinear magnons},
  author = {Corones, James},
  journal = {Phys. Rev. B},
  volume = {16},
  pages = {1763--1764},
  year = {1977},
  doi = {10.1103/PhysRevB.16.1763}
}

@article{its1976explicit,
  title={Explicit formulas for the solutions of a nonlinear Schrödinger equation},
  author={Its, O. R. and Kotlyarov, V. P.},
  journal = {Dokl. Akad. Nauk Ukrainian SSR Ser. A},
  pages={965--968},
  year={1976}
}

@article{kotlyarov1976periodic,
  title={Periodic problem for the Schrödinger nonlinear equation},
  author={Kotlyarov, V. P.},
  journal = {Voprosy matematicheskoi fiziki i funkcionalnogo analiza},
  volume={1},
  pages={121--131},
  year={1976}
}

@article{landau1935equation,
  title={On the theory of the dispersion of magnetic permeability in ferromagnetic bodies},
  author={Landau, Lev Davidovich and Lifshitz, E},
  journal = {Phys. Z. Sowjet.},
  volume = {8},
  pages={153},
  year={1935}
}

@article{jost1947,
  title={Über die falschen Nullstellen der Eigenwerte der S-Matrix},
  author={Res Jost},
  journal = {Helv. Phys. Acta},
  volume = {20},
  pages={256},
  year={1947}
}

@article{Beisert:2003xu,
    author = "Beisert, N. and Minahan, J. A. and Staudacher, M. and Zarembo, K.",
    title = "{Stringing spins and spinning strings}",
    eprint = "hep-th/0306139",
    archivePrefix = "arXiv",
    reportNumber = "AEI-2003-044, UUITP-09-03, ITEP-TH-31-03",
    doi = "10.1088/1126-6708/2003/09/010",
    journal = "JHEP",
    volume = "09",
    pages = "010",
    year = "2003"
}

@article{fogedby1980solitons,
  title={Solitons and magnons in the classical Heisenberg chain},
  author={Fogedby, Hans C},
  journal={J. Phys. A},
  volume={13},
  number={4},
  pages={1467},
  year={1980},
  doi={10.1088/0305-4470/13/4/035}
}

@article{demontis2019continuous,
  title={The continuous classical Heisenberg ferromagnet equation with in-plane asymptotic conditions. I. Direct and inverse scattering theory},
  author={Demontis, Francesco and Ortenzi, Giovanni and Sommacal, Matteo and van der Mee, Cornelis},
  journal={Ricerche Mat.},
  volume={68},
  pages={145--161},
  year={2019},
  doi={10.1007/s11587-018-0394-8}
}

\end{bibtex}
\fi

\bibliographystyle{nb}
\bibliography{ismcurve}

\end{document}